\documentclass[letterpaper]{article}
\usepackage[12pt]{extsizes}
\usepackage{authblk}
\usepackage{setspace}
\usepackage[margin=0.75in]{geometry}
\usepackage{float}
\usepackage{amsmath, enumitem, dsfont}            
\usepackage{graphicx}
\usepackage{amssymb}
\usepackage{bbm}
\usepackage{amsfonts}
\usepackage{esint}
\usepackage{caption}
\usepackage{wrapfig}
\usepackage{subcaption}
\usepackage{cleveref}
\usepackage[mathscr]{euscript}
\usepackage{mdframed}
\usepackage{stix}
\usepackage{cases}
\usepackage[bottom]{footmisc}

\numberwithin{equation}{section}

\newenvironment{proof}[1][Proof]{\noindent\textbf{#1.} }{\ \rule{0.5em}{0.5em}}

\newcommand{\be}{\begin{equation}}
\newcommand{\ee}{\end{equation}}

\newtheorem{innercustomgeneric}{\customgenericname}
\providecommand{\customgenericname}{}

\newenvironment{customlemma}[1][]{\noindent\textbf{#1} }

\newenvironment{customthm}[1][]{\noindent\textbf{#1} }

\newcommand{\comment}[1]{}

\begin{document}

\begin{center}
\textbf{\Large{Mathematical Analysis of a Model of Blood Flow \\ through a Channel with Flexible Walls}\\~\\
\large{Marianna A. Shubov \footnote{Corresponding Author}
 and Madeline M. Edwards}}\\~\\
\small{
Department of Mathematics and Statistics\\
 University of New Hampshire\\
 33 Academic Way\\
Durham, NH 03825\\
\begin{align*}
\text{Emails:}&~\text{marianna.shubov@gmail.com}\\
&~\text{madelineedwards379@gmail.com}
\end{align*}
}

\large{\today}

\end{center}
	\begin{abstract}
	\onehalfspacing{
	The present research is devoted to the problem of stability of the fluid flow moving in a channel with flexible walls and interacting with the walls, which are subject to traveling waves. Experimental data shows that the energy of the flowing fluid can be transferred and consumed by the structure (the walls), which induces "traveling wave flutter." The problem of stability of fluid--structure interaction splits into two parts: $(i)$ stability of fluid flow in the channel with harmonically moving walls and $(ii)$ stability of solid structure participating in the energy exchange with the flow. Stability of fluid flow is the main focus of the research. It is shown that using the mass conservation and the incompressibility condition one can obtain the initial boundary value problem for the \textit{stream function}. The boundary conditions reflect the facts that $(i)$ for the axisymmetrical flow, there is no movement in the vertical direction along the axis of symmetry, and $(ii)$ there is no relative movement between the near boundary flow and the structure ("no--slip" condition). The closed form solution is derived and represented in the form of an infinite functional series.}

	\end{abstract}
	
\newpage
\onehalfspacing{
\section{Introduction}
\setcounter{section}{1}

The present paper is the first one of a two--paper set devoted to the mathematical analysis of the model describing incompressible fluid flowing through a relatively long channel with flexible walls. The model has its origin in a specific biological setting of blood moving through large arteries (or veins). Our choice of a model is motivated by the fact that the walls of the channel conveying fluid are subject to traveling waves, which comprises a complicated fluid--structure interaction. Experimental analysis shows that under specific conditions, the energy of the fluid flow can be transferred to the structure (the walls), which initiate ``traveling wave flutter''\cite{Huang1998}. Flutter is a specific type of instability that may occur due to fluid--structure interactions. Flutter is known to occur in many different areas of applied sciences such as marine and aerospace propulsion, outer skin panels of aircrafts, missiles, and aerospace vehicles, shell structures in jet pumps, heat exchanges and storage tanks.

In the present set of works, we focus on the problem of stability, which in turn splits into two parts: the first one is related to the stability of the fluid flow in the channel, whose walls undergo axisymmetric harmonic movement, and the second part is related to the stability of the channel wall structure, participating in the energy exchange with the flow. The present paper deals with the first part of the problem, i.e. we provide a closed form solution of the problem of \textit{stability of the fluid flow} in the channel with harmonically moving walls. The second paper will be devoted to the structure response on the pressure changes from the flow and will appear in our forthcoming work. The main achievement of the present paper is the derivation of the explicit formulas for the flow velocities, which yields the answer to the question of flow stability. We expect that the results obtained for two--dimensional configuration will be relevant to the investigation of flutter of partially collapsed tubes in a three--dimensional setting.

All fluid--conveying vessels in a human body are highly elastic and substantially deformable in their response to the pressure and viscous stress that the fluid exerts on them.  In propagation of the \textit{pulse wave} through the arterial system under normal conditions, the arteries are subject to \textit{positive} transmural (internal minus external) pressure, which keeps them inflated and stiff during the pulse cycle.  However, in the situation when fluid--conveying vessels are under the \textit{negative} (compressive) transmural pressure, the vessels often buckle and collapse. Buckled vessels are very flexible and small changes in fluid pressure induce large changes in their cross--sectional areas, which in turn induces such phenomena as \textit{flow limitation} and large amplitude self--excited oscillations. Self--excited oscillations of collapsible airways in lungs are responsible for respiratory wheezes during forced expiration, for speech production during flow--induced vibrations of the vocal chords, and for snoring sounds during deformation of the soft palate and pharyngeal wall \cite{Aittokallio1999, Auregan1995, Beck1995, Grotberg1989, Heil2003, Liu2007, Pevernagie2010, Wang2007, Whittaker2010}. Medical observations show that human snoring is due to the soft palate vibrations induced by the inspiratory flow: in some cases the soft palate becomes aeroelastically unstable and, when it bumps against the pharingeal walls, it closes the upper airways causing large changes of pressure in inspiratory airflow and producing its characteristic noise \cite{Auregan1995, Liu2007, Pevernagie2010}. A remarkable characteristic of snoring is that it occurs while the inspiratory flow rate is still increasing \cite{Huang1998} where as wheezing \cite{Grotberg1989} starts after the volume flow rate becomes limited. Thin--walled circular \textit{shell structures} containing or immersed in flowing fluid may be found in many engineering and biomechanical systems  \cite{Karagiozis2008, Paidoussis1972, Paidoussis1985, Paidoussis2005a}. Numerous studies on the aeroelasticity of cylindrical shells reflect the enormous interest on the effect of high--speed flow on the outer--skin panels of aircraft, missiles, and aerospace vehicles.

Now we briefly discuss some research works, where either new important results are presented or new techniques are developed. The authors of \cite{Case1960, Case1962, Shivamoggi1982}, Case and Shivamoggi, consider the stability problem for inviscid incompressible fluid flowing between infinite \textit{parallel plates}. The solution of the corresponding initial--value problem is represented as a normal mode decomposition. Carpenter and Garrad \cite{Garrad1986} investigate flow--induced surface instabilities using as a model a thin \textit{elastic plate} supported on a spring elastic foundation. The authors show that shear flow in the boundary layer gives rise to a fluctuating pressure, which leads to the \textit{energy transfer} from the main stream \textit{to the compliant surface}, and possibly to traveling--wave flutter. The important question raised by the authors is: "Why is the flow over compliant surfaces susceptible to so many types of instabilities? The simple answer is that the dynamic system in question consists of two coupled wave--bearing media: the flowing fluid and the solid flexible wall." 

Kumaran \cite{Kumaran1996, Kumaran1998} investigates the stability problem for the fluid flowing through \textit{flexible tubes} such as blood vessels. The author derives in \cite{Kumaran1996} the results on hydrodynamic stability for inviscid flow through a flexible tube, which are similar to the \textit{classical} results of Drazin and Reid \cite{Drazin1981}. Kumaran \cite{Kumaran1998} performs a linear stability analysis of a \textit{viscous} flow in a tube with viscoelastic walls and examines the mechanism that causes unstable oscillations \textit{in the wall}.

Heil and Jensen \cite{Heil2003} indicate that their investigations of steady flows in \textit{three--dimensional} collapsible tubes have already revealed many features, which cannot be predicted by the lower--dimensional models. Asymptotic analysis of flows in slightly buckled elastic tubes provides useful insight into the \textit{flow separation} and the development of flow instabilities.

The authors of \cite{Aittokallio1999}, Aittokallio, et.\ al., propose a mathematical model describing the operation of the lungs (both the inspiratory and the expiratory phases). The model can reasonably explain the normal behavior of the upper airways as well as the partial collapse and snoring. 

Beck, et.\ al.\ \cite{Beck1995} discuss the acoustic properties of snoring sounds and identify two distinctly different patterns: the "simple--waveform" and the "complex--waveform". The complex--waveform snore is characterized by repetitive, equally--spaced, train of sound structures, with decaying in time wave amplitude. Simple--waveform snores have a quasi--sinusoidal waveform, with a range of variants and no secondary internal oscillations. Grotberg and Gavriely \cite{Grotberg1989} analyze the model of the flow through a flexible channel focusing on flow--induced flutter oscillations producing wheezing breath sounds. The model predicts the critical fluid speed that initiates \textit{flutter of the wall.} 

In many applications, including the lung airways, the length of the flexible tube is shorter than the entrance length of the flow. Larose and Grotberg \cite{Larose1997} study the fluid--elastic (flutter) instability with a \textit{developing flow} in a compliant channel. They introduce the mathematical model and present analytical (long wave) study and numerical approximations for the solutions, which are in good agreement with representative experiments.

Pa\"{i}doussis and Li \cite{Paidoussis1993} present a comprehensive study of different types of \textit{pipes conveying} fluid (straight and curved pipes, cantilevered and supported pipes). The study of pipes conveying fluid was initiated by Ashley and Haviland \cite{Ashley1950} in an attempt to explain the vibrations observed in the Trans--Arabain Pipeline. The authors of \cite{Paidoussis1999, Paidoussis2005},  Pa\"{i}doussis, Semler, and Wadham-Gagnon, present an extended review on the stability of \textit{aspiring pipes}. To support their theoretical analysis of \cite{Paidoussis2005}, the authors conduct an experiment, involving two identical vertical, cantilevered, flexible pipes, submerged in a water tank; at the free end, each pipe is fitted with a $90^{\circ}$ elbow. One pipe discharges water at the free end, while the other aspirates. When the flow is turned on, the discharging pipe bent backward in the plane of the elbow as a result of the centrifugal force; however, the aspirating pipe remained vertical and undeformed. This observation supports the theoretical conclusion on \textit{the stability} of aspiring pipes.

It has been observed that subsonic incompressible flows were associated with loss of stability due to small amplitude \textit{divergence}, in contrast to the flutter occurring in supersonic flows. Pa\"{i}doussis and Denise \cite{Paidoussis1972} provide an analytical model and experimental results for clamped--clamped and clamped--free \textit{shells} conveying inviscid fluid. Karagiozis, et.\ al.\ \cite{Karagiozis2008} consider the model based on \textit{Fl\"{u}gge's shell equations} and potential flow theory. The experimental results and numerical simulations for clamped--clamped shells show that at sufficiently high flow velocity the system develops \textit{flutter}, with shell amplitudes much larger than the shell thickness. Amabili, et.\ al.\ \cite{Amabili2002} discuss the non--linear dynamics and stability of circular \textit{cylindrical shells} containing fluid flow. The non--linearities due to large amplitude shell motions are taken into account by using Donnell's non--linear shallow shell theory. Numerical results show that the system \textit{loses stability by divergence}.

Jensen and Heil \cite{Jensen2003} derive high Reynolds number asymptotics and perform numerical simulations to describe two--dimensional, unsteady, pressure--driven flow in a finite--length channel, one wall of which contains a \textit{section of membrane} under longitudinal tension. The authors show that oscillations can grow by extracting kinetic energy from the mean Poiseuille flow faster than by losing energy due to viscous dissipation. Ellis, et.\ al.\ \cite{Ellis1993} suggest \textit{a mechanical model}, which simulates the palate and upper airways. It comprises a round tube divided into two channels by a rigid partition equivalent to the hard palate, and a short length of flexible material analogous to the soft palate. When air is sucked through the tube by a pump above a critical speed, the flexible membrane vibrates and a sound similar to snoring is produced.  The vibrating membrane enables transfer of energy from flow to sound, which is an inevitable by-product of unsteady flow.

Huang and Williams \cite{Huang1999} study effective \textit{neuromuscular} functioning of the upper airways during sleep. They demonstrate that upper airway neural receptors sense the negative pressure generated by inspiration and "trigger" reflex muscle activation to sustain the airway that might otherwise collapse. In the authors' model the mechanics are coupled to the neuromuscular physiology through the generation of reflex wall stiffening proportional to gas pressure. Huang \cite{Huang1998} studies the collapse and subsequent self--excited oscillation of compliant tubes conveying fluids. The author considers a two--dimensional, inviscid, shear flow in a flexible channel of infinite length subject to linear \textit{traveling varicose waves}. One of the main findings is \textit{the reversal} of the collapsing tendency of compliant fluid passages proven for the flow having moderate velocity gradient near the wall. Huang \cite{Huang1998} also observes that the fluid pressure always has a component in phase with the wave slope causing wave drag and energy transfer from the flow to the perturbation waves, which is suggested as a mechanism for traveling wave flutter. In \cite{Huang2001}, Huang presents a linear analysis of a coupled system describing the Poiseuille flow interacting with a tensioned membrane of finite length. The author shows that $(i)$ flutter and divergence may occur at similar flow velocities, and \textit{may coexist}. Flutter and divergence exchange dominance when the wall properties change. $(ii)$ The elastic waves over a finite membrane have a standing wave pattern, which can be decomposed into \textit{upstream} traveling wave and \textit{downstream} traveling waves. The component of the downstream traveling wave is responsible for \textit{energy transfer} from the fluid to the wall.


\textit{Now we are in a position to outline briefly the content of the present paper.} Section 2 is devoted to the derivation of the mathematical model describing pressure perturbation in the horizontal channel. We formulate the linearized version of the Naiver--Stokes equations using the mass conservation law and incompressibility condition. We derive the boundary conditions for the model. The boundary conditions reflect an assumption about the axial symmetry of the unperturbed problem with respect to the centerline of the channel. We consider quite general behavior of the boundary of the channel and derive the equation that relates the vertical component of the flow velocity and the velocity of the points of a solid boundary of the channel (see equation (\ref{eq:2.9})). We emphasize here that we do not impose strong restrictions on the function describing the elevation of the lower boundary, which means that equation (\ref{eq:2.9}) can be used to analyze non--harmonic movement of the channel boundary.

In the present work, we assume that the channel (the artery) is a horizontal tube with the following properties of the walls reflecting physical origin of the model. There exists a large positive number, $R$ ($R\gg 1$) such that within a symmetric interval $(-R, R)$ the wall is flexible and the lower boundary moves "almost harmonically" with the amplitude that changes according to the law
\be \label{eq:1.1} g(x,t) = C_0 ~e^{i\omega(x-ct)}, \quad \omega >0, \quad x \in [-R,R],\ee
where $\omega$ is the wave number, $c$ is the speed of the wave crest propagation (see \cite{Huang1998}, \cite{Kumaran1996}, \cite{Kumaran1998}), and $g(x,t)$ denotes the transverse displacement of the channel wall at location $x$ and the moment in time, $t$; $C_0$ is a small, positive constant. We also assume that (1) here exists a positive number, $r:~r\ll R,$ such that outside the interval $[-R-r,~R+r]$, the wall is rigid, i.e. the vertical displacement on $\mathbb{R} \backslash [-R-r,~R+r]$ is reduced to zero, and (2) there exists a smooth function $g_0(x)$ that governs the transition of displacement $g(x,t)$ from formula (\ref{eq:1.1}) to zero, i.e. we set 
\be \label{eq1.4} g(x,t) = g_0(x) e^{i\omega(x-ct)} \ee
with $supp\{g_0(x)\} \in [-R-r, R+r]$ and $g_0(x) = g_0(0)$ for $x \in[-R,R]$ (see Fig. \ref{fig:0} below)

\begin{figure}[H]
\begin{center}
\includegraphics[scale=0.33] {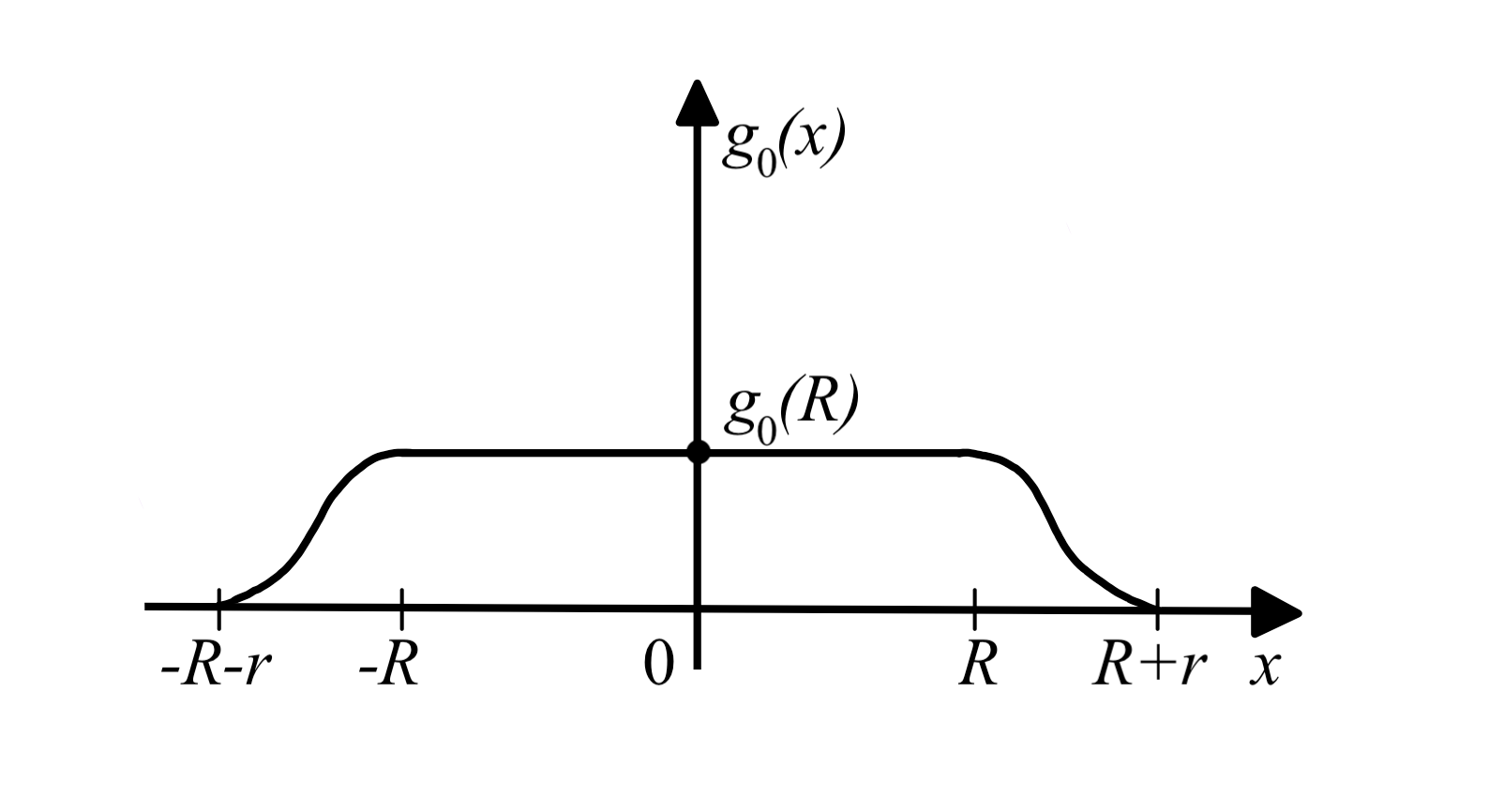} 
\caption{The amplitude profile function}
\label{fig:0}
\end{center}
\end{figure}
For more precise description of the amplitude profile function, see Assumption 2.1 below.

In Section 2, we present the initial-boundary value problem, IBVP: (\ref{eq:2.14}) -- (\ref{eq:2.17}) for the stream function $\psi(x,y,t)$, in which an "almost harmonic" behavior of the wall us taken into account.

In Section 3, we provide a modification of the IBVP given by (\ref{eq:2.14}) -- (\ref{eq:2.17}) and reduce it to a more tractable form. First, we apply the Fourier transformation to the equation and the boundary and initial conditions. As the result, we obtain an equation in which the partial derivatives, with respect to the $x$-variable, are replaced with polynomials with respect to the Fourier transform parameter, $k$ (see equation (\ref{eq:3.6})). Second, we apply, to the new equation, the Laplace transformation with respect to the time variable, $t$. The second integral transformation allows us to take into account the explicit expression for the unperturbed flow velocity profile, $U(y)$. By applying the double integral transformations to the stream function, we reduce IBVP (\ref{eq:2.14}) -- (\ref{eq:2.17}) to the new formulation (\ref{eq:3.9}) -- (\ref{eq:3.12}). The main equation (\ref{eq:3.9}) is an ordinary differential equation with respect to the variable $y$ (the width of the channel) with the complex parameter $k$ (the result of the Fourier transformation) entering both the equation and the boundary conditions. This equation has time--dependent right--hand side and non--homogeneous boundary conditions. We conclude this section with introducing and explicitly solving the boundary problem (\ref{eq:3.19}) -- (\ref{eq:3.21}) for the corresponding stationary solution (\ref{eq:3.21}). Taking into account the stationary solution allows us to specify the set of functions that we use as the initial data for the IBVP (\ref{eq:3.9}) -- (\ref{eq:3.12}) and reduce this problem to a new one having homogeneous boundary conditions (\ref{eq:3.25}) -- (\ref{eq:3.27}).

The rest of Section 3 is devoted to the construction of the inverse operator to the second order differential operator with the Dirichlet boundary conditions. The inverse operator is an integral operator, whose kernel is the Green's function. We derive an explicit formula for the Green's function and present a closed--form expression for the Fourier transform of the stream function denoted by $\widetilde \psi(k,y,t)$ (see formula (\ref{eq:3.30})). To examine the behavior of the stream function as a function of  $x$, we have to evaluate the inverse Fourier transform of (\ref{eq:3.30}).  Sections 4 -- 6 contain the main technical results of the paper. 

In the conclusion of the Introduction, we briefly explain the technical difficulties one has to overcome to get to these results. To this end, we need some explicit formulae. As follows from (\ref{eq:3.30}), the Fourier transform of the stream function denoted by, $\widetilde \psi(k,y,t)$, is represented as a sum of three terms. The inverse Fourier transform, which is the stream function $\psi$ itself, also contains three terms denoted by $I_1$, $I_2$, and $I_3$. To demonstrate our technical results let us consider the first term $I_1$, whose detailed analysis is given in Section 4. We outline the analytical tools needed to investigate the structure of $I_1$. Using Fourier and Laplace transforms we obtain the following formula for $I_1$ (see (\ref{eq:4.2})):
\be \label{eq:1.5} I_1 =- ~\frac{c\omega}{\pi} e^{-i\omega c t} \int_R^{R+r}  d\xi~ g_0'(\xi)  \int_{-\infty}^{\infty}dk~ e^{ikx} ~ \dfrac{ ~\sinh(k(1-y))}{\sinh(k)}  ~ \frac{\sin(\xi(k-\omega))}{k-\omega},\ee
where $R$ and $r$ are introduced in (\ref{eq:1.5}) in the description of amplitude function. Formula (\ref{eq:1.5}) is not convenient for the analysis of the qualitative behavior of $I_1$ as a function of $x$ and $y$. For this reason, we investigate the improper integral from (\ref{eq:1.5}) further and evaluate it using techniques of complex analysis. To carry out the integration in (\ref{eq:1.5}) with respect to $k$, we split $\sin(\xi(k-\omega))~$ into two exponential functions and prove that each of the two resulting integrals, which we denote by $\mathcal{I}$ and $\widetilde{\mathcal{I}}$ respectively, converges in the sense of the \textit{principal value}.  Each integral, $\mathcal{I}$ and $\widetilde{\mathcal{I}}$, can be approximated by sequences of integrals $\mathcal{I}_n$ and $\widetilde{\mathcal{I}}_n$, $n\rightarrow \infty$, i.e. $I_1 =\lim_{n\rightarrow \infty} (\mathcal{I}_n + \widetilde{\mathcal{I}}_n)$.  If we fix $n$, then the domain of integration for $\mathcal{I}_n$ (as well as $\widetilde{\mathcal{I}}_n$) is given by $(-\infty, \omega -\varepsilon_n) \cup (\omega + \varepsilon_n, \infty)$, with $\{ \varepsilon_n\}_{n=1}^{\infty}$ being a sequence of positive numbers converging to zero. For each $n$, the domain consists of two disjoint subdomains. In turn, $\mathcal{I}_n$ can be represented as a limit of a sequence of integrals $\mathcal{I}_n^S$, where each $\mathcal{I}_n^S$ is defined on the domain $(-S, \omega -\varepsilon_n) \cup(\omega + \varepsilon_n, S)$, i.e. $\mathcal{I}_n = \lim_{S\rightarrow \infty} \mathcal{I}_n^S$. For a given pair $(n, S)$ we consider a closed contour on the complex $k$-plane, obtained by connecting two segments $(-S, \omega -\varepsilon_n)$ and $(\omega +\varepsilon_n, S)$ by two semi-circles (see Fig. \ref{fig:intro_contour} below): $C_S(0)$ and $C_{\varepsilon_n} (\omega)$, where $C_S(0)$ is is a semi-circle centered at the origin of radius $S$ and $C_{\varepsilon_n}(\omega)$ is a semi-circle centered at $x = \omega$ of radius $\varepsilon_n$.
 
\begin{figure}[H]
\begin{center}
\includegraphics[scale=0.3] {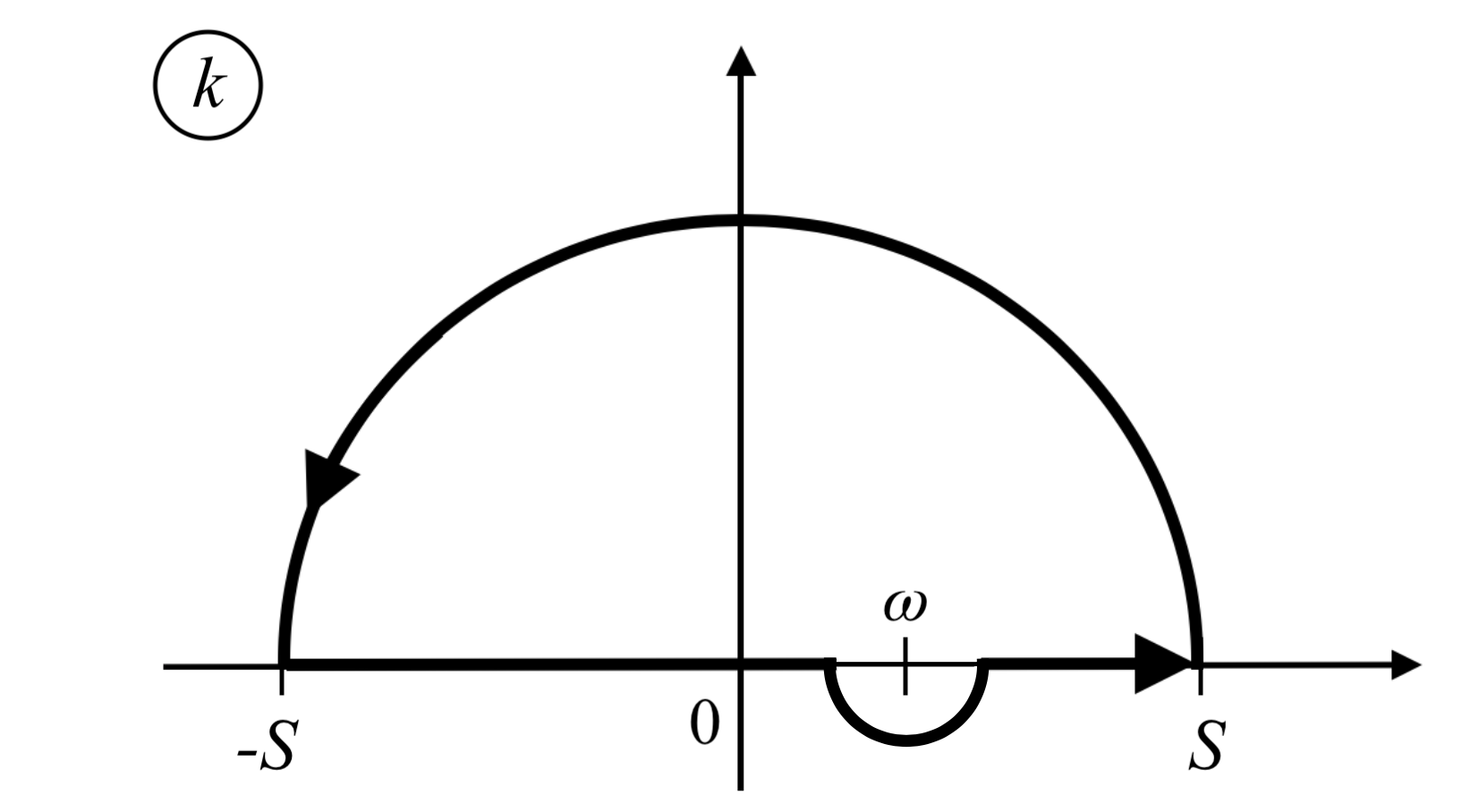} 
\caption{Closed semi-circle contour in the complex $k$-plane}
\label{fig:intro_contour}
\end{center}
\end{figure}

\noindent We show that the integral along $C_S(0)$ tends to zero as $S\rightarrow \infty$ and the integral along $C_{\varepsilon_n}(\omega)$ tends to zero as $\varepsilon_n \rightarrow 0$. Thus, we show that the integral $\mathcal{I}_n$ can be obtained as a limit of a sequence of the closed contour integrals. Similar steps can be carried out for evaluation of the integral $\widetilde{\mathcal{I}}_n$ by closing the contour of its integration in the lower half-plane. In turn, closed contour integrals can be evaluated by using the Residue theorem. As the result, we obtain that the term $I_1$ can be represented in the following form of an infinite series with respect to the residues:
\begin{align*} I_1 = \frac{\omega c}{\pi}~e^{-i\omega c t} \bigg\{ &e^{i\omega x}~\frac{\sinh(\omega (1-y))}{2i~\sinh(\omega)} \\+ &~i\sum_{m=0}^{\infty} e^{\pi m x}~\sin(\pi m y)
 \bigg( \int_{R}^{R+r}d\xi~g_0'(\xi)~e^{-\pi m \xi}~ \frac{im~\cos(\omega \xi)-\omega~\sin(\omega \xi)}{m^2 + \omega ^2}\bigg)\bigg\}. \end{align*}
We notice that for $|x|<R$ and $\xi >x$, the series converges at an exponential rate. Thus to analyze the behavior of $I_1$ as a function of $x,y,t,$ one can keep a finite number of terms in the sum and obtain the required accuracy.

The remaining part of the paper, Sections 5 and 6, are devoted to a detailed derivation of the formulas for the terms $I_2$ and $I_3$. We carry out the evaluation of the improper integrals entering $I_2$ and $I_3$ and represent them as infinite series making use of the Residue theorem. Collecting together the results of Sections 4 -- 6, we obtain that the stream function $\psi(x,y,t) = I_1 + I_2 + I_3$ can be represented in the form of an infinite series with an exponential rate of convergence (see Theorems 4.6, 5.3, and 6.2 below).

\section{Statement of the problem}

We consider a two--dimensional, inviscid, incompressible shear flow through a long channel with partially elastic and partially rigid walls.  The flexible parts of the channel walls undergo varicose heaving motions of small amplitudes in the form of a wave traveling in the direction of the flow.  We assume that the unperturbed flow is moving in the horizontal direction along the $x$-axis (see Fig.\ref{fig:1}). Using dimensionless variables, we set the channel height to equal two units and assume that the undisturbed flow is the symmetrization of the Couette flow with respect to the axis $y=1$.  Channel wall movements induce pressure disturbances within the fluid flow. \textit{Our goal is to find the analytical representation of the pressure distribution.}

Making use of the axial symmetry of the model, we reduce the 3--dimensional problem to 2--dimensional and introduce the following 2--dimensional vector--field. Let $\mathbf{q} (x,y,t) = \big(U(y) + u(x,y,t), ~v(x,y,t)\big)^T$ be the velocity vector with $U(y)$ being the given profile function (unperturbed flow) and $u(x,y,t)$ and $v(x,y,t)$ being the perturbations. The Euler momentum equation
written for $\mathbf{q}(x,y,t)$ generates a non--linear system of two partial differential equations with respect to $u(x,y,t)$ and $v(x,y,t)$. Eliminating the nonlinear terms from this system, we obtain the linearized version corresponding to the Euler momentum equation:
\be
\begin{aligned} \label{eq:2.1}
&u_t(x,y,t) + U(y)~ u_x(x,y,t)  +  v(x,y,t) ~U'(y) +p_x(x,y,t) = 0, \\
&v_t(x,y,t) + U(y)~v_x(x,y,t) + p_y(x,y,t) = 0.
\end{aligned}
\ee

\begin{figure}[H]
\begin{center}
\includegraphics[scale=0.2] {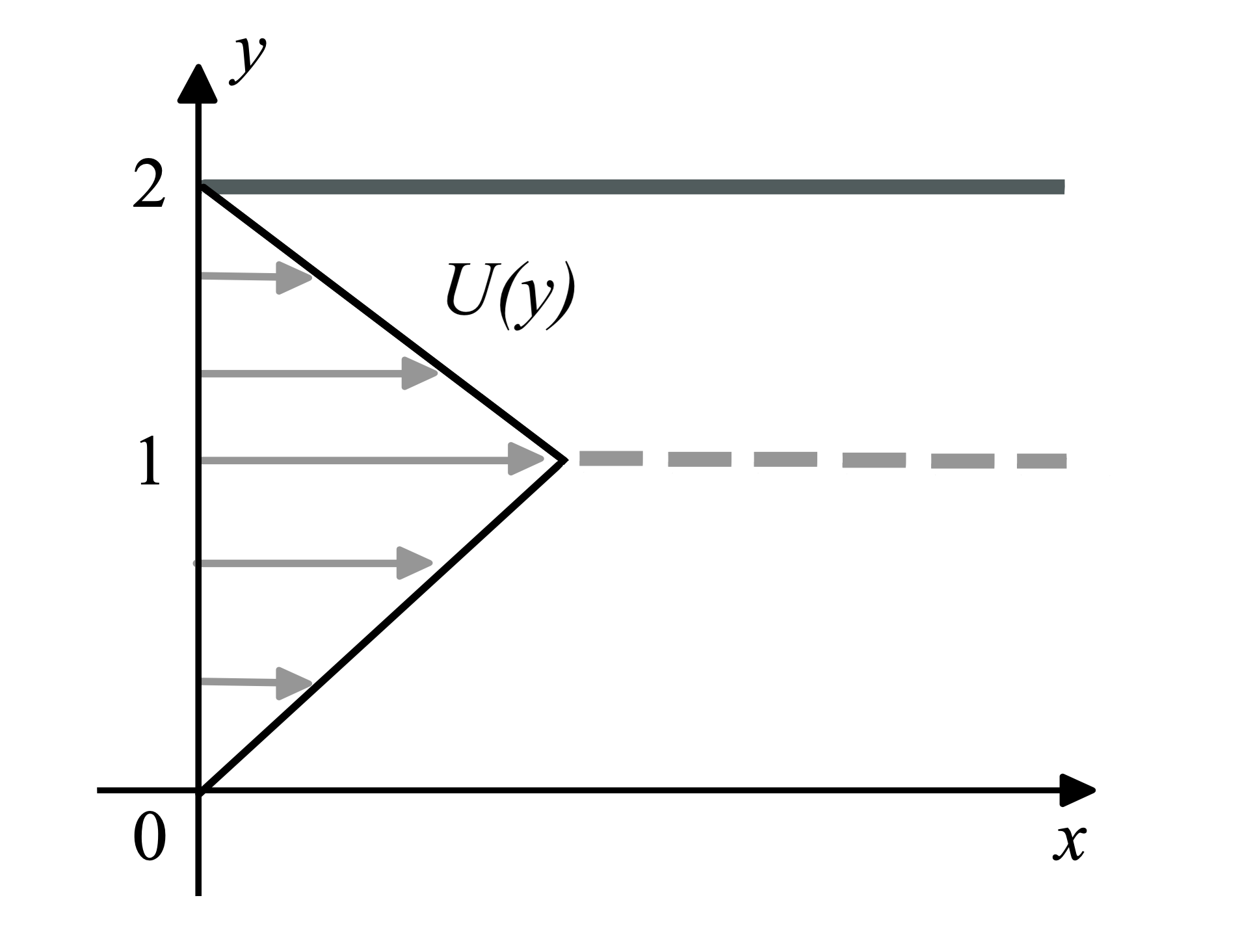} 
\caption{Symmetric undisturbed flow profile, $U(y).$}
\label{fig:1}
\end{center}
\end{figure}

The incompressibility condition, $(\nabla \cdot \mathbf{q}) (x,y,t) = 0, $  written in terms of $u(x,y,t)$ and $v(x,y,t)$, has the form: $u_x(x,y,t) + v_y(x,y,t) = 0.$ As is well--known \cite{Drazin1981}, this equation means that there exists a scalar function  $\psi(x,y,t) $ (called the \textit{stream function}) defined on the domain $(x,y)\in \Omega \in \mathbb{R} \times [0,1]$ such that 
\be \label{eq:2.2} 
u(x,y,t)=\psi_y(x,y,t)\quad \text{ and} \quad v(x,y,t) = -\psi_x(x,y,t).\ee

\noindent Rewriting system (\ref{eq:2.1})  in terms of the stream function and taking into account the symmetric modification of the Couette flow, i.e $U(y) =y$, $0\leq y\leq1$, we obtain a system of two equations,
\be \begin{aligned}
\psi_{yt}(x,y,t) + y ~\psi_{xy}(x,y,t) - \psi_x(x,y,t)+ p_x(x,y,t) =&~0, \label{eq:2.3}\\
\psi_{xt}(x,y,t) + y ~\psi_{xx} (x,y,t)- p_y(x,y,t)  =&~0.
\end{aligned} \ee 

\noindent Eliminating pressure terms from system (\ref{eq:2.3}), we obtain the following partial differential equation for the stream function: 
\be \label{eq:2.4} \bigg( \frac{\partial}{\partial t} + y \frac{\partial}{\partial x} \bigg)\bigg( \psi_{xx}(x,y,t) + \psi_{yy}(x,y,t) \bigg) = 0.\ee

\noindent We look for a particular solution for this equations satisfying the boundary conditions presented below.

 \textbf{The boundary conditions.} Since we consider an axially symmetric model, there is no vertical flow across the centerline $y=1$, i.e., the following boundary condition holds for $y=1$:
\be v(x,1,t) = -~\psi_x(x,1,t) =0 \label{eq:2.5}. \ee
The situation with the lower boundary of the channel $y=0$ is more complicated and the lower boundary condition requires derivation. We consider quite general behavior of the lower boundary of the channel and derive the equation that relates the vertical component of the flow velocity and the velocity of the points of a solid boundary of the channel (see equation (\ref{eq:2.10}) below). We emphasize here that we do not impose any restrictions on the function describing the elevation of the lower boundary, which means that equation (\ref{eq:2.10}) can be used to analyze non--harmonic movement of the channel boundary. 

Let the vertical displacement of the lower wall at a position, $x,$ and at a moment in time, $t,$ is given by the equation $y=g(x,t)$. Let $\vec{\nu}$ be the unit tangent vector at the point $(x, g(x,t))$ and $\mathbf{n}$ be the unit vector normal to the wall.  We have
$\vec{\nu}=\big(\nu_1,\nu_2 \big)$, $\nu_1^2+\nu_2^2=1$, and $\nu_2=\nu_1 g_x(x,t)$. Hence, the following representations are valid:
\be \label{eq:2.6}
\vec{\nu}=\frac{1}{\sqrt{1+g_x^2(x,t)}}\Big(1, g_x(x,t) \Big) 
\qquad \text{and} \qquad 
\mathbf{n}=\frac{1}{\sqrt{1+g_x^2(x,t)}}\Big(-g_x(x,t),1 \Big).
\ee

\begin{figure}[H]
\center
\includegraphics[scale=0.2] {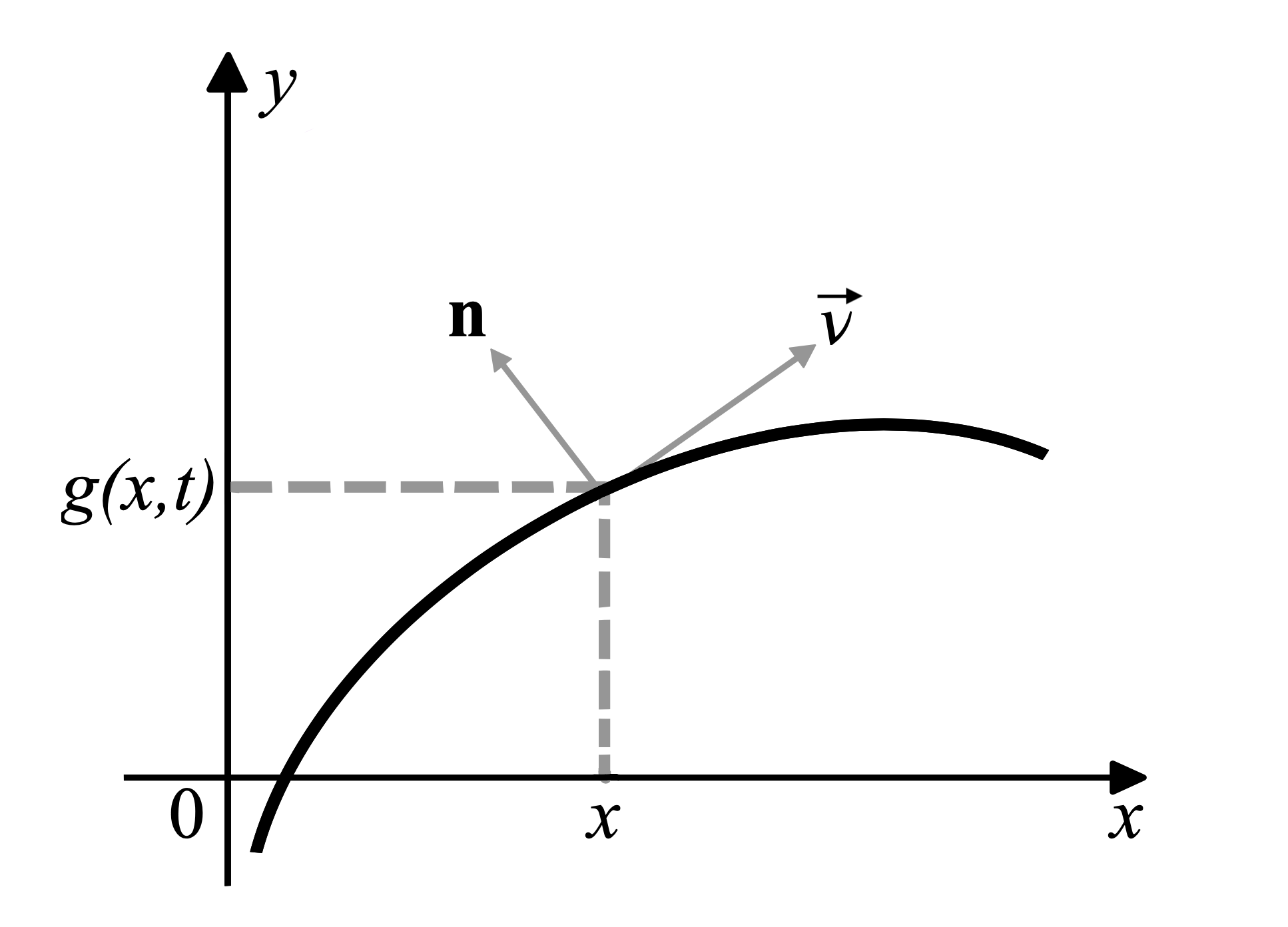}
\label{fig:1.1}
\caption{Geometry of the lower boundary}
\end{figure}

\noindent Assuming that the wall points move only in the vertical direction, for the velocity $\mathbf{w}(x,t)$ of the point $(x,g(x,t))$ we get
\be \label{eq:2.7} \mathbf{w}(x,t)=\big(0, g_t(x,t) \big).\ee
Since the normal component of the fluid velocity and the wall velocity are equal, i.e. $\mathbf{w}\cdot \mathbf{n} = \mathbf{q}\cdot \mathbf{n}$, we obtain that for small values of $y$ the following relations hold:
\[\mathbf{w} \cdot \mathbf{n}=
\frac{g_t(x,t)}{\sqrt{1+g_x^2(x,t)}}~,\quad
\mathbf{q} \cdot \mathbf{n}=
\frac{-\big[U(y)+u(x,y,t) \big]g_x(x,t)+v(x,y,t)}
{\sqrt{1+g_x^2(x,t)}},
\]
which yields the desired boundary condition:
\be  \label{eq:2.8}
g_t(x,t)=-\Big[U(y)+u(x,y,t) \Big]g_x(x,t)+v(x,y,t).
\ee

\noindent Let us simplify this relation by taking into account that $g(x,t)$ is a function of a small amplitude. For small values of $y$, we can use the following asymptotic approximations: 
\begin{align*} U(y) =~U(0) + y ~U_y(0) + \mathcal{O}(y^2), \qquad
&u(x,y,t) =~ u(x,0,t) + y~ u_y(x, 0, t ) + \mathcal{O}(y^2),\\
& v(x,y,t) = ~v(x,0,t) + y~ v_y(x,0,t) +  \mathcal{O}(y^2).
 \end{align*}
Keeping zero order terms, we obtain the approximation for the boundary condition of (\ref{eq:2.8}):
\be \label{eq:2.9}
g_t(x,t)+\Big[U(0)+u(x,0,t) \Big]g_x(x,t)=v(x,0,t).
\ee
Since in our model $u(x,0,t)=0$, the following modification of (\ref{eq:2.9}) holds:
\be \label{eq:2.10} g_t(x,t)+U(0)~ g_x(x,t)=v(x,0,t). \ee
In the present work, we consider a specific function, $g(x,t)$, describing the lower boundary elevation.

 \textbf{Assumption 2.1.} \textit{The lower boundary of the channel has a long flexible part beyond which the wall is rigid. The vertical displacement of the wall, $y=g(x,t)$, satisfies the following conditions: ($i$) for $x\in [-R, R]$ with $R$ being a large positive constant, the following representation is valid:}
\be \label{eq:2.11} g(x,t) =C_0 e^{i\omega(x-ct)}, \quad C_0 \ll 1, \quad \omega >0,\ee
\textit{where $C_0$ is a small positive amplitude, $\omega$ is a positive wave number, and $c$ is the speed of wave propagation;  ($ii$) there exist a positive constant $r$ ($r\ll R)$ and a twice continuously differentiable even function $g_0(x)$ such that (see Fig. \ref{fig:0} above):}
\be \begin{aligned} \label{eq:2.12}  
&g(x,t) = g_0(x) e^{i\omega(x-ct)},\quad x\in[-R-r, -R)\cup(R,R+r], \\
&g_0(-R-r) = g_0(R+r) = 0, \quad g_0(-x) = g_0(x) = C_0,~~x\in[-R,R],\\
&g_0(x) = 0 \text{~~for~~} x\in (-\infty, -R-r) \bigcup (R+r, \infty).
\end{aligned}\ee

\noindent Taking into account that $U(0)=0$, we rewrite (\ref{eq:2.10}) in terms of the stream function as
\be \label{eq:2.13}\psi_x(x,0,t) = i \omega c~ g(x,t). \ee

\noindent Note that in the problem with a rigid wall channel, we would have had the condition $\psi_x(x,0,t) =0.$ In the sequel, we consider (\ref{eq:2.13}) as the boundary condition at $y=0$. Thus, in the present paper we focus on the following initial boundary--value problem (IBVP) for the stream function:
\begin{numcases}{}
 \bigg(\frac{\partial}{\partial t} + y \frac{\partial}{\partial x} \bigg) \bigg( \psi_{xx}(x,y,t) + \psi_{yy}(x,y,t)\bigg) = 0,\quad x\in(-\infty, \infty),~~0\leq y\leq 1,~~t\geq0,\label{eq:2.14}\\
 \psi_x(x,0,t) = i \omega c ~ g(x,t), \label{eq:2.15}\\
 \psi_x(x,1,t) = 0, \label{eq:2.16}\\
 \psi(x,y,0) = F(x,y). \label{eq:2.17}
\end{numcases}


\section{Reformulation of IBVP using Fourier and Laplace \\integral transformations}

In this section, we apply two integral transformations to IBVP (\ref{eq:2.14}) -- (\ref{eq:2.17}), which will allow us to reduce IBVP involving a partial differential equation to the boundary--value problem involving a parameter--dependent ordinary differential equation.

Assuming that the stream function and its higher order derivatives tend to $0$ as $x \rightarrow \pm ~\infty$, we use the following formulae for the Fourier transforms for the function and its derivatives 
\be
\int_{-\infty}^{\infty}e^{-ikx} \bigg(\frac{\partial}{\partial x}\bigg)^n~\psi(x,y,t) dx = ~(-ik)^n ~\widetilde \psi(k,y,t),~~~ n= 0, 1, ....,
\label{eq:3.1}
\ee
where $\widetilde \psi(k,y,t)$ is the Fourier transform of $\psi(x,y,t)$. Applying the Fourier transformation to equation (\ref{eq:2.14}) yields a new equation
\begin{align}
\label{eq:3.2}
-k^2~ \widetilde\psi_t(k,y,t) + \widetilde \psi_{yyt}(k,y,t) -ik^3y~ \widetilde \psi (k,y,t)+ iky~ \widetilde \psi_{yy} (k,y,t)=&~0.
\end{align}
Let $\Psi(k,y, \lambda)$ be the Laplace transform of $\widetilde\psi(k,y,t)$ with respect to time variable, $t$. Applying the Laplace transformation to both sides of equation (\ref{eq:3.2}) we obtain
\be \label{eq:3.3}
-ik^3y~ \Psi(k,y,\lambda) + iky~ \Psi_{yy}(k,y,\lambda) -\lambda k^2~ \Psi(k,y,\lambda) + \lambda~ \Psi_{yy}(k,y,\lambda) = \bigg(\frac{\partial^2}{\partial y^2} ~-~ k^2 \bigg) \widetilde \psi(k,y,0), 
\ee
where $\widetilde \psi(k,y,0)$ is the Fourier transform of the initial state. This equation can be written in the form
\be
 \bigg(\frac{\partial^2}{\partial y^2} - k^2\bigg)\Psi(k,y,\lambda) = \frac{1}{\lambda + iky}\bigg(\frac{\partial^2}{\partial y^2} - k^2 \bigg) \widetilde \psi(k,y,0), \label{eq:3.4}
\ee
 Notice, the only influence of the Couette--like shape of the flow is the $y-$term in the denominator on the right--hand side of (\ref{eq:3.4}). If instead of Couette--like profile, we have an arbitrary axisymmetric profile, $U(y)$, then in place of $y$, we would have had the function, $U(y)$, and thus have a factor of $\big(\lambda + ik U(y)\big)^{-1}$ in equation (\ref{eq:3.4}). 

The unknown function, $\Psi(k,y,\lambda)$ in (\ref{eq:3.4}), being a function of $y$, $0\leq y\leq 1$, depends on two complex parameters, $k$ and $\lambda$. Applying the inverse Laplace transformation to both sides of equation (\ref{eq:3.4}) yields 
\be \label{eq:3.5}
\frac{1}{2\pi i} \int_{\gamma} \bigg(\frac{\partial^2}{\partial y^2} - k^2\bigg)~ \Psi(k,y,\lambda) ~e^{\lambda t} ~d\lambda = \frac{1}{2\pi i } \int_{\gamma} \frac{e^{\lambda t}}{\lambda + i k y} \bigg( \frac{\partial^2}{\partial y^2} -k^2\bigg)~ \widetilde \psi(k,y,0) ~d\lambda,
\ee
where as the contour of integration can be taken as any vertical line, located in the open right half--plane of the complex $\lambda$-plane (see Fig.\ref{fig:2.1} below).
\begin{figure}[H]
\begin{center}
\includegraphics[scale=0.25] {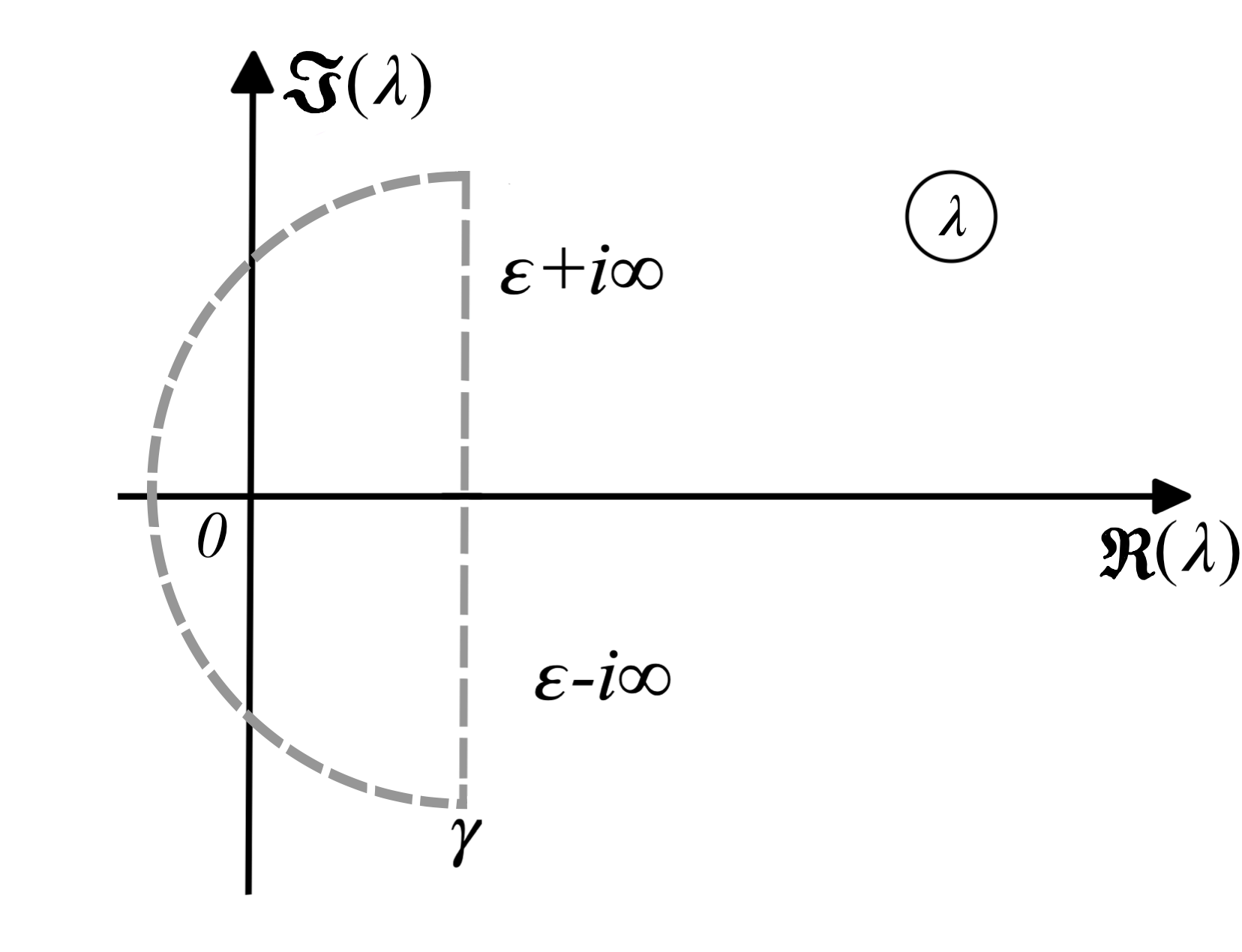} 
\end{center}
\caption{Contour for inversion of Laplace transform}
\label{fig:2.1}
\end{figure}

\noindent Changing of order of integration with respect to $\lambda$ and differentiation with respect to $y$, we reduce (\ref{eq:3.5}) to the following:
\be \label{eq:3.6}
\bigg(\frac{\partial^2}{\partial y^2} - k^2\bigg)\widetilde \psi(k,y,t) = \bigg(\frac{\partial^2}{\partial y^2} - k^2\bigg)\widetilde \psi(k,y,0) ~\bigg[ \frac{1}{2\pi i } \int_{\gamma} \frac{e^{\lambda t}}{\lambda + i ky}~ d\lambda\bigg].
\ee
Closing the contour of integration in the $\lambda$-complex plane and using the Residue theorem to evaluate the contour integral at the right--hand side of equation (\ref{eq:3.6}), we obtain the desired form of the differential equation:
\be \label{eq:3.7}
\bigg(\frac{\partial^2}{\partial y^2} - k^2\bigg) \widetilde \psi (k,y,t) = e^{-ikyt} \bigg(\frac{\partial^2}{\partial y^2} - k^2\bigg) \widetilde \psi (k,y,0).
\ee
Notice that at $t=0$, the left and right sides of the equation are identical.  Equation (\ref{eq:3.7}) is an \textit{ordinary differential equation} with respect to $y$ parametrically depending on $k\in \mathbb{C}$. However, we do not have the boundary conditions for $\widetilde \psi(k,y,t)$; we have boundary conditions only for $\psi_x(k,y,t)$ when $y=0$ and $y=1$ (see (\ref{eq:2.15}) and (\ref{eq:2.16})). Since $\psi_x(x,1,t) =0$, we immediately obtain the boundary condition at $y=1$,
$ \widetilde \psi(k,1,t) =0.$

To derive the boundary condition at $y=0$, we have to take into account the properties of the force profile function, $g_0(x)$ (see Assumption 2.1). We recall that according to the \textit{physical origin of the model}, the channel walls are flexible for $|x|\leq R+r$ and rigid for $|x|>R+r.$ Let us evaluate the Fourier transformation of the boundary condition (\ref{eq:2.15}). Taking into account that $g_0(x)$ is an even function with compact support, we have
\begin{align}
ik~\widetilde \psi(k,0,t) =& ~i\omega c \int_{-\infty}^{\infty} g_0(\xi)~ e^{i\omega (\xi-ct)} e^{-ik\xi} d\xi \nonumber \\
=&~ -2i \omega c ~e^{-i \omega c t} \int_{R}^{R+r}g_0'(\xi) ~\frac{\sin(\xi(\omega - k))}{\omega - k}~d\xi.~\label{eq:3.8}
\end{align}

\noindent Collecting equation (\ref{eq:3.7}), boundary condition at $y=1$ and condition (\ref{eq:3.8}) we obtain the following reformulation of IBVP (\ref{eq:2.14}) -- (\ref{eq:2.17}) for $\widetilde \psi(k,y,t)$:
\begin{numcases}{}
 \bigg(\frac{\partial^2}{\partial y^2} -k^2  \bigg)\widetilde \psi(k,y,t) = e^{-ikyt} \bigg(\frac{\partial^2}{\partial y^2} -k^2  \bigg)\widetilde \psi(k,y,0),\quad 0\leq y\leq,~~k\in \mathbb{C},~~ t\geq 0,\label{eq:3.9}\\
 \widetilde \psi(k,0,t) =-~\frac{2\omega c}{k}~e^{-i\omega c t} \int_{R}^{R+r} g'_0(\xi) ~\frac{\sin(\xi(k-\omega))}{k-\omega} d\xi, \label{eq:3.10} \\
\widetilde \psi(k,1,t) = 0,\label{eq:3.11} \\
\widetilde \psi(k,y,0) = \widetilde F(k,y). \label{eq:3.12}
\end{numcases}

\noindent \textit{IBVP (\ref{eq:3.9}) -- (\ref{eq:3.12}) is the main object of interest for the rest of the paper.}

\textbf{Reduction to the problem with time--independent boundary conditions.} Let us introduce a new function
\be \label{eq:3.13} Y(k,y,t) = e^{i\omega c t } ~ \widetilde \psi(k,y,t).\ee
Since $Y(k,y,0) =\widetilde \psi(k,y,0),$ the problem for $Y(k,y,t)$ can be represented in the form
\begin{numcases}{}
 \bigg(\frac{\partial^2}{\partial y^2} -k^2  \bigg)Y(k,y,t) = e^{-i(ky-\omega c)t} \bigg(\frac{\partial^2}{\partial y^2} -k^2  \bigg)Y(k,y,0), \label{eq:3.14}\\
 Y(k,0,t) =-~\frac{2\omega c}{k}~\int_{R}^{R+r} g'_0(\xi)~ \frac{\sin(\xi(k-\omega))}{k-\omega} d\xi, \label{eq:3.15}\\
Y(k,1,t) = 0, \label{eq:3.16}\\
Y(k,y,0) = \widetilde F(k,y). \label{eq:3.17}
\end{numcases}
To reduce IBVP (\ref{eq:3.14}) -- (\ref{eq:3.17}) to IBVP with homogeneous boundary conditions, we introduce the steady state solution, $Y_0(k,y)$, that satisfies the following boundary--value problem:
\noindent
\begin{numcases}{}
 \bigg(\frac{\partial^2}{\partial ^2y} -k^2  \bigg)Y_0(k,y) = 0, \label{eq:3.18} \\
Y_0(k,1)= 0, \label{eq:3.19} \\
Y_0(k,0)= -~\frac{2\omega c}{k} \int^{R+r}_{R} g'_0(\xi) ~\frac{\sin(\xi(k-\omega))}{k-\omega} ~d\xi. \label{eq:3.20}
\end{numcases}
This problem has a unique solution that can be given by the following explicit formula:
\be \label{eq:3.21} Y_0(k,y) = -~\frac{2 \omega c ~\sinh(k(1-y))}{k~\sinh(k)}\int_R^{R+r} g'_0(\xi) ~\frac{\sin(\xi(k-\omega))}{k-\omega} ~d\xi. \ee
It can be readily seen that the steady state solution satisfies the initial--boundary value problem (\ref{eq:3.14}) -- (\ref{eq:3.18}), in which the initial state is simply $Y_0(k,y)$.  

Now we are in a position to describe the class of initial data.

\textbf{Assumption 3.1.} \textit{The initial state of the system is a function, whose Fourier transform is a small perturbation of the steady--state solution, i.e.}
\be \label{eq:3.22} \widetilde F(k,y) = \widetilde f(k,y) + Y_0(k,y),\ee
\textit{where $\widetilde f(k,y)$ is a smooth compactly supported function of $y$. }

\indent \textbf{Reduction to the problem with homogenous boundary condition.} Let us introduce a new function, $Z(k,y,t)$, by the formula
\be \label{eq:3.23} Z(k,y,t) = Y(k,y,t) - Y_0(k,y), \ee 

\noindent Taking into account (\ref{eq:3.14}) and (\ref{eq:3.18}), we obtain the equation for $Z(k,y,t)$:
\begin{align}
\bigg(\frac{\partial^2}{\partial y^2} - k^2 \bigg) Z(x,y,t) =&~e^{-i(ky-\omega c)t}  \bigg( \frac{\partial^2 }{\partial y^2} - k^2 \bigg)Y(k,y,0) \nonumber \\
=& ~e^{-i(ky-\omega c)t}  \bigg( \frac{\partial^2 }{\partial y^2} - k^2 \bigg)\widetilde \psi(k,y,0). \label{eq:3.24}
\end{align}
Using (\ref{eq:3.15}) -- (\ref{eq:3.20}), we obtain that the boundary conditions at $y=0$ and $y=1$ are  \[Z(k,0,t) =~ Y(k,0,t) - Y_0(k,0) = 0,~~ Z(k,1,t) =~0,\] and from (\ref{eq:3.22}) and (\ref{eq:3.23}), we obtain the initial condition $ Z(k,y,0) = Y(k,y,0) - Y_0(k,y) = \widetilde f(k,y).$

Thus, the IBVP for $Z(k,y,t)$ can be written in the following form:
\begin{numcases}{}
\bigg(\frac{\partial^2}{\partial y^2} - k^2 \bigg) Z(k,y,t) =~ e^{-i(ky - \omega c) t} \bigg(\frac{\partial^2}{\partial y^2} -k^2 \bigg) \widetilde \psi(k,y,0),\label{eq:3.25}\\
Z(k,0,t)= ~Z(k,1,t) =~0, \label{eq:3.26}\\
Z(k,y,0) =~\widetilde f(k,y). \label{eq:3.27}
\end{numcases}

\textbf{Solving problem (\ref{eq:3.25}) -- (\ref{eq:3.27}) via Green's function.} We construct an explicit solution of equation (\ref{eq:3.25}) satisfying the Dirichlet boundary conditions (\ref{eq:3.27}). The initial conditions (\ref{eq:3.27}) is incorporated into equation (\ref{eq:3.25}) in which
$\widetilde \psi(k,y,0) = \widetilde f(k,y).$

As is known \cite{Roach1995, Stakgold1998}, the solution of the boundary problem
\[ W''(k,y) - k^2 ~W(k,y) = \Phi(k,y),~~~~W(k,0) = W(k,1) = 0,\]
can be given in terms of the Green's function as
\[ W(k,y) = \int_0^1 G(\eta,y)~ \Phi(\eta,k) ~d\eta.\]
\noindent The following properties must be satisfied for $G(\eta, y)~\cite{Roach1995, Stakgold1998}$: $(i)$ symmetry, i.e. $G(\eta,y) = G(y,\eta)$; $(ii)$ continuity, i.e. $G(\eta,y) \rightarrow G(y,y)$ as $\eta \rightarrow y$; $(iii)$ a unit jump of the derivative, i.e. $G_y(\eta,y)|_{y=\eta +0} - G_y(\eta,y)|_{y=\eta - 0} = 1$; $(iv)$ the boundary conditions: $G(\eta, 0) = G(\eta,1) = 0$; $(v)$ the equation: $G_{yy}(\eta,y) - k^2 G(\eta,y) = \delta(\eta -y)$ with $\delta(\cdot)$ being the Dirac delta function.
\noindent It can be verified directly that the function given by explicit formula 
\be\label{eq:3.28}G(\eta, y) =  \frac{1}{k~\sinh(k)}\begin{cases} 
     -\sinh(k(1-\eta)) \sinh(ky), \quad& y<\eta \\
     - \sinh(k\eta) \sinh(k(1-y)), \quad& y>\eta \\
    \end{cases}
\ee
satisfies all properties $(i)$ -- $(v)$. Using this Green's function, we obtain the following representation for a solution of system (\ref{eq:3.25}) and (\ref{eq:3.27}):
\begin{align}
Z(k,y,t) =&~ e^{i\omega c t} \int_0^1 e^{-ik\eta t}~G(y,\eta)~ \bigg(\frac{\partial^2}{\partial \eta^2} - k^2 \bigg) \widetilde \psi(k,\eta,0)~d\eta \nonumber \\
=&~-e^{i\omega c t}~ \frac{\sinh(k(1-y))}{k~\sinh(k)}~\int_0^y e^{-ik\eta t}~\sinh(k\eta)~\bigg(\frac{\partial^2}{\partial \eta^2} - k^2 \bigg) \widetilde \psi(k,\eta,0) ~d\eta \nonumber \\
&~ -e^{i\omega c t} ~\frac{\sinh(ky)}{k~\sinh(k)}~\int_y^1 e^{-ik\eta t}~\sinh(k(1-\eta))~\bigg(\frac{\partial^2}{\partial \eta^2} - k^2 \bigg) \widetilde \psi(k,\eta,0) ~d\eta \label{eq:3.29}
\end{align}
which yields the desired result for $\widetilde \psi(k,y,t)=e^{-i\omega c t} \big[Z(k,y,t) + Y_0(k,y) \big]$:
\begin{align}
\widetilde \psi(k,y,t) =&-\frac{\sinh(k(1-y))}{k~ \sinh(k)}~ \int_0^y e^{-ik\eta t}~ \sinh(k\eta)~\bigg(\frac{\partial^2}{\partial \eta^2} - k^2\bigg) \widetilde f(k,\eta)~d\eta \nonumber \\
&-\frac{\sinh(ky)}{k~\sinh(k)}~\int_y^1 e^{-ik\eta t}~\sinh(k(1-\eta))~\bigg(\frac{\partial^2}{\partial \eta^2} - k^2 \bigg) \widetilde f(k,\eta) ~d\eta \nonumber \\
&- e^{-i\omega c t}~ \frac{2\omega c~\sinh(k(1-y))}{k~\sinh(k)}~\int_R^{R+r} g_0'(\xi)~\frac{\sin(\xi(k - \omega))}{k-\omega}~d\xi. \label{eq:3.30}
\end{align}

\section{Evaluation of the inverse Fourier transform of \\the integral involving amplitude function $g_0(x)$}

Our goal is to find the space--time representation for the vertical component of the velocity perturbation $v(x,y,t)$, which is related to the derivative of the stream function by formula (\ref{eq:2.2}). It means that, in fact, we are looking for the inverse Fourier transform of the function $~k~\widetilde \psi(k,y,t)$. It is convenient to introduce the following notation for the inverse Fourier transform:
\be \label{eq:4.1}
i\psi_x(x,y,t) = \frac{1}{2\pi}\int_{-\infty}^{\infty} e^{ikx}~k~ \widetilde \psi(k,y,t)~dk \equiv I_1+I_2+I_3,
\ee
where
\begin{align} \label{eq:4.2} I_1 =& -~\frac{c\omega}{\pi} e^{-i\omega c t} \int_{-\infty}^{\infty} e^{ikx} \bigg[ \dfrac{ ~\sinh(k(1-y))}{\sinh(k)} \int_R^{R+r} g_0'(\xi)~ \frac{\sin(\xi(k-\omega))}{k-\omega} d\xi\bigg]~dk,\\
 I_2 =& -~\frac{1}{2\pi} \int_{-\infty}^{\infty} e^{ikx} \bigg[ \dfrac{\sinh(k(1-y))}{\sinh(k)} \int_0^y e^{-ik\eta t} \sinh(k\eta)\bigg(\frac{\partial^2}{\partial \eta^2} -k^2 \bigg) \widetilde f(k,\eta)~d\eta \bigg] dk, \label{eq:4.3}\\
 I_3 =& -~\frac{1}{2\pi} \int_{-\infty}^{\infty} e^{ikx} \bigg[ \dfrac{\sinh(ky)}{\sinh(k)} \int_y^1 e^{-ik\eta t} \sinh(k(1-\eta))\bigg(\frac{\partial^2}{\partial \eta^2} -k^2 \bigg) \widetilde f(k,\eta)~d\eta \bigg] dk.  \label{eq:4.4}
\end{align}

\noindent In this section, we focus on the evaluation of the double integral (\ref{eq:4.2}), which can be represented in the form 
\[ I_1 =~ -\frac{\omega c}{\pi}~e^{-i\omega ct}~\int_{R}^{R+r} d\xi~ g_0'(\xi)~ \mathbb{I}_1(\xi,x),\]
where
\be \label{eq:4.5} \mathbb{I}_1 = \int_{-\infty}^{\infty} dk~e^{ikx}~\frac{\sinh(k \alpha)}{\sinh(k)}~\frac{\sin(\xi(k-\omega))}{k-\omega}, \qquad \omega >0,~~~ 0<\alpha \equiv 1-y<1.\ee
Our first statement is the following technical result.

\indent \begin{customlemma}{}\label{lemma4.1}\textbf{Lemma 4.1.} 
\textit{Let $\mathscr{A}(k)$ be a function defined by the formula}
\be \label{eq:4.6}
\mathscr{A}(k) =~\frac{\sinh(\alpha k)}{\sinh(k)}, \qquad \alpha = 1-y<1. 
\ee
\textit{Let $\Gamma_N$ be a semi--circle in the upper--half plane of the $k$-complex plane centered at the origin and of radius $r = \pi \big( N + \frac{1}{2} \big)$, $N \in \mathbb{N}^+$. Then for each $N$ the following estimate holds:}

\be |\mathscr{A}(k)|\leq C_0 < \infty,~~~~~ k\in \Gamma_N, ~~N \in \mathbb{N}^+, \label{eq:4.7} \ee
\textit{with $C_0$ being an absolute constant. A similar estimate holds when $k$ belongs to a symmetric semi--circle located in the lower half--plane.}
\end{customlemma}

\indent \begin{proof}
Since $\mathscr{A}(-\overline{k}) = \mathscr{A}(\overline{k}) = \overline{\mathscr{A}(k)},$ it suffices to prove the lemma when $k$ belongs to the quarter of the circle located in the first quadrant of the complex $k$-plane.

\begin{figure}[H]
\begin{center}
\includegraphics[scale=0.30] {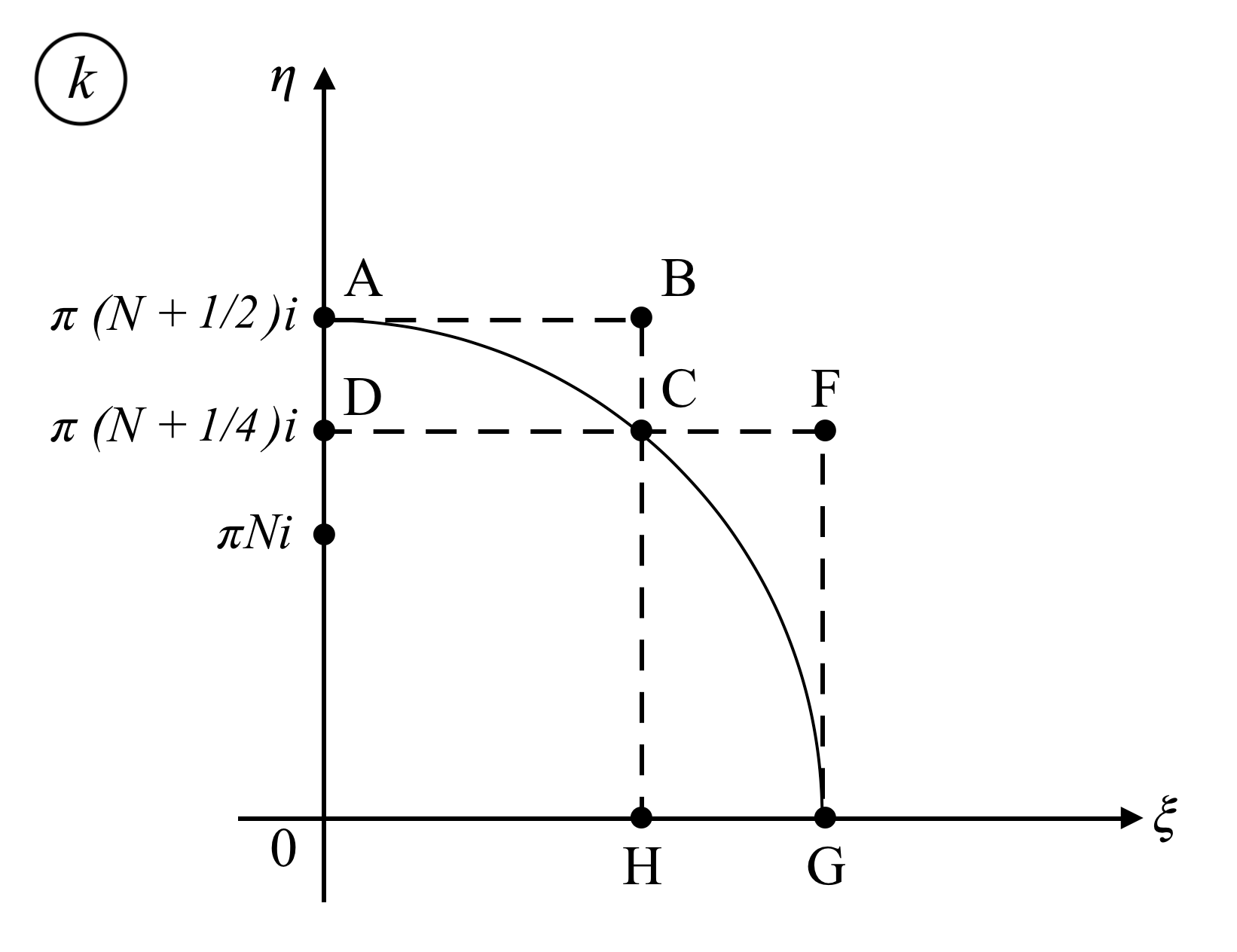} 
\caption{Geometrical setting for evaluation of $\mathscr{A}(k)$}
\label{fig:2.3}
\end{center}
\end{figure}

\noindent To show that $|\mathscr{A}(k)|$ is bounded along \textit{arc} $AC$, we prove that $|\mathscr{A}(k)|$ is bounded along the sides of the rectangle $ABCD$, and then using analyticity of $\mathscr{A}(k)$ we obtain that $|\mathscr{A}(k)|$ is bounded inside this domain. Similarly, to show that $|\mathscr{A}(k)|$ is bounded along \textit{arc} $CG$, we prove that $|\mathscr{A}(k)|$ is bounded along the sides of the rectangular domain $CFGH$. (It is technically convenient to evaluate $|\mathscr{A}(k)|$ along sides of a rectangle than along an arc). In the sequel, we need the following formula for $\sinh(\alpha k)$. If $k=\xi+i\eta$, then  
\be \label{eq:4.8} |\sinh(\alpha k)|^2 = \frac{1}{2}~[\cosh(2\alpha \xi) - \cos(2\alpha \eta)]. \ee
Indeed,  $|\sinh(\alpha(\xi + i\eta))|^2 =~|\sinh(\alpha \xi)\cos(\alpha \eta) - i \cosh(\alpha \xi) \sin(\alpha \eta)|^2 = $\\
$ \sinh^2(\alpha \xi) \cos^2(\alpha \eta) + \cosh^2(\alpha \xi) \big(1 - \cos^2(\alpha \eta) \big)
=~ \cosh^2(\alpha \xi) - \cos^2(\alpha \eta),$
which yields (\ref{eq:4.8}).

Let us prove that $|\mathscr{A}(k)|$ is bounded along each side of $ABCD$. If $k=\xi+i\eta \in AB$, then $0\leq \xi \leq \pi \sqrt{N/2 + 3/16}$ and $\eta = \pi(N +1/2)$. Using (\ref{eq:4.8}), one gets the estimate:
\be \label{eq:4.9} |\mathscr{A}(k)|^2 = ~\frac{\cosh(2\alpha \xi) - \cos(2\alpha \eta)}{\cosh(2\xi) -\cos(2\eta)} \leq~ \frac{\cosh(2\alpha \xi)+1}{\cosh(2\xi) + 1} \leq ~1. \ee

\noindent If $k\in DF$, then $0\leq \xi\leq \pi(N+1/2)$ and $\eta = \pi(N+1/4)$. Using (\ref{eq:4.8}) one gets
\be \label{eq:4.10} |\mathscr{A}(k)|^2 =~ \frac{\cosh(2\alpha \xi) - \cos(2 \alpha \eta)}{\cosh(2\xi)} \leq~\frac{\cosh(2 \alpha \xi)}{\cosh(2\xi)} + \frac{1}{\cosh(2\xi)} \leq~2. \ee

\noindent If $k \in DA$, then $k=i\eta$, and $\pi (N + 1/4) \leq \eta\leq \pi(N + 1/2)$ using $\cos(2\eta) \leq 0$ one gets 
\be \label{eq:4.11} |\mathscr{A}(k)|^2 =~\frac{1 - \cos(2\alpha \eta)}{1-\cos(2 \eta)} \leq~ \frac{1 + |\cos(2\alpha \eta)|}{1} \leq~2.\ee

\noindent If $k\in BH$, then $\xi = \pi \sqrt{N/2 + 3/16}$ and $0\leq \eta \leq \pi(N+1/2)$. Using (\ref{eq:4.6}) one gets
\begin{align}
|\mathscr{A}(k)|^2 
 \leq&~ \frac{\cosh(2\pi ~\sqrt{N/2+3/16}~\alpha) +1}{\cosh(2\pi ~\sqrt{N/2+3/16}~) -1}
 \leq ~ 1 + \frac{2}{\cosh(2\pi\sqrt{N/2+3/16}~) -1} \leq~ 2. \label{eq:4.12}
\end{align}

\noindent If $k\in FG$, then $\xi= \pi(N+1/2)$ and $0\leq \eta \leq \pi(N+1/4)$. Using (\ref{eq:4.6}) one gets
\be
|\mathscr{A}(k)|^2 =~\frac{\cosh(2\pi~(N+1/2)~\alpha) - \cos(2 \alpha \eta)}{\cosh(2\pi~(N+1/2)) - \cos(2 \eta)} \leq~ 1 + \frac{2}{\cosh(2\pi(N+1/2)) -1} \leq~2. \label{eq:4.13} 
\ee

\noindent If $k\in HG$, then $|\mathscr{A}(k)|$ is bounded by 1 due to the monotonic behavior of the $sinh$ function for real $k$. Collecting together (\ref{eq:4.9}) -- (\ref{eq:4.13}), we obtain that $|\mathscr{A}(k)|$ is bounded on $\Gamma_N$ uniformly with respect to $N$. \\
\indent The lemma is proven. \end{proof}

Our goal is to evaluate integral $\mathbb{I}_1$ of (\ref{eq:4.5}) using the Residue theorem. To this end, we split the derivation of the desired result into several steps.

\indent \textit{Step 1: Reduction of the improper integral $\mathbb{I}_1(\xi,x)$ to a principal value integral.} It can be readily seen that $\mathbb{I}_1(\xi,x)$ is an absolutely convergent integral. Indeed, for each $y>0$, the following estimate holds for $\mathscr{A}(k)$, $k\in (-\infty,\infty)$:
\be \label{eq:4.14} \mathscr{A}(k) \leq ~\text{exp}\{-y|k|\},\quad k\in \mathbb{R}.\ee
Since $\mathscr{A}(k)$ is an even function it suffices to show (\ref{eq:4.14}) for $k\geq0.$ The result follows from the estimates: $\mathscr{A}(k) = \text{exp}\{-(1-\alpha)k\}~(1-e^{-2\alpha k})(1-e^{-2k})^{-1} \leq \text{exp}\{-(1-\alpha)k\} = e^{-yk}$.
Taking into account that the function $\dfrac{\sin(\xi(k-\omega))}{k-\omega}$ is bounded at the vicinity of $k=\omega$ and behaves as $\dfrac{1}{k}$ for large enough $k$, we obtain that the integral in (\ref{eq:4.5}) converges absolutely.

 \textbf{Definition 4.2.} \textit{Let P.V.($\omega$) be the notation for the integral that converges in the sense of the principal value "centered" at the point $\omega$, i.e for any absolutely integrable function $h(k)$, we define}

\be \label{eq:4.15} \text{P.V.}(\omega) \int_{-\infty}^{\infty} h(k) ~dk= \lim_{n\rightarrow \infty} \Bigg[\int_{-\infty}^{\omega-\delta_n}+\int_{\omega + \delta_n}^{\infty} \Bigg] h(k) dk, \ee

\noindent \textit{for any sequence $\delta_n \rightarrow 0$ as $n\rightarrow \infty$. }

 \noindent Using absolute convergence of the integral $\mathbb{I}_1(\xi, x)$ and Definition 4.2, we obtain
 \be \label{eq:4.16} \mathbb{I}_1(\xi,x) = \text{P.V.}(\omega) \int_{-\infty}^{\infty} dk~e^{ikx} ~\mathscr{A}(k)~\dfrac{\sin(\xi(k-\omega))}{k-\omega}.\ee
\indent \textit{Step 2: Reduction of the principal value integral to the limit of a sequence of contour integrals.} Let $\gamma_n$ be a semi--circle in the lower half--plane centered at the point $k=\omega$ of radius $\delta_n$, i.e.,  $\gamma_n = \{k\!:k=\delta_n e^{i\varphi}, ~-\pi\leq \varphi \leq 0\}$;~ let $\ell_n$ be the following contour on the $k$-plane:

\be \label{eq:4.17} \ell_n = (-\infty, \omega -\delta_n)~ \bigcup~\gamma_n~ \bigcup~(\omega + \delta_n, \infty). \ee

\begin{figure}[H]
\begin{center}
\includegraphics[scale=0.25] {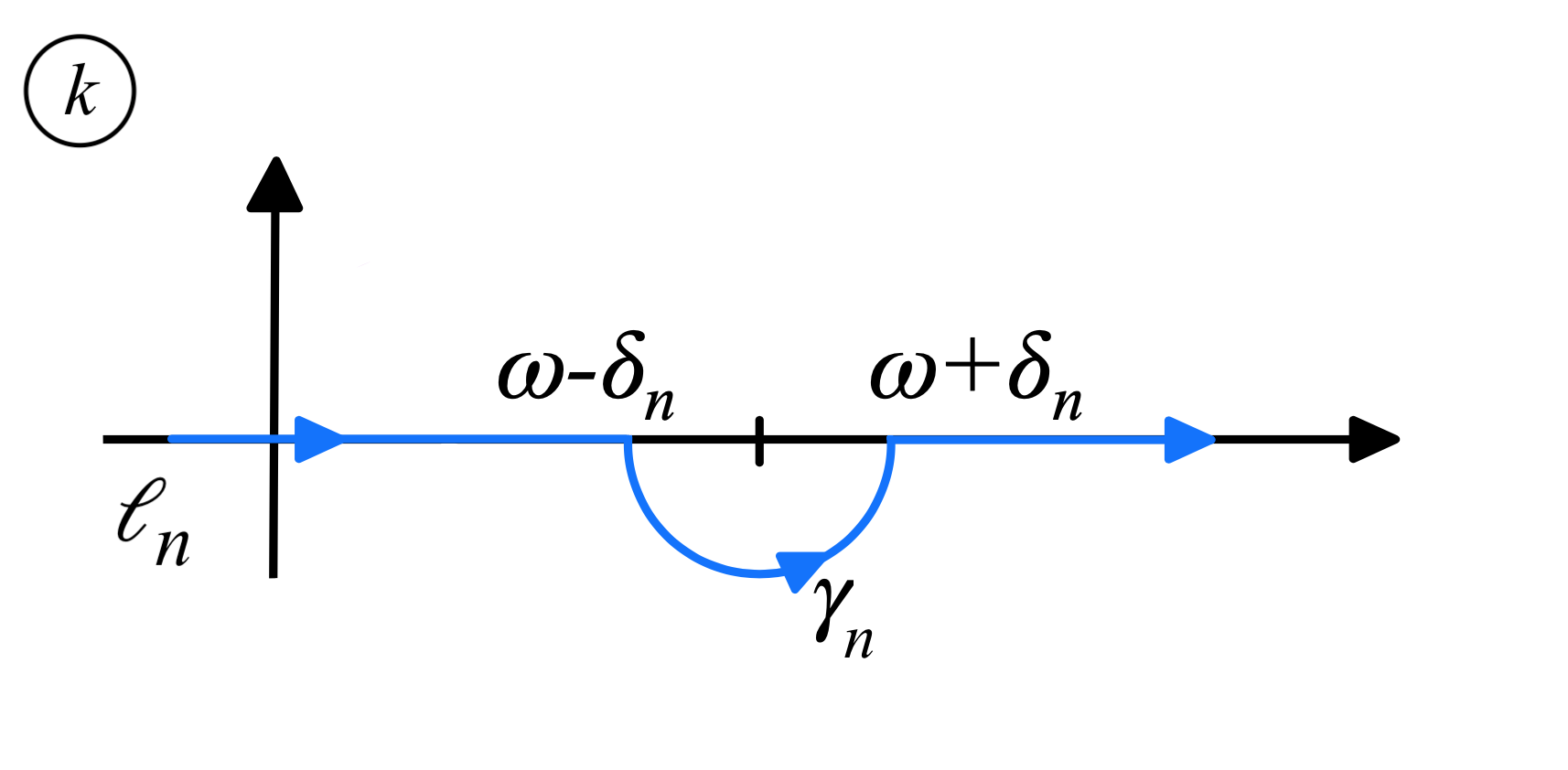} 
\caption{Contour of integration $\ell_n$}
\label{fig:2.3}
\end{center}
\end{figure}
\indent \begin{customlemma}{}\label{lemma4.3}\textbf{Lemma 4.3.} 
\textit{Integral $\mathbb{I}_1(\xi,x)$ of (\ref{eq:4.16}) can be represented as the following limit of a sequence of contour integrals:}
\be \label{eq:4.18}
\text{P.V.} (\omega) \int_{-\infty}^\infty dk~e^{ikx} ~\mathscr{A}(k)~\dfrac{\sin(\xi(k-\omega))}{k-\omega}=\lim_{n \rightarrow \infty}~ \int_{\ell_n} dk~e^{ikx} ~\mathscr{A}(k)~\dfrac{\sin(\xi(k-\omega))}{k-\omega}. 
\ee
\end{customlemma}

\indent \begin{proof} For each $n$, the following decomposition holds:
\begin{align} \label{eq:4.20}
\int_{\ell_n} dk~e^{ikx} ~&\mathscr{A}(k)~\dfrac{\sin(\xi(k-\omega))}{k-\omega}
&= \Bigg[ \int_{-\infty}^{\omega - \delta_n} + \awint_{\gamma_n} + \int_{\omega +\delta_n}^{\infty} \Bigg] dk~e^{ikx} ~\mathscr{A}(k)~\dfrac{\sin(\xi(k-\omega))}{k-\omega},
\end{align}
where $\awint_{\gamma_n}$ is the notation for the integral along a semi--circle $\gamma_n$. Let us show that the sequence of integrals along semi--circular contours $\gamma_n$ tends to 0 as $n\rightarrow \infty$ (i.e. $\delta_n \rightarrow 0$):
\be \label{eq:4.20}
\awint_{\gamma_n} dk~e^{ikx}~\mathscr{A}(k)~\dfrac{\sin(\xi(k-\omega))}{k-\omega} \rightarrow 0\qquad \text{as}~~n\rightarrow \infty.
\ee
The proof is based on the fact that the integrand is an analytic function of $k$ at the vicinity of $k=\omega$. Thus, if $k = \omega + \delta_n e^{i \varphi}$, then 
\begin{align*}
\bigg| \awint_{\gamma_n} dk~e^{ikx}~\mathscr{A}(k)~\dfrac{\sin(\xi(k-\omega))}{k-\omega}\bigg| =& \bigg| \int_{-\pi}^0 e^{ix(\omega + \delta_n e^{i\varphi})}~\mathscr{A}(\omega + \delta_n e^{i\varphi})~\sin(\xi ~\delta_n e^{i\varphi}) ~d\varphi\bigg|\leq C_0 \delta_n,
\end{align*}
which completes the proof. \\
\indent The lemma is shown. \end{proof}

\indent \textit{Step 3: Introducing closed contour integrals.}
Since each integral along $\ell_n$ has the domain of integration that does not pass through the point $k=\omega$, each contour integral can be split up into two integrals denoted by $\mathcal{I}_n(\xi, x)$ and $\widetilde \mathcal{I}_n(\xi,x)$, where:
  \be
  \mathcal{I}_{n}(\xi,x)=~ \frac{1}{2i}~ e^{-i\xi \omega} J_n(\xi,x),~~~~~J_n(\xi,x) \equiv \landdownint_{\ell_n} dk~\mathscr{A}(k) ~\dfrac{e^{ik(x+\xi)}}{k-\omega}, \label{eq:4.21}
  \ee
  \be
  \widetilde \mathcal{I}_{n}(\xi,x) =~ -\frac{1}{2i}~ e^{i\xi \omega} \widetilde J_n(\xi,x),~~~~~\widetilde J_n(\xi,x) \equiv \landdownint_{\ell_n} dk~\mathscr{A}(k)~ \dfrac{e^{ik(x-\xi)}}{k-\omega}. \label{eq:4.22}
  \ee
  
\textbf{Assumption 4.4.} \textit{In what follows we assume that }
 \be x \in [-R + \varepsilon, R -\varepsilon],~~~~~ \varepsilon\ll 1, \label{eq:4.23} \ee
 \textit{which is consistent with the physical origin of the model.}
 
 Taking into account that $\xi \in [R, R+r]$ (see (\ref{eq:4.2})) and (\ref{eq:4.23}), we obtain the bounds on $(x + \xi)$ and $(x-\xi)$ as 
 \be 0<\varepsilon \leq x + \xi \leq 2R+r-\varepsilon,~~~~~-2R-r+\varepsilon \leq x-\xi \leq -\varepsilon<0. \label{eq:4.24} \ee
  \noindent In the sequel, these estimates will allow us to use the Jordan lemma appropriately. Let us show that each integral, $J_n(\xi,x)$ and $\widetilde J_n(\xi,x)$, can be evaluated as a limit of a specifically chosen sequence of contour integrals. We provide a detailed proof only for the integral $J_n(\xi,x)$ since the second integral, $\widetilde J_n(\xi,x)$, can be treated in a similar manner. 
 
 \indent \begin{customlemma}\label{lemma4.5}\textbf{Lemma 4.5.} \textit{Let $\Gamma_N$ be a semi--circle in the upper half--plane centered at the origin of radius $\pi (N + 1/2)$, with $N$ being a positive integer (see Fig.\ref{fig:3} below). Then on a sequence of expanding semi--circles, the following result is valid:}
 \be \lim_{N\rightarrow \infty}\bigg( \intclockwise_{\Gamma_N} dk~ \mathscr{A}(k)~\frac{e^{ik(x+\xi)}}{k-\omega}\bigg) = 0. \label {eq:4.25} \ee
 
 \noindent \textit{Let $\widetilde \Gamma_N$ be a symmetric semi--circle in the lower half--plane. Then on a sequence of expanding semi--circles, the following result is valid:}
 \be \lim_{N\rightarrow \infty}  \awint_{\widetilde \Gamma_N} dk~ \mathscr{A}(k)~\frac{e^{ik(x-\xi)}}{k-\omega} = 0. \label {eq:4.26} \ee
\end{customlemma}

\indent \begin{proof}
The proof is based on the straightforward application of estimates (\ref{eq:4.24}), the Jordan lemma \cite{Jeffrey2006, Saff2003}, and Lemma 4.1.
 \begin{figure}[H]
\begin{center}
\includegraphics[scale=0.3] {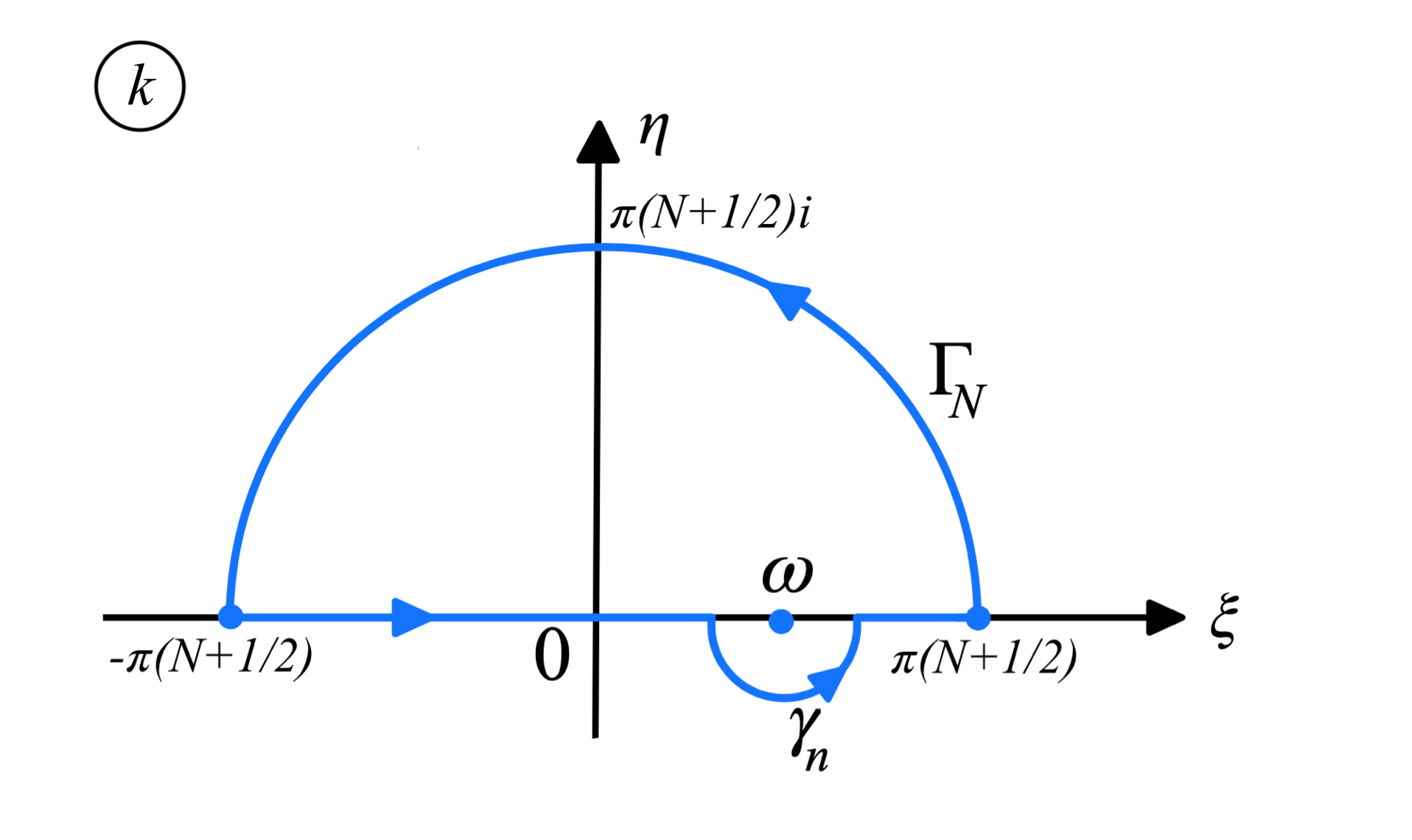} 
\caption{Closed contour of integration in the upper half--plane}
\label{fig:3}
\end{center}
\end{figure}
The proof of the lemma is complete.  \end{proof}

Let $n$ be fixed and $N$ be a positive integer. Consider a sequence of closed contours in the upper half--plane denoted by $\{\mathbb{\Gamma}_{N,n}\}^{\infty}_{N=1}$ and defined as
\be \label{eq:4.27} \mathbb{\Gamma}_{N,n} = \Gamma_N~ \bigcup~[-R_N, \omega-\delta_n]~\bigcup~\gamma_n ~\bigcup~ [\omega+\delta_n, R_N], \quad R_N = \pi(N+1/2).
\ee
Let us consider the following integral along contour $\mathbb{\Gamma}_{N,n}$ :
\be \label{eq:4.28}
\ointctrclockwise_{\mathbb{\Gamma}_{N,n}} dk~\frac{e^{ik(x+\xi)}}{k-\omega}~\mathscr{A}(k) 
=\bigg[ \int_{-R_N}^{\omega-\delta_n}+ \awint_{\gamma_n} + \int_{\omega+\delta_n}^{R_N} -\intclockwise_{\Gamma_N} \bigg]~ dk~ \frac{e^{ik(x + \xi)}}{k-\omega}~\mathscr{A}(k).
\ee
By using the Residue theorem \cite{Jeffrey2006, Saff2003}, we obtain the following series representation:
\begin{align} \label{eq:4.29}
\frac{1}{2\pi i } \ointctrclockwise_{\mathbb{\Gamma}_{N,n}} dk~\mathscr{A}(k)~\frac{e^{ik(x+\xi)}}{k - \omega} =& \sum_{m=1}^N \text{Res}\bigg(\frac{e^{ik(x+\xi)}}{k - \omega}~\frac{\sinh(k(1-y))}{\sinh(k)},~\pi m i\bigg)\nonumber \\&~~+~\text{Res}\bigg(\frac{e^{ik(x+\xi)}}{k - \omega}~\frac{\sinh(k(1-y))}{\sinh(k)},~\omega\bigg)\nonumber \\
=&\sum_{m=1}^N \frac{e^{-\pi m (x+\xi)}}{\pi m i - \omega}~\frac{i~\sin(\pi m (1-y))}{\cos(\pi m)} + \frac{e^{i \omega(x+\xi)}\sinh(\omega \alpha)}{\sinh(\omega)} \nonumber \\
=&-\sum_{m=1}^N e^{-\pi m (x+\xi)}~\frac{\sin(\pi m y)}{\pi m + i\omega} + \frac{e^{i \omega(x+\xi)}\sinh(\omega \alpha)}{\sinh(\omega)}.
\end{align}
Passing to the limits as $N\rightarrow \infty$ in (\ref{eq:4.29}) and using Lemma 4.5, from the decomposition \ref{eq:4.28}, we arrive at the following formula for $J_n(\xi,x)$:
\be \label{eq:4.30} \frac{1}{2i}~ J_n(\xi,x) = ~- \pi~\sum_{m=1}^{\infty} e^{-\pi m(x+\xi)}~\frac{\sin(\pi m y)}{\pi m + i \omega}+ \frac{\pi ~e^{i\omega (x + \xi)} \sinh(\omega \alpha)}{\sinh(\omega)}. \ee

Now we turn to the integral $\widetilde J_n(\xi,x)$ from (\ref{eq:4.22}). Let us introduce a sequence of closed contours $\{\widetilde \mathbb{\Gamma}_{N,n} \}_{N=1}^{\infty}$ in the lower half--plane defined as
\be \label{eq:4.31}\widetilde \mathbb{\Gamma}_{N,n} = ~\widetilde \Gamma_N ~\bigcup~[-R_N, ~\omega-\delta_n]~\bigcup~\gamma_n~\bigcup~[\omega+\delta_n, R_N], \quad R_N=\pi(N+1/2),\ee
where $\widetilde \Gamma_N$ is a semi--circle in the lower half--plane (in the clockwise direction).

 \begin{figure}[H]
\begin{center}
\includegraphics[scale=0.3] {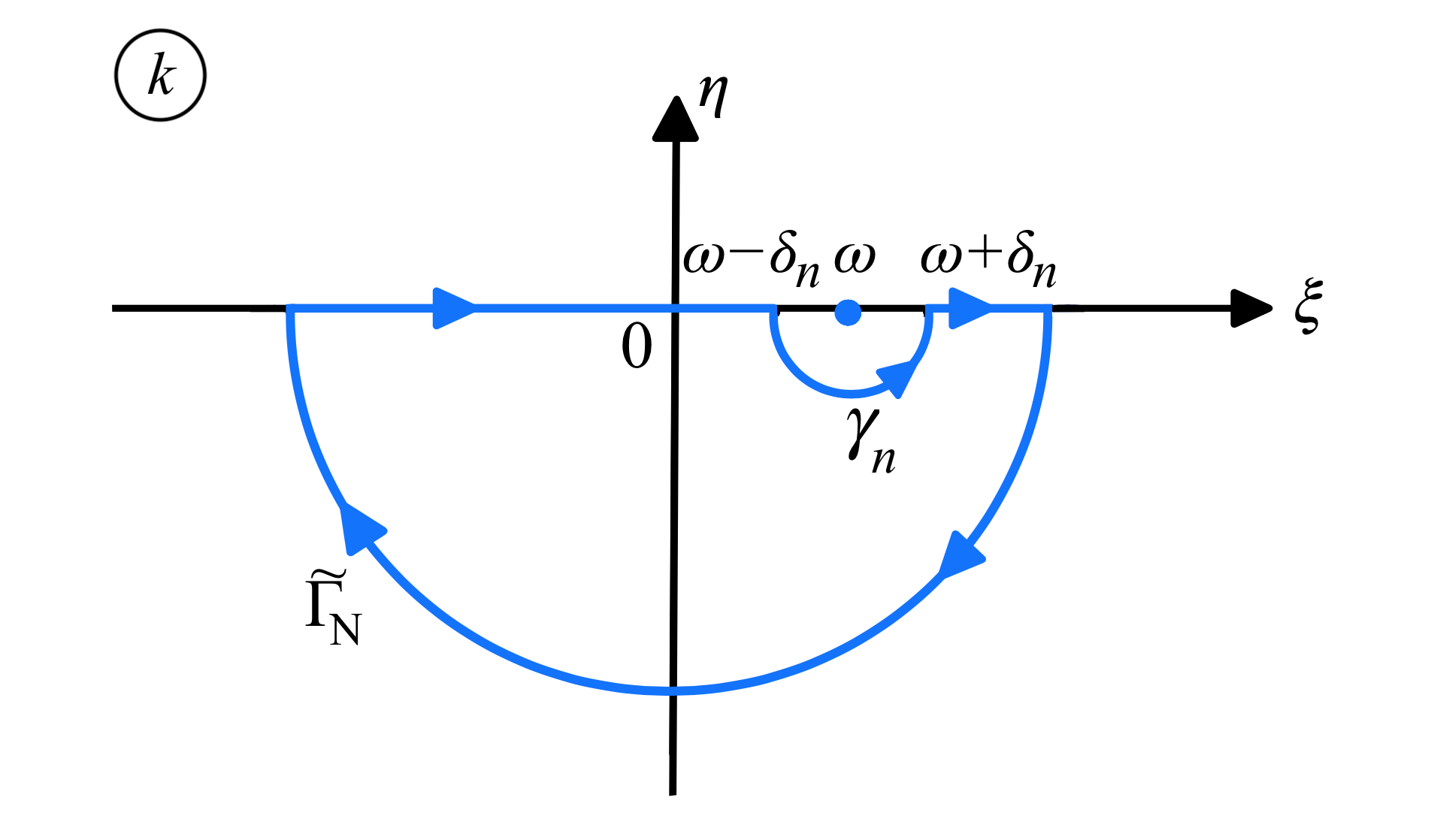} 
\caption{Closed contour of integration in the lower half--plane}
\label{fig:5}
\end{center}
\end{figure}

\noindent Evaluation of the integral along the closed contour $\widetilde \mathbb{\Gamma}_{N,n}$ yields the decomposition:
\begin{align} \label{eq:4.32}
\ointclockwise_{\mathbb{\widetilde \Gamma}_{N,n}} dk~\mathscr{A}(k)~\frac{e^{ik(x-\xi)}}{k-\omega} \equiv&-\ointctrclockwise_{\mathbb{\widetilde \Gamma}_{N,n}} dk~\mathscr{A}(k)~\frac{e^{ik(x-\xi)}}{k-\omega} \nonumber\\
=&-\bigg[ \int_{-R_N}^{\omega-\delta_n}+ \awint_{\gamma_n} + \int_{\omega+\delta_n}^{R_N} - \awint_{\widetilde\Gamma_N} \bigg] dk~\mathscr{A}(k)~ \frac{e^{ik(x - \xi)}}{k-\omega}.
\end{align}
Using the Residue theorem, we obtain for this integral 
\begin{align} \label{eq:4.33}
\frac{1}{2\pi i } \ointctrclockwise_{\mathbb{\Gamma}_{N,n}} dk~\mathscr{A}(k)~\frac{e^{ik(x-\xi)}}{k - \omega}=& \sum_{m=1}^N \text{Res}\bigg(\frac{e^{ik(x-\xi)}}{k - \omega}~\frac{\sinh(k(1-y))}{\sinh(k)},~-\pi m i\bigg)=\nonumber \\
=&\sum_{m=1}^N \frac{e^{\pi m (x-\xi)}}{\pi m i + \omega}~\frac{i~\sin(\pi m (1-y))}{\cos(\pi m i)}
=-\sum_{m=1}^N e^{\pi m (x-\xi)}~\frac{\sin(\pi m y)}{\pi m - i\omega}. 
\end{align}
Passing to the limit when $N\rightarrow \infty$, we have from (\ref{eq:4.32}) and (\ref{eq:4.33})
\be \label{eq:4.34} \frac{1}{2i}~\widetilde J_n(\xi,x) = ~\pi \sum_{m=1}^{\infty} e^{\pi m(x-\xi)}~\frac{\sin(\pi m y)}{\pi m- i\omega}. \ee
Substituting (\ref{eq:4.30}) into formula (\ref{eq:4.21}) for $\mathcal{I}_n(\xi,x)$ and (\ref{eq:4.34}) into formula (\ref{eq:4.22}) for $\widetilde \mathcal{I}_n(\xi,x)$, we obtain
\begin{align} \label{eq:4.35}
(\mathcal{I}_n + \widetilde \mathcal{I}_n) &=~\pi~ e^{i\omega x}~\frac{\sinh(\omega(1-y))}{\sinh(\omega)} - \pi\sum_{m=1}^{\infty} e^{-\pi m \xi} ~\sin(\pi m y) \bigg\{ \frac{e^{-\pi m x}~e^{-i\xi\omega}}{\pi m + i \omega}+ \frac{e^{\pi m x}~e^{i\xi \omega}}{\pi m - i \omega}\bigg\}\nonumber \\
=&~\pi~ e^{i\omega x}~\frac{\sinh(\omega(1-y))}{\sinh(\omega)} -2~\sum_{m=1}^{\infty} e^{-\pi m \xi}\sin(\pi m y)~\times\nonumber \\
&\qquad \qquad \qquad \qquad \qquad  \bigg\{\frac{\pi m~\cosh(\pi m x +i\xi \omega) - i\omega~\sinh(\pi m x+ i \xi \omega)}{\pi^2m^2 + \omega^2} \bigg\}.
\end{align}
Now we use formulae (\ref{eq:4.16}), (\ref{eq:4.18}), and (\ref{eq:4.35}) to obtain the formula for $\mathbb{I}_1(\xi,x)$
\begin{align} \label{eq:4.36} 
\mathbb{I}_1(\xi,x) =&~\lim_{n\rightarrow\infty} (\mathcal{I}_n + \widetilde \mathcal{I}_n)(\xi,x) =~\pi~ e^{i\omega x}~~\frac{\sinh(\omega(1-y))}{\sinh(\omega)}\nonumber \\
&-2~\sum_{m=1}^{\infty} e^{-\pi m \xi}\sin(\pi m y) \bigg\{\frac{\pi m~\cosh(\pi m x +i\xi \omega) - i\omega~\sinh(\pi m x+ i \xi \omega)}{\pi^2m^2 + \omega^2} \bigg\}.
\end{align}
Summarizing the results of this section, we formulate the following statement.

\indent \begin{customthm}{}\label{theorem4.6}\textbf{Theorem 4.6.} 
\textit{The following explicit formula holds the for the term $I_1$ from the decomposition (\ref{eq:4.1}):}
\begin{align}\label{eq:4.37}
I_1 =& -\omega c~e^{i\omega(x-ct)}~g_0(R)~\frac{\sinh(\omega(1-y))}{\sinh(\omega)} - 2\omega c~e^{-i\omega ct}~\sum_{m=1}^{\infty} \frac{\sin\pi m y}{\pi^2m^2 + \omega^2}~\times\nonumber \\
&\quad \int_R^{R+r} d\xi~g_0'(\xi)~e^{-\pi m \xi}\big[ \pi m ~\cosh(\pi m x + i \xi \omega) - i \omega~ \sinh(\pi m x + i \xi \omega)\big],
\end{align}
\textit{with the series being absolutely convergent for any $x\in(-R,R)$, $0<y<1$, and $-\infty<t<\infty$.}
\end{customthm}


\section{Evaluation of integrals  $I_2$ and $I_3$ from (\ref{eq:4.3}) and (\ref{eq:4.4})}
Our goal in this section is to evaluate the integrals $I_2$ and $I_3$ using the Residue theorem. To simplify further calculations, we denote
\be \label{eq:5.1} \mathbb{I}_2 \equiv -2\pi ~I_2 \qquad \text{and} \qquad \mathbb{I}_3\equiv-2\pi~ I_3, \ee
where
\be \label{eq:5.2}
\mathbb{I}_2 = \int_{-\infty}^{\infty} dk~e^{ikx}~ \bigg[ \frac{\sinh(k(1-y))}{\sinh(k)} \bigg] \int_0^y d\eta~e^{-ik\eta t}~\sinh(k\eta) \bigg[ \widetilde f_{\eta \eta}(k,\eta) - k^2 \widetilde f(k,\eta)\bigg],
\ee
\be \label{eq:5.3}
\mathbb{I}_3 = \int_{-\infty}^{\infty} dk~e^{ikx}~ \bigg[ \frac{\sinh(ky)}{\sinh(k)} \bigg]\int_y^1 d\eta~e^{-ik\eta t}~\sinh(k(1-\eta)) \bigg[ \widetilde f_{\eta \eta}(k,\eta) - k^2 \widetilde f(k,\eta)\bigg].
\ee
Integrating by parts twice the integral along $[0,y)$, we modify $\mathbb{I}_2$ to the following form:
\begin{align} \label{eq:5.4}
\mathbb{I}_2& = \int_{-\infty}^{\infty} dk~ \frac{e^{ikx}}{\sinh(k)} \sinh(k(1-y))\bigg\{ e^{-ikyt} \sinh(ky) ~\widetilde f_y(k,y)\nonumber \\
& \qquad + (ikt)~e^{-ikyt}  \sinh(ky) ~ \widetilde f(k,y) - k~e^{-ikyt}  \cosh(ky)~\widetilde f(k,y)   + k~\widetilde f(k,0) \nonumber \\
& \qquad +  \int_0^y d \eta~\widetilde f(k,\eta)~\bigg( \frac{\partial^2}{\partial \eta^2} -k^2\bigg) \big[ e^{-ik \eta t} \sinh(k\eta) \big]  \bigg\}.
\end{align}
Integrating by parts twice the integral along $(y,1)$, we modify $\mathbb{I}_3$ to the following form:
\begin{align} \label{eq:5.5}
\mathbb{I}_3 &= \int_{-\infty}^{\infty} dk ~\frac{e^{ikx}}{\sinh(k)} ~\sinh(ky)\bigg\{ -e^{-ikyt}\sinh(k(1-y))~\widetilde f_y(k,y) \nonumber\\
&\qquad  +k~ e^{-ikt} ~\widetilde f(k,1)   - ikt~ e^{-ikyt}~\sinh(k(1-y))  \widetilde f(k,y) \nonumber \\
&\qquad  -k~e^{-ikyt} ~\cosh(k(1-y)) \widetilde f(k,y)\nonumber \\
&\qquad +\int_y^1 d \eta~ \widetilde f(k,\eta)~\bigg( \frac{\partial^2}{\partial \eta^2} -k^2\bigg) \big[ e^{-ik \eta t} \sinh(k(1-\eta)) \big]  \bigg\}.
\end{align}

\noindent Without loss of generality to simplify further calculations, we assume that $f(x,0)=f(x,1)=0$ and obtain the following representation for the sum $\mathbb{I}_2+\mathbb{I}_3$:
\begin{align}
\mathbb{I}_2 + \mathbb{I}_3 =& -\int_{-\infty}^{\infty} dk~k~e^{ikx} e^{-ikyt}~\widetilde f (k,y) + \mathbf{I}, \quad \text{where}  \label{eq:5.6}\\
\mathbf{I} = &\int_{-\infty}^{\infty} dk ~\frac{e^{ikx}}{\sinh(k)}\bigg\{ \sinh(k(1-y)) \int_0^y d \eta~\widetilde f(k,\eta)~\bigg( \frac{\partial^2}{\partial \eta^2}\ - k^2 \bigg) \big[ e^{-ik\eta t }\sinh(k\eta) \big] \nonumber \\
&~+ \sinh(ky) \int_y^1 d \eta ~\widetilde f(k,\eta)~\bigg( \frac{\partial^2}{\partial \eta^2}\ - k^2 \bigg) \big[ e^{-ik\eta t }\sinh(k(1-\eta)) \big] \bigg\}. \label{eq:5.7}
\end{align}

 \textbf{Remark 5.1.} One can readily check that the integral of (\ref{eq:5.6}) generates the following result:
\[ -\int_{-\infty}^{\infty} dk~k~e^{ik(x-yt)}~\widetilde f (k,y) =~i f_x(x-yt,y). \]
\noindent If we fix $y$ for $0<y<1$ and consider $x$ and $t$ such that  $x-yt = C$, where $C$ is some constant, then the function $f_x(x-yt,y)$ is also equal to a constant. It means that the perturbation velocity moves with the speed $y$ along the $x$-axis.

The integral $\mathbf{I}$ defined in (\ref{eq:5.7}) can be represented as a sum: $\mathbf{I} \equiv \sum_{j=1}^4 \mathbf{I}_j$, where
\begin{align}
\begin{split}\label{eq:5.8}
\mathbf{I}_1 =& \int_0^y d\eta \int_{-\infty}^{\infty} dk~e^{ik(x-\eta t)}~ \widetilde f(k,\eta)~(-k^2 t^2)~ \frac{\sinh(k(1-y))}{\sinh(k)}~\sinh(k\eta), \\
\mathbf{I}_2 =& \int_0^y d\eta \int_{-\infty}^{\infty} dk~ e^{ik(x-\eta t)} ~\widetilde f(k,\eta)~(-2i~k^2 t)~ \frac{\sinh(k(1-y))}{\sinh(k)}~\cosh(k\eta),\\
\mathbf{I}_3 =& \int_y^1 d\eta \int_{-\infty}^{\infty} dk~e^{ik(x-\eta t)}~ \widetilde f(k,\eta)~(-k^2 t^2)~ \frac{\sinh(ky)}{\sinh(k)}~\sinh(k(1-\eta)), \\
\mathbf{I}_4 =& -\int_y^1 d\eta \int_{-\infty}^{\infty} dk~e^{ik(x-\eta t)}~ \widetilde f(k,\eta)~(-2i~k^2 t) ~\frac{\sinh(ky)}{\sinh(k)}~\cosh(k(1-\eta)).
\end{split}
\end{align}
In this work, we make the following assumptions on the perturbation of $f(x,y)$.

\textbf{Assumption 5.2.}\textit{Let $f(\cdot, y)$ be a smooth function with compact support, i.e.,}

\be \label{eq:5.9} \text{supp}\{f(\cdot, y)\}\in (0,r_0), \qquad r_0>0. \ee
To simplify further calculations, we also assume that $f(x,y)$ has the first three partial derivatives in $x$ equal to zero for $x=0$ and $x= r_0$, i.e., 
\be \label{eq:5.10}
f_x(x,y) = f_{xx}(x,y) = f_{xxx}(x,y) = 0,\quad x=0~~~~\text{and}~~~~x=r_0.
\ee
Using Assumption 5.2, one can obtain that 
\be \label{eq:5.11} \widetilde f(k,y) =~\frac{i}{k^3} \int_0^{r_0} dx~e^{-ikx} f_{xxx}(x,y) \equiv ~\frac{1}{k^3}~\widetilde \chi(k,y),\quad \text{~~~where~~~~~~} \chi(x,y) =~i~f_{xxx}(x,y). \ee
It is convenient to present the integrals $\mathbf{I}_j$, $j=1,2,3,4$ in the forms 
\be \label{eq:5.12} \mathbf{I}_{2j-1} = -t^2~\widehat{\mathbf{I}}_{2j-1, j=1,2} \quad \text{and}\quad \mathbf{I}_{2j} = -2it~\widehat{\mathbf{I}}_{2j, j=1,2}.\ee
Taking into account (\ref{eq:5.11}) we obtain the following representations for $~\widehat{\mathbf{I}}_j$, $j=1,2,3,4:$
\begin{align}
\widehat{\mathbf{I}}_1 =&\int_0^{r_0} d\xi ~ \int_0^yd\eta~\chi(\xi,\eta)~\int_{-\infty}^{\infty}~dk~ e^{ik(x-\eta t-\xi)}~\frac{\sinh(k(1-y))\sinh(k\eta)}{k~\sinh(k)} \label{eq:5.13},\\
\widehat{\mathbf{I}}_2 =& \int_0^{r_0} d\xi ~\int_0^y d\eta ~\chi(\xi,\eta)~ ~\text{P.V.} \int_{-\infty}^{\infty}~dk~e^{ik(x-\eta t-\xi)}~ \frac{\sinh(k(1-y))\cosh(k\eta)}{k~\sinh(k)} \label{eq:5.14},\\
\widehat{\mathbf{I}}_3 =& \int_0^{r_0} d\xi ~\int_y^1d\eta ~\chi(\xi,\eta)~\int_{-\infty}^{\infty}~dk~ e^{ik(x-\eta t-\xi)} ~\frac{\sinh(k(1-\eta))\sinh(ky)}{k~\sinh(k)} \label{eq:5.15},\\
\widehat{\mathbf{I}}_4 =& -\int_0^{r_0} d\xi ~\int_y^1d\eta ~\chi(\xi,\eta)~~ \text{P.V.} \int_{-\infty}^{\infty}~dk~ e^{ik(x-\eta t-\xi)} ~\frac{\sinh(k y)\cosh(k(1-\eta))}{k~\sinh(k)}. \label{eq:5.16}  
\end{align}

To evaluate the improper integrals from (\ref{eq:5.13}) -- (\ref{eq:5.16}), we will represent each of them as a limit of a sequence of closed contour integrals on the complex $k$-plane and then use the Residue theorem. It turns out that the derivation of the desired result for the sum $\big(~\widehat{\mathbf{I}}_1+\widehat{\mathbf{I}}_3\big)$ is technically different from the derivation for the sum $\big(~\widehat{\mathbf{I}}_2+ \widehat{\mathbf{I}}_4\big)$. For the rest of this section we will focus on the results for $\big(~\widehat{\mathbf{I}}_1+\widehat{\mathbf{I}}_3\big)$.  

In what follows, we provide a detailed analysis for the case $x \geq r_0$. The case $x\leq r_0$ can be analyzed in a similar manner. 

 \textbf{Theorem 5.3.} \textit{For $x\geq r_0$, the following representation is valid for $(~\widehat{\mathbf{I}}_1 + \widehat{\mathbf{I}}_3$):}
\be \label{eq:5.17} \widehat{\mathbf{I}}_1 + \widehat{\mathbf{I}}_3 = ~2\pi \int_0^{r_0} d\xi \int_0^1 d\eta~\chi(\xi,\eta)~\sum_{m=1}^{\infty} e^{-\pi m|x-\eta t-\xi|}~\frac{\sin(\pi m y)~\sin(\pi m \eta)}{\pi m}.\ee
\indent \begin{proof}
To prove the result we consider three different cases. The case when $~\infty>t\geq x/y$, will be called Case A; the case when $~x/y> t\geq (x-r_0)/y$, will be called Case B, and the case when $~(x-r_0)/y>t\geq (x-r_0)$, will be called Case C.

\textbf{Case A.} For this case, we provide a very detailed proof. We consider separately the subcases ($i$) $t=x/y$ and ($ii$) $t>x/y$. Let us evaluate the integral $\widehat{\mathbf{I}}_1$  for $t=x/y$. The domain of integration with respect to $\xi$ and $\eta$ in $\widehat{\mathbf{I}}_1$ of (\ref{eq:5.13}) can be given as $[0,r_0]\times [0,y]$. A straight line defined as $x-\eta t-\xi=0$ passes through the points $B$ with coordinates $(\xi = x, \eta = 0)$ and $D$ with the coordinates $(\xi=0, \eta=y)$ (see Fig.\ref{fig:BKL}).
\begin{figure}[H]
\begin{center}
\includegraphics[scale = 0.3] {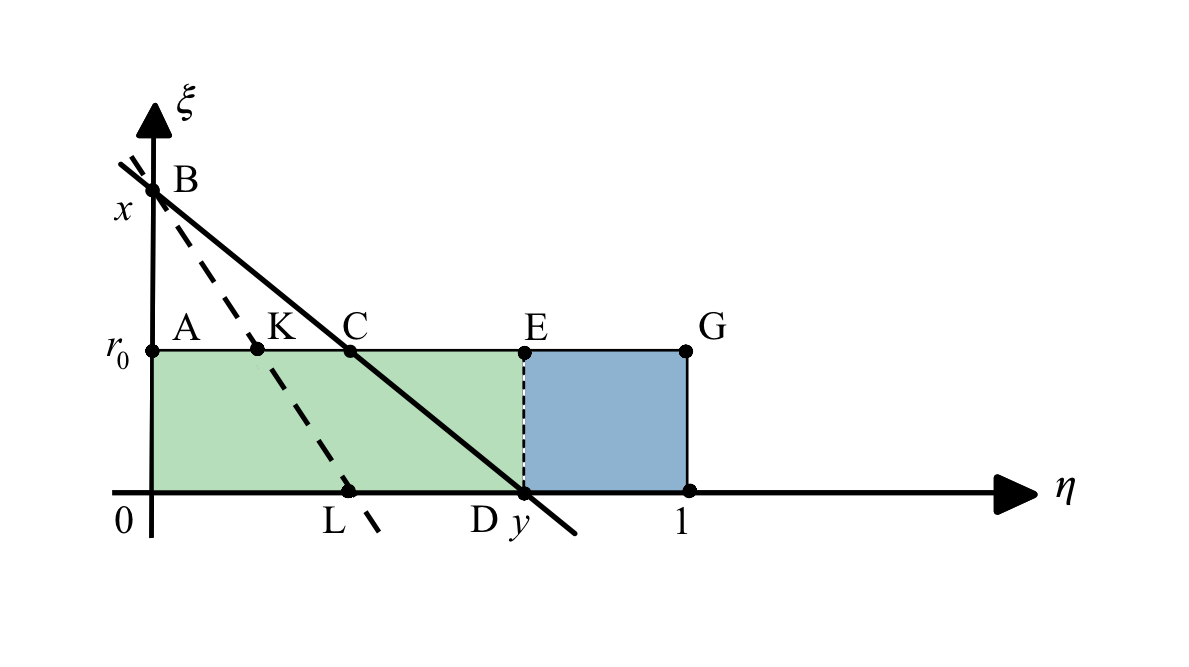} 
\caption{Positions of the straight lines for Case A}
\label{fig:BKL}
\end{center}
\end{figure}

\noindent For all points located above the line $BD$, one has $x-\eta t - \xi<0$ and for all points located below $BD$, one has $x-\eta t -\xi>0.$ This straight line splits the domain of integration into two subdomains: the triangular subdomain $CDE$ and the trapezoidal subdomain$ACD0$. Therefore, we have for the integral of (\ref{eq:5.13})
\begin{align} \label{eq:5.18}
\frac{1}{2\pi i } ~\widehat{\mathbf{I}}_1 =& ~\widehat{\mathcal{I}}_1 + \widehat{\mathcal{I}}_2, \quad \text{where} \nonumber \\
 \widehat{\mathcal{I}}_1=&~ \frac{1}{2\pi i} \int_0^{r_0} d\xi~ \int_0^{\frac{x-\xi}{t}} d\eta ~\chi(\xi, \eta)~ \int_{-\infty}^{\infty}~dk~ e^{ik(x-\eta t - \xi)}~\frac{\sinh(k(1-y))\sinh(k\eta)}{k~\sinh(k)}, \nonumber \\
  \widehat{\mathcal{I}}_2=&~\frac{1}{2\pi i} \int_0^{r_0} d\xi~ \int_{\frac{x-\xi}{t}}^y d\eta ~\chi(\xi, \eta)~ \int_{-\infty}^{\infty}~dk~ e^{ik(x-\eta t - \xi)}~\frac{\sinh(k(1-y))\sinh(k\eta)}{k~\sinh(k)}.
\end{align}

\noindent To evaluate $\widehat{\mathcal{I}}_1$ (where $x-\eta t - \xi \geq 0$), we will show that the improper integral with respect to $k$ can be represented as a limit of a sequence of contour integrals in the closed upper half--plane, and each of the aforementioned integrals can be evaluated by using the Residue theorem. 

\indent Let $\mathscr{B}(k)$ be an analytic function defined by 
\be \label{eq:5.19} \mathscr{B}(k)= \frac{\sinh(k(1-y))~\sinh(k\eta)}{k~\sinh(k)}. \ee

\noindent Let $\Gamma_N$ be a semi--circle in the upper half--plane centered at the origin of radius $r = \pi(N+1/2)$, where $N$ is a large positive integer. It can be readily shown that when $\eta \leq y,$
\be \label{eq:5.20} |\mathscr{B}(k)|\rightarrow 0\text{~~when~~} k\in \Gamma_N \text{~~and~~} N\rightarrow \infty. \ee
Using the Jordan lemma and (\ref{eq:5.20}) yields
\be \label{eq:5.21} \lim_{N\rightarrow\infty} \intclockwise_{\Gamma_N}e^{ik(x-\eta t-\xi)}~\mathscr{B}(k) \rightarrow 0 \text{~~as~~} N\rightarrow\infty,\ee
where $\intclockwise_{\Gamma_N}$ denotes the integral along $\Gamma_N$. Consider a sequence $\big \{\mathbb{\Gamma}_N\big\}^{\infty}_{N=1}$ of closed contours defined by $\mathbb{\Gamma}_N =\Big[-\pi \big(N+1/2 \big),~ \pi\big(N+1/2\big)\Big] \bigcup \Gamma_N$. Based on (\ref{eq:5.21}) we obtain the following relation for the improper integral from $\widehat{\mathcal{I}}_1$ :

\be \label{eq:5.22}
\int_{-\infty}^{\infty} dk~e^{ik(x-\eta t- \xi)} ~\mathscr{B}(k)
= \lim_{N\rightarrow\infty} \ointctrclockwise_{\mathbb{\Gamma}_N}dk~e^{ik(x-\eta t -\xi)} ~\mathscr{B}(k).
\ee

\begin{figure}[h]
\begin{subfigure}[t]{0.5\textwidth}
\includegraphics[scale = 0.29] {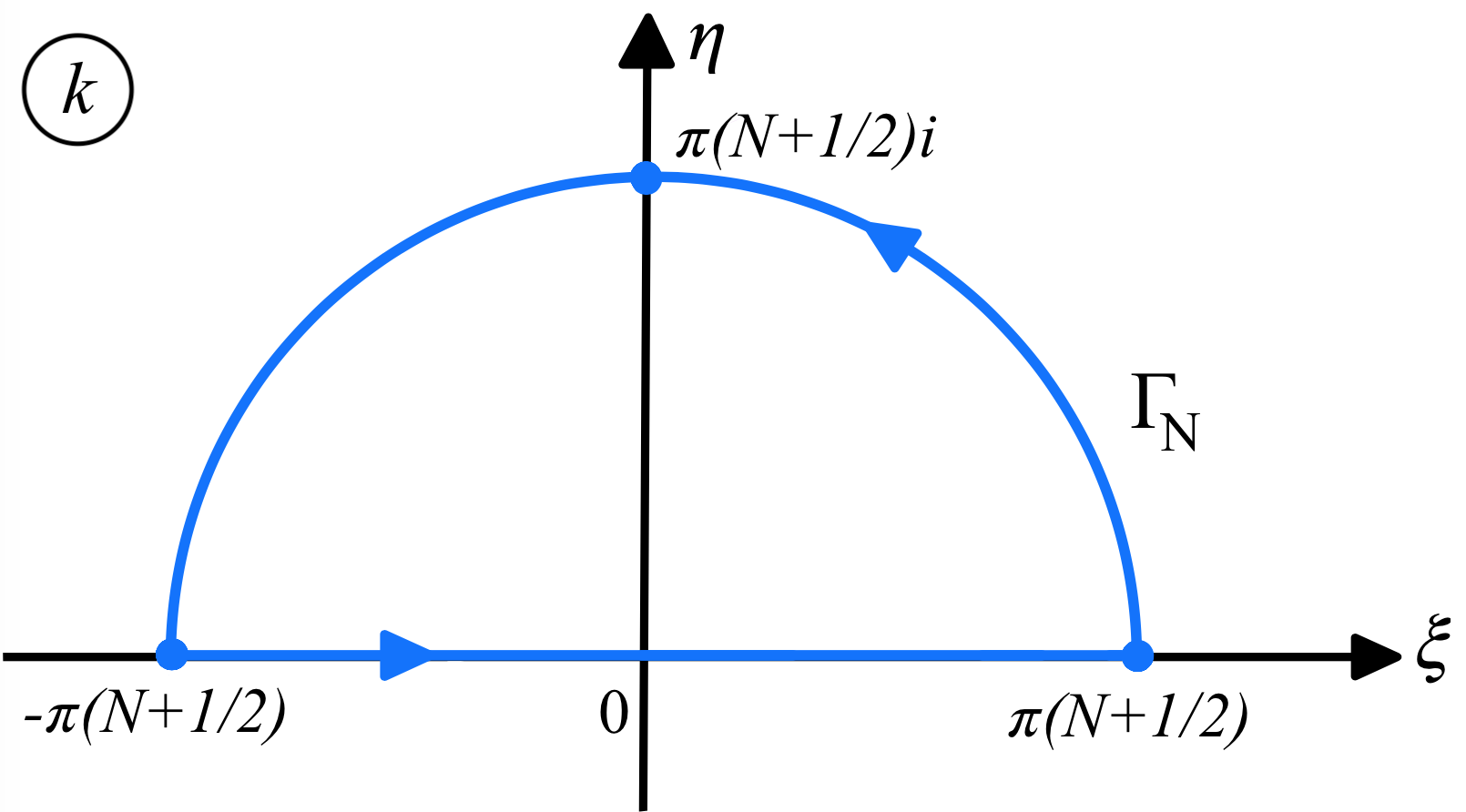} 
\caption{}
\label{fig:semicirc_a}
\end{subfigure}
\begin{subfigure}[t]{0.5\textwidth}
\includegraphics[scale = 0.27] {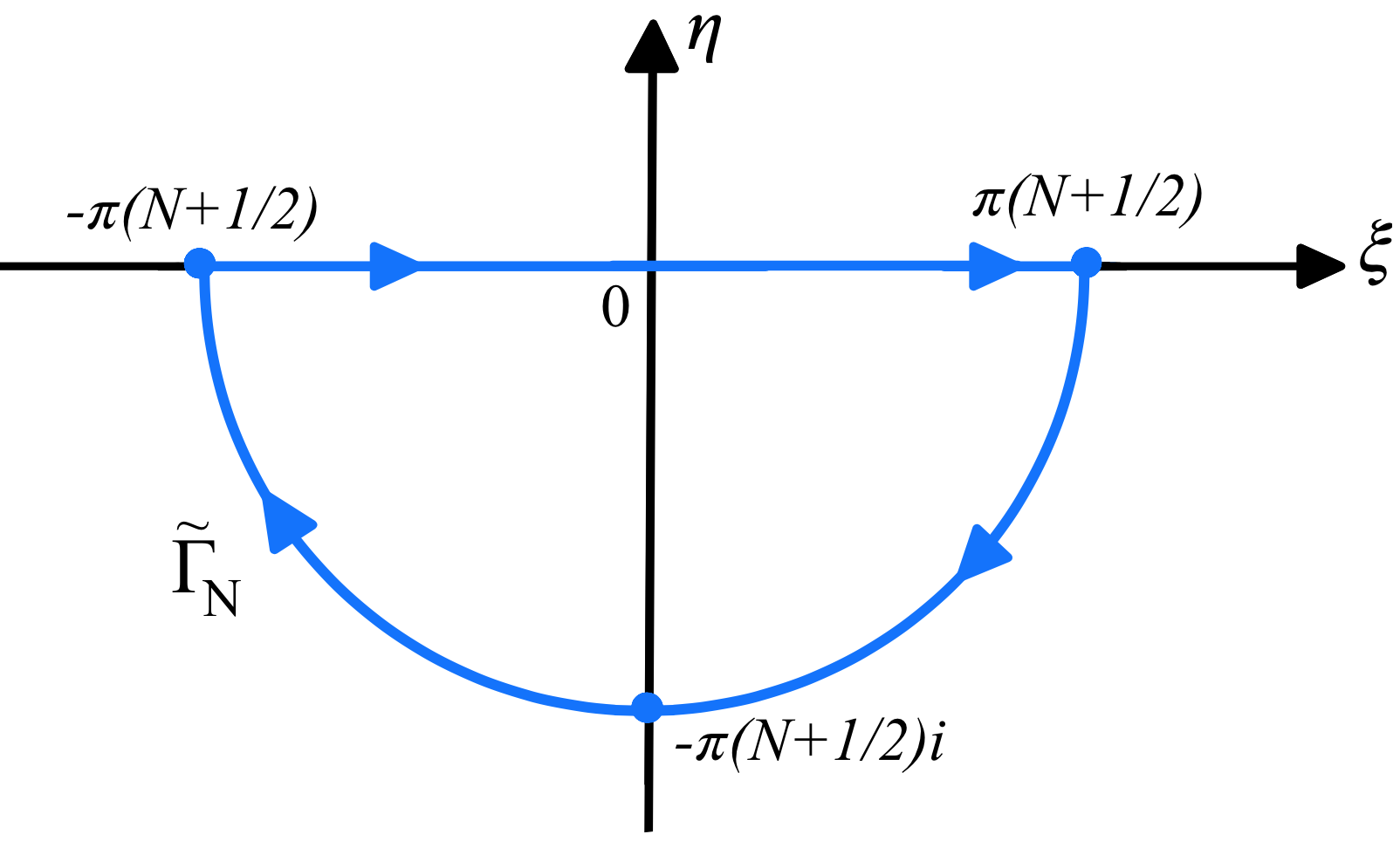} 
\caption{}
\label{fig:semicirc_b}
\end{subfigure}
\caption{Closed contours of integration for the integrals $\widehat{\mathcal{I}}_1$ (a) and $\widehat{\mathcal{I}}_2$ (b).}\label{fig:up_low_contours}
\label{fig:up_low_contours}
\end{figure}

\noindent Using the Residue theorem and relation (\ref{eq:5.22}), we evaluate $\widehat\mathcal{I}_1$ and have
\begin{align}\label{eq:5.23}
\widehat{\mathcal{I}}_1 =&~ \int_0^{r_0} d\xi~ \int_0^{\frac{x-\xi}{t}} d\eta~\chi(\xi, \eta)~ \sum_{m=1}^{\infty} \text{Res}\bigg\{ e^{ik(x-\eta t - \xi)}~\frac{\sinh(k(1-y))\sinh(k\eta)}{k~\sinh(k)}, ~\pi m i \bigg\} \nonumber\\
=&~ -i \int_0^{r_0} d\xi~ \int_0^{\frac{x-\xi}{t}} d\eta ~\chi(\xi, \eta)~ \sum_{m=1}^{\infty} e^{-\pi m(x-\eta t -\xi)}~\frac{\sin(\pi m y)~\sin(\pi m \eta)}{\pi m}.
\end{align}

For the second integral $\widehat{\mathcal{I}}_2$ (where $x - \eta t - \xi \leq0$), we will consider a  sequence $\big \{\widetilde{\mathbb{\Gamma}}_N\big\}^{\infty}_{N=1}$ of contours in the lower half--plane which is symmetric to the sequence $\big \{\mathbb{\Gamma}_N\big\}^{\infty}_{N=1}$ (see Fig.\ref{fig:up_low_contours} (b)). Evaluating a limit as $N\rightarrow \infty$, we obtain
\begin{align} \label{eq:5.24}
\widehat{\mathcal{I}}_2=&~ - \int_0^{r_0} d\xi ~ \int_{\frac{x-\xi}{t}}^y d \eta~ \chi(\xi, \eta)~ \sum_{m=1}^{\infty} \text{Res}\bigg\{e^{ik(x-\eta t - \xi)}~\frac{\sinh(k(1-y))\sinh(k\eta)}{k~\sinh(k)}, ~-\pi m i\bigg\} \nonumber \\
=&~-i \int_0^{r_0} d\xi~ \int_{\frac{x-\xi}{t}}^y d \eta~ \chi(\xi, \eta)~ \sum_{m=1}^{\infty} e^{\pi m (x - \eta t -\xi)}~ \frac{\sin(\pi m y)~\sin(\pi m \eta)}{\pi m}. 
\end{align}
 Summing the integrals (\ref{eq:5.23}) and (\ref{eq:5.24}) together, we obtain
 \be \label{eq:5.25} \widehat{\mathbf{I}}_1 = 2\pi \int_0^{r_0} d\xi~\int_0^y d\eta~ \chi(\xi, \eta)~ \sum_{m=1}^{\infty} e^{-\pi m |x - \eta t -\xi|}~~\frac{\sin(\pi m y)~\sin(\pi m \eta)}{\pi m}. \ee

On the next step, we show that this formula is valid for any moment of time $t>x/y$. Let $BL$ (the dashed line on Fig.\ref{fig:BKL}) be the straight line $x-\eta t -\xi = 0 $ corresponding to such $t$.  The domain of integration for $\widehat{\mathbf{I}}_1$ consists of two trapezoidal subdomains, $AKL0$ (where $x-\eta t-\xi >0$) and $KEDL$ (where $x-\eta t-\xi <0$). Therefore, we have
\begin{align}  \label{eq:5.26}
\frac{1}{2\pi i}~ \widehat{\mathbf{I}}_1  =& ~\frac{1}{2\pi i} \int_0^{r_0} d\xi~\int_0^{\frac{x-\xi}{t}} d\eta~\chi(\xi,\eta)~ \int_{-\infty}^{\infty} dk~e^{ik(x-\eta t -\xi)}~\frac{\sinh(k(1-y))~\sinh(k\eta)}{k~\sinh(k)} \nonumber \\
    &+~\frac{1}{2\pi i} \int_0^{r_0} d\xi~\int_{\frac{x-\xi}{t}}^y d\eta~\chi(\xi,\eta)~ \int_{-\infty}^{\infty}dk~ e^{ik(x-\eta t -\xi)}~\frac{\sinh(k(1-y))~\sinh(k\eta)}{k~\sinh(k)}.
\end{align}

\noindent To evaluate the first integral from (\ref{eq:5.26}), we consider a sequence of expanding contours in the upper half--plane of the complex $k$-plane (see Fig.\ref{fig:up_low_contours} (a)). Passing to the limit as $N\rightarrow \infty$, we get the result
\be \label{eq:5.27} -i\int_0^{r_0}d\xi~\int_0^{\frac{x-\xi}{t}}d\eta~g(\xi,\eta) ~ \sum_{m=1}^{\infty} e^{-\pi m(x-\eta t -\xi)}~~\frac{\sin(\pi m y)~\sin(\pi m \eta)}{\pi m}. \ee
To evaluate the second integral from (\ref{eq:5.26}), we consider a sequence of expanding contours in the lower half--plane (see Fig.\ref{fig:up_low_contours} (b)). Passing to the limit as $N\rightarrow \infty$, we get the result
\be \label{eq:5.28} -i\int_0^{r_0}d\xi~\int_{\frac{x-\xi}{t}}^yd\eta~\chi(\xi,\eta) ~ \sum_{m=1}^{\infty} e^{\pi m(x-\eta t -\xi)}~~\frac{\sin(\pi m y)~\sin(\pi m \eta)}{\pi m}. \ee
Summing up (\ref{eq:5.27}) and (\ref{eq:5.28}), we obtain formula (\ref{eq:5.25}).

Now we turn to the evaluation of the integral $\widehat{\mathbf{I}}_3$ (where $x-\eta t - \xi <0$). We introduce a sequence of expanding contours closed in the lower half--plane (see Fig.\ref{fig:up_low_contours} (b)). Passing to the limit as $N\rightarrow \infty$, we get
\begin{align} \label{eq:5.29}
~\frac{1}{2 \pi i} ~\widehat{\mathbf{I}}_3 =&~- \int_0^{r_0} d\xi~ \int_y^1d\eta~ \chi(\xi, \eta)~ \sum_{m=1}^{\infty} \text{Res}\bigg\{e^{ik(x-\eta t - \xi)}~\frac{\sinh(k(1-\eta))\sinh(k y)}{k~\sinh(k)}, ~-\pi m i\bigg\} \nonumber \\
=&~-i \int_0^{r_0} d\xi~ \int_y^1 d \eta~ \chi(\xi, \eta)~ \sum_{m=1}^{\infty} e^{-\pi m |x - \eta t -\xi|}~ \frac{\sin(\pi m y)~\sin(\pi m \eta)}{\pi m}. 
\end{align}
Using formulae (\ref{eq:5.27}) for $\widehat{\mathbf{I}}_1$ and (\ref{eq:5.29}) for $\widehat{\mathbf{I}}_3$, we obtain the result (\ref{eq:5.17}). 

The Case A is complete.

  \textbf{Case B.}
Let us evaluate the integrals $\widehat{\mathbf{I}}_1$ and $\widehat{\mathbf{I}}_3$ for $t = (x-r_0)/y.$ A straight line defined as $x-\eta t -\xi =0$ passes through the points $B$ and $E$ with the coordinates $(\xi = r_0, \eta = y)$ (see Fig.\ref{fig:BMNP}).
\begin{figure}[H]
\begin{center}
\includegraphics[scale = 0.3] {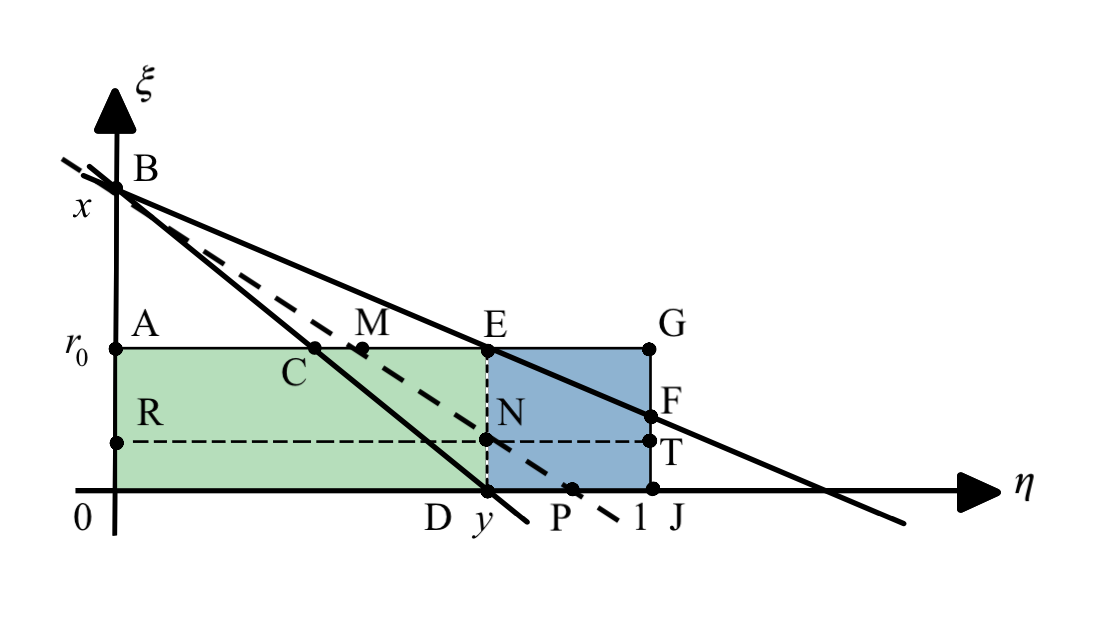} 
\caption{Positions of the straight lines for the Case B}
\label{fig:BMNP}
\end{center}
\end{figure}

In what follows it is convenient to introduce notations
\begin{align}  
\frak{P}(\xi,\eta) =&~\frac{1}{2\pi i}~ \chi(\xi,\eta) \int_{-\infty}^{\infty} dk~ e^{ik(x-\eta t -\xi)}~\frac{\sinh(k(1-\eta)) \sinh(ky)}{k~\sinh(k)}, \label{eq:5.30} \\
\frak{S}(\xi,\eta) =&~-i~ \chi(\xi,\eta) \sum_{m=1}^{\infty} e^{-\pi m|x-\eta t -\xi|}~\frac{\sin(\pi m\eta) \sin(\pi m y)}{\pi m}. \label{eq:5.31}
\end{align}

To evaluate $\widehat{\mathbf{I}}_1$, we notice that the domain of integration is located below the line $BE$ (where $x - \eta t - \xi >0$).  We choose a sequence of closed contours in the upper half--plane (see Fig.\ref{fig:up_low_contours} (a)). Passing to the limit as $N\rightarrow \infty$, we get 
\begin{align} \label{eq:5.32}
\frac{1}{2\pi i }~\widehat{\mathbf{I}}_1
=&~ \int_0^{r_0} d\xi~ \int_0^y d\eta~g(\xi, \eta)~ \sum_{m=1}^{\infty} \text{Res}\bigg\{ e^{ik(x-\eta t - \xi)}~\frac{\sinh(k(1-y))\sinh(k\eta)}{k~\sinh(k)}, ~\pi m i \bigg\} \nonumber\\
=& \int_0^{r_0} d\xi~ \int_0^y d\eta ~\frak{S}(\xi, \eta),
\end{align}
which coincides with formula (\ref{eq:5.25}) for $\widehat{\mathbf{I}}_1$.

To evaluate $\widehat{\mathbf{I}}_3$ we have to consider two subdomains: the trapezoidal subdomain $EFJD$ and triangular subdomain $EGF$, which yields the decomposition
\begin{align} \label{eq:5.33}
\frac{1}{2 \pi i}~ \widehat{\mathbf{I}}_3 \equiv &~\widehat{\mathcal{I}}_3 + \widehat{\mathcal{I}}_4  
=  \int_0^{r_0} d\xi~ \int_y^{\frac{x - \xi}{t}} d\eta ~\frak{P}(\xi,\eta)  + \int_0^{r_0} d\xi~ \int_{\frac{x - \xi}{t}}^1 d\eta ~ \frak{P}(\xi,\eta).
\end{align}
For the first integral $\widehat{\mathcal{I}}_3$ (where $x - \eta t - \xi>0$) we choose a sequence of closed contours in the upper half--plane (see Fig.\ref{fig:up_low_contours} (a)).  For the second integral $\widehat{\mathcal{I}}_4$ (where $x-\eta t - \xi<0$) we choose a sequence of closed contours in the lower half--plane (see Fig.\ref{fig:up_low_contours} (b)). Evaluating the integral $\widehat{\mathcal{I}}_3$, we find
\begin{align} \label{eq:5.34}
\widehat{\mathcal{I}}_3=&~\frac{1}{2 \pi i} \int_0^{r_0} d\xi~ \int_y^{\frac{x - \xi}{t}} d\eta ~\chi(\xi, \eta)~ \sum_{m=1}^{\infty} \text{Res} \bigg\{e^{ik(x-\eta t - \xi)}~\frac{\sinh(k(1-y))\sinh(k\eta)}{k~\sinh(k)}, \pi m i \bigg\}  \nonumber \\
= & \int_0^{r_0} d\xi~ \int_y^{\frac{x - \xi}{t}} d\eta ~\frak{S}(\xi,\eta). 
\end{align}

\noindent Evaluating the integral $\widehat{\mathcal{I}}_4$, we find
\begin{align} \label{eq:5.35}
\widehat{\mathcal{I}}_4=&~\frac{1}{2 \pi i} \int_0^{r_0} d\xi~ \int_{\frac{x - \xi}{t}}^1 d\eta ~\chi(\xi, \eta)~ \sum_{m=1}^{\infty} \text{Res} \bigg\{e^{ik(x-\eta t - \xi)}~\frac{\sinh(k(1-y))\sinh(k\eta)}{k~\sinh(k)}, -\pi m i \bigg\}  \nonumber \\
= & \int_0^{r_0} d\xi~ \int_{\frac{x - \xi}{t}}^1 d\eta ~\frak{S}(\xi,\eta). 
\end{align}

Summing (\ref{eq:5.34}) and (\ref{eq:5.35}), we obtain the result for $\widehat{\mathcal{I}}_3$ similar to (\ref{eq:5.29}). Using formulae (\ref{eq:5.32}) for $\widehat{\mathcal{I}}_1$ and (\ref{eq:5.34}), (\ref{eq:5.35}) for $\widehat{\mathcal{I}}_3$, we obtain the result of (\ref{eq:5.17}).

On the next step we show that the formula for $(\widehat{\mathbf{I}}_1 + \widehat{\mathbf{I}}_3)$ is valid for any $t$ such that $x/y>t\geq (x-r_0)/y$. Let $BP$ (the dashed line on Fig.\ref{fig:BMNP}) be the straight line $x-\eta t \xi=0$ located between $BD$ and $BF$. The domain of integration for $\widehat{\mathbf{I}}_1$ consists of ($i$) a rectangular subdomain $RND0$, ($ii$) a trapezoidal subdomain $RAMN$, and ($iii$) a triangular subdomain $MEN$.  Evaluating the integral over the rectangle $RND0$, denoted by $\widehat{\mathcal{I}}_5$, yields 
\be \label{eq:5.36}
\widehat{\mathcal{I}}_5 \equiv \int_0^R d\xi~\int_0^y d\eta~\frak{P}(\xi,\eta)= \int_0^R d\xi~\int_0^y d\eta~\frak{S}(\xi,\eta).\ee

\noindent Evaluating the integral over the rectangle $RAMN$, denoted by $\widehat{\mathcal{I}}_6$, yields 
\be \label{eq:5.37}
\widehat{\mathcal{I}}_6 \equiv \int_R^{r_0} d\xi~\int_0^{\frac{x-\xi}{t}} d\eta~\frak{P}(\xi,\eta)= \int_R^{r_0} d\xi~\int_0^{\frac{x-\xi}{t}} d\eta~\frak{S}(\xi,\eta).\ee

\noindent Evaluating the integral over the triangle $MEN$, denoted by $\widehat{\mathcal{I}}_7$,  yields
\be \label{eq:5.38}
\widehat{\mathcal{I}}_7 \equiv \int_R^{r_0} d\xi~\int_{\frac{x-\xi}{t}}^y d\eta~\frak{P}(\xi,\eta)=\int_R^{r_0} d\xi~\int_{\frac{x-\xi}{t}}^y d\eta~\frak{S}(\xi,\eta).\ee

\noindent Summing results of (\ref{eq:5.36}) -- (\ref{eq:5.38}), we obtain that for $~\widehat{\mathbf{I}}_1 = \widehat{\mathcal{I}}_5 + \widehat{\mathcal{I}}_6 + \widehat{\mathcal{I}}_7~$ the formula (\ref{eq:5.32}) holds.

To evaluate $\widehat{\mathbf{I}}_3$, we notice that the domain of integration is a union $(i)$ a rectangular subdomain $NEGT$, $(ii)$ a trapezoidal subdomain $NTJP$, and $(iii)$ a triangular subdomain $DNP$. To deal with the triangular subdomain, we consider a sequence of expanding contours in the upper half--plane and to deal with the two remaining subdomains, we consider sequences of expanding contours in the lower half--plane. Evaluating the integral over the triangle $DNP$, denoted by $\widehat{\mathcal{I}}_ 8$, yields
\be \label{eq:5.39}
\widehat{\mathcal{I}}_8 \equiv \int_0^N d\xi \int_y^{\frac{x-\xi}{t}} d\eta~\frak{P}(\xi,\eta)=\int_0^N d\xi \int_y^{\frac{x-\xi}{t}} d\eta~\frak{S}(\xi,\eta).
\ee
\noindent Evaluating the integral over the trapezoid $NTJP$, denoted by $\widehat{\mathcal{I}}_9$, yields
\be \label{eq:5.40}
\widehat{\mathcal{I}}_9 \equiv \int_0^N d\xi \int_{\frac{x-\xi}{t}}^1 d\eta~\frak{P}(\xi,\eta)=\int_0^N d\xi \int_{\frac{x-\xi}{t}}^1 d\eta~\frak{S}(\xi,\eta).\ee

\noindent Summing (\ref{eq:5.39}) and (\ref{eq:5.40}) we get
\be \label{eq:5.41}
\widehat{\mathcal{I}}_{10} \equiv~\widehat{\mathcal{I}}_8 + \widehat{\mathcal{I}}_9= \int_0^N d\xi \int_y^1 d\eta~\frak{S}(\xi,\eta).\ee

\noindent Evaluating the integral over the rectangle $NEGT$ yields
\be \label{eq:5.42}
\widehat{\mathcal{I}}_{11}  \equiv \int_N^E d\xi \int_y^1 d\eta~\frak{P}(\xi,\eta)
= \int_N^E d\xi \int_y^1 d\eta~\frak{S}(\xi,\eta).\ee

\noindent Summing (\ref{eq:5.40}) -- (\ref{eq:5.42}) we obtain that for
$~\widehat{\mathbf{I}}_3 =\widehat{\mathcal{I}}_{10}+\widehat{\mathcal{I}}_{11}$ the formula (\ref{eq:5.29}) holds. Summing up formula (\ref{eq:5.32}) for $\widehat{\mathbf{I}}_1$ and (\ref{eq:5.29}) for $\widehat{\mathbf{I}}_3$ we obtain the result of (\ref{eq:5.17}). 

The Case B is complete.

 \textbf{Case C.} Let us evaluate the integrals $\widehat{\mathbf{I}}_1$ and $\widehat{\mathbf{I}}_3$ for the moment $t = x - r_0$. A straight line $x - \eta t -\xi = 0$ passes through the points $B$ and $G$ with the coordinates $(\xi = r_0,~ \eta = 1)$ (see Fig.\ref{fig:BQS}).
\begin{figure}[H]
\begin{center}
\includegraphics[scale = 0.28] {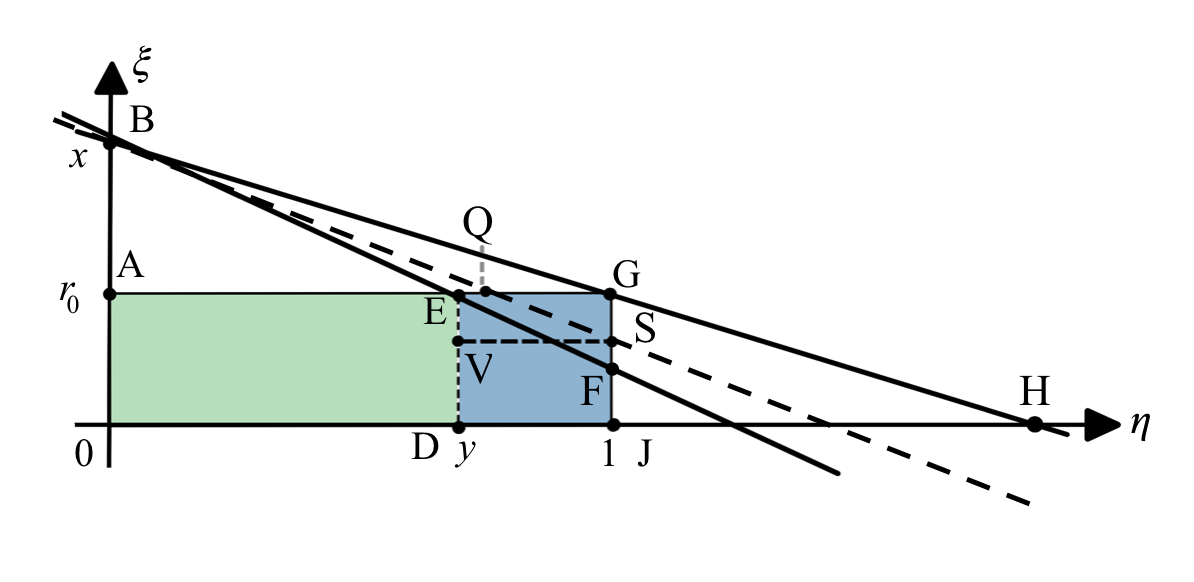} 
\caption{Positions of straight lines for the Case C}
\label{fig:BQS}
\end{center}
\end{figure}
We notice that for both $\widehat{\mathbf{I}}_1$ and $\widehat{\mathbf{I}}_3$ all points in their respective subdomains are below the line $x - \eta t -\xi=0$ (where  $x - \eta t -\xi>0$). To approximate both integrals we choose sequences of closed contours with in the upper half--plane. We obtain the results similar to (\ref{eq:5.29}) and (\ref{eq:5.32}), which yields the result similar to (\ref{eq:5.17}) for the sum $(~\widehat{\mathbf{I}}_1 + \widehat{\mathbf{I}}_3)$.

Now we show that the formula for $(~\widehat{\mathbf{I}}_1 + \widehat{\mathbf{I}}_3)$ is valid for $(x-r_0)/y > t \geq x-r_0$. Let $BS$ (the dashed line on Fig.\ref{fig:BQS}) be the straight line $x-\eta t -\xi = 0$ located between $BF$ and $BH$. It can be readily checked that the evaluation of~$\widehat{\mathbf{I}}_1$, for any position of the line $x - \eta t - \xi = 0$ above $BF$, yields the result similar to (\ref{eq:5.32}).

The domain of integration for $\widehat{\mathbf{I}}_3$  is a union of $(i)$ a rectangular subdomain $DVSJ$, $(ii)$ a trapezoidal subdomain $VEQS$, and $(iii)$ a triangular subdomain $QGS$ (Fig.\ref{fig:BQS}). To evaluate $\widehat{\mathbf{I}}_3$ on the subdomains $DVSJ$ and $VEQS$, we consider sequences of expanding contours in the upper half--plane and for the subdomain $QGS$, we consider sequences of expanding contours in the lower--half plane. Evaluating the integral on the rectangle $DVSJ$  and the the trapezoid $VEQS$, denoted by $\widehat{\mathcal{I}}_{12}$ and $\widehat{\mathcal{I}}_{13}$, yields
\be \label{eq:5.43} 
\widehat{\mathcal{I}}_{12} =\int_0^V d\xi \int_y^1 d\eta~\frak{S}(\xi,\eta), \qquad 
\widehat{\mathcal{I}}_{13} = \int_V^E d\xi \int_y^{\frac{x-xi}{t}}  d\eta~\frak{S}(\xi,\eta).
\ee
\noindent Evaluating the integral over the triangle $QGS$, denoted by $\widehat{\mathcal{I}}_{14}$,  yields
\be\label{eq:5.44}
\widehat{\mathcal{I}}_{14} \equiv  \int_V^{r_0} d\xi \int_{\frac{x-\xi}{t}}^1 d\eta~\frak{P}(\xi,\eta) =\int_V^{r_0} d\xi \int_{\frac{x-\xi}{t}}^1  d\eta~\frak{S}(\xi,\eta).\ee

\noindent Taking the sum $~\widehat{\mathcal{I}}_{12} +\widehat{\mathcal{I}}_{13} +\widehat{\mathcal{I}}_{14}$, we obtain the result similar to (\ref{eq:5.29}).  Summing the results for $\widehat{\mathbf{I}}_1$ and $\widehat{\mathbf{I}}_3$ we find the results similar to (\ref{eq:5.17}). 

The Case C is complete.

The combination of the results for the Cases A, B, and C yields the proof.

The theorem is proven. \end{proof}

\section{Evaluation of integrals $\widehat{\mathbf{I}}_2$ and $\widehat{\mathbf{I}}_4$ of (\ref{eq:5.14}) and (\ref{eq:5.16})}

We focus on the case $x>r_0$. Our goal is to show that both integrals can be evaluated by using the Residue theorem. Without misunderstanding we will use the notation $\ell_n$ as in (\ref{eq:4.17}) for a contour on the complex $k$-plane defined by
\be \label{eq:6.1} \ell_n = \{ k:k\in(-\infty,-\delta_n)~ \bigcup ~\gamma_n ~\bigcup~ (\delta_n,\infty), ~\gamma_n = \delta_n e^{i\varphi},~ -\pi\leq \varphi \leq 0\}. \ee

\indent \begin{customlemma}{}\label{lemma6.1}\textbf{Lemma 6.1.} 
\textit{1) The following representation holds for the integrals $\widehat {\mathbf{I}}_2$  defined in (\ref{eq:5.14}):}
\begin{align} \label{eq:6.2}
\widehat {\mathbf{I}}_2 
=& \lim_{\delta_n \rightarrow 0} \int_0^{r_0} d\xi \int_0^y d\eta~ \chi(\xi,\eta)\landdownint_{\ell_n}  dk ~e^{ik(x-\eta t -\xi)}~\frac{\sinh(k(1-y))\cosh(k\eta)}{k~\sinh(k)}\nonumber \\&-\pi i ~(1-y) \int_0^{r_0} d\xi \int_0^y d\eta~ \chi(\xi,\eta).
\end{align}
\textit{ 2) The following representation holds for the integral $\widehat{\mathbf{I}}_4$ defined in (\ref{eq:5.16}):}
\begin{align} \label{eq:6.3}
\widehat {\mathbf{I}}_4 =& -\lim_{\delta_n \rightarrow 0} \int_0^{r_0} d\xi \int_y^1 d\eta~ \chi(\xi,\eta)\landdownint_{\ell_n}  dk ~e^{ik(x-\eta t -\xi)}~\frac{\sinh(ky)\cosh(k(1-\eta))}{k~\sinh(k)}\nonumber \\&+\pi i~ y  \int_0^{r_0} d\xi \int_y^1 d\eta~ \chi(\xi,\eta).
\end{align}
\end{customlemma}

\indent \begin{proof} Using the definition of a principal value integral, we represent $\widehat{\mathbf{I}}_2$ of (\ref{eq:5.14}) as follows:
\begin{align} \label{eq:6.4}
\widehat {\mathbf{I}}_2 =& \lim_{\delta_n\rightarrow 0} \int_0^{r_0} d\xi \int_0^y d\eta~\chi(\xi,\eta) \landdownint_{\ell_n} dk ~e^{ik (x-\eta t -\xi)}~\frac{\sinh(k(1-y)) \cosh(k\eta)}{k~\sinh(k)} \nonumber \\
&~- \lim_{\delta_n \rightarrow 0} \int_0^{r_0} d\xi \int_0^y d\eta~ \chi(\xi,\eta)\awint_{\gamma_n}  dk ~e^{ik(x-\eta t -\xi)}~\frac{\sinh(k(1-y))\cosh(k\eta)}{k~\sinh(k)},
\end{align}
where $\awint_{\gamma_n}$ denotes the integral along $\gamma_n$. To evaluate the integral over $\gamma_n$ at the vicinity of $k=0$, we set $k=\delta_n e^{i\varphi}$ and use the approximations: $\sinh(k(1-y)) = \delta_n e^{i\varphi}(1-y)(1+\mathcal{O}(\delta_n^2))$,  $\cosh(k\eta)=(1+\mathcal{O}(\delta_n^2))$. Substituting these formulae into (\ref{eq:6.4}) we obtain for the second limit of (\ref{eq:6.4}) 
\[\lim_{\delta_n\rightarrow 0} \int_0^{r_0} d\xi\int_0^y d\eta ~\chi(\xi,\eta)\int_{-\pi}^0 
(1-y)(1+\mathcal{O}(\delta_n^2))~i~ d\varphi 
=~ \pi i~(1-y) \int_0^{r_0} d\xi\int_0^y d\eta ~\chi(\xi,\eta).\]
Substituting the result into formula (\ref{eq:6.4}) we obtain the result of (\ref{eq:6.2}). It can be readily seen that $\widehat{\mathbf{I}}_4$ can be represented in the form
\begin{align} \label{eq:6.5}
\widehat {\mathbf{I}}_4 =&- \lim_{\delta_n\rightarrow 0} \int_0^{r_0} d\xi \int_y^1 d\eta~\chi(\xi,\eta) \landdownint_{\ell_n} dk ~e^{ik (x-\eta t -\xi)}~\frac{\sinh(ky) \cosh(k(1-\eta))}{k~\sinh(k)} \nonumber \\
&~+ \lim_{\delta_n \rightarrow 0} \int_0^{r_0} d\xi \int_y^1 d\eta~ \chi(\xi,\eta)\awint_{\gamma_n}  dk ~e^{ik(x-\eta t -\xi)}~\frac{\sinh(ky)\cosh(k(1-\eta))}{k~\sinh(k)}.
\end{align}
Evaluating the integral over $\gamma_n$, we obtain
\[ \lim_{\delta_n\rightarrow 0} \int_0^{r_0} d\xi\int_y^1 d\eta ~\chi(\xi,\eta)~\int_{-\pi}^0  y~(1+\mathcal{O}(\delta_n^2))~i~ d\varphi = \pi i ~y \int_0^{r_0} d\xi \int_y^1 d\eta ~\chi(\xi,\eta),\]
which yields (\ref{eq:6.3}). 

The lemma is shown. \end{proof}

The main result of this section is the following statement.

\begin{customthm}{}\label{theorem6.2}\textbf{Theorem 6.2.} \textit{Let $\frak{G}(x,y,t)$ be defined as}
\be \label{eq:6.6}
\frak{G}(x,y,t) = ~-2\pi i \int_0^{r_0} d\xi \int_0^1 d\eta~\chi(\xi,\eta)~\frac{x-\eta t -\xi}{|x - \eta t -\xi|}~\sum_{m=1}^{\infty} e^{-\pi m|x - \eta t -\xi|}~\frac{\sin(\pi m y)\cos(\pi m \eta)}{\pi m}.
 \ee
\textit{For $x\geq r_0$ the following representations are valid for the sum $(\widehat{\mathbf{I}}_2 + \widehat{\mathbf{I}}_4$).\\
\noindent (a) When $~t\geq x/y$, then}
\begin{align}\label{eq:6.7} 
\widehat{\mathbf{I}}_2 + \widehat{\mathbf{I}}_4 =& ~\frak{G}(x,y,t) + 2\pi i~(1-y) \int_0^{r_0} d\xi \int_0^{\frac{x-\xi}{t}} d\eta~\chi(\xi,\eta) \nonumber \\
&-\pi i~(1-y) \int_0^{r_0}d\xi \int_0^y d\eta~\chi(\xi,\eta) + \pi i~ y \int_0^{r_0} d\xi \int_y^1 d\eta~ \chi(\xi,\eta). 
\end{align}

\noindent \textit{(b) When $~x/y > t \geq (x-r_0)/y$, then}
\begin{align}\label{eq:6.8} 
\widehat{\mathbf{I}}_2 + \widehat{\mathbf{I}}_4 =&~ \frak{G}(x,y,t) - 2\pi i~y \int_0^{r_0} d\xi \int_y^{\frac{x-\xi}{t}} d\eta~\chi(\xi,\eta) \nonumber \\
&+\pi i~y \int_0^{r_0}d\xi \int_y^1 d\eta~\chi(\xi,\eta) + \pi i~ (1-y) \int_0^{r_0} d\xi \int_0^y d\eta~ \chi(\xi,\eta). 
\end{align}

\noindent \textit{(c) When $~(x-r_0)/y > t\geq (x-r_0)$, then}
\be \label{eq:6.9}
\widehat{\mathbf{I}}_2 + \widehat{\mathbf{I}}_4 =~ \frak{G}(x,y,t) -\pi i~y \int_0^{r_0} d\xi \int_y^1 d\eta~\chi(\xi,\eta) + \pi i~(1-y) \int_0^{r_0}d\xi \int_0^y d\eta~\chi(\xi,\eta) .\ee
\end{customthm}

\indent \begin{proof} To prove the result we consider three different time intervals, that we call Cases A, B, and C respectively (see the beginning of the proof of Theorem 5.3).

\textbf{Case A.} We consider separately the subcases ($i$) $t = x/y$ and ($ii$) $t>x/y$. Let us evaluate $\widehat{\mathbf{I}}_2$ for $t=x/y$. To evaluate the integral along $\ell_n$, we use Lemma 6.1 and notice that straight line $BD$ (see Fig.\ref{fig:BKL}) splits the domain of integration into two subdomains resulting in the following decomposition:
\begin{align} \label{eq:6.10}
 \frac{1}{2\pi i}~\widehat{\mathbf{I}}_2 \equiv &~\widetilde{\mathcal{I}}_1 + \widetilde{\mathcal{I}}_2 - \frac{(1-y)}{2} \int_0^{r_0} d\xi\int_0^y d\eta ~\chi(\xi,\eta), \quad \text{where}\nonumber\\
\widetilde{\mathcal{I}}_1=&~ \frac{1}{2\pi i}~\lim_{n\rightarrow \infty} \int_0^{r_0} d\xi \int_0^{\frac{x-\xi}{t}} d\eta~\chi(\xi,\eta)~\landdownint_{\ell_n} dk~ e^{ik(x-\eta t -\xi)}~\frac{\sinh(k(1-y)) \cosh(k\eta)}{k~\sinh(k)},\nonumber \\
\widetilde{\mathcal{I}}_2 =&~ \frac{1}{2\pi i}~\lim_{n\rightarrow \infty} \int_0^{r_0} d\xi \int_{\frac{x-\xi}{t}}^y d\eta~\chi(\xi,\eta)~\landdownint_{\ell_n} dk~ e^{ik(x-\eta t -\xi)}~\frac{\sinh(k(1-y)) \cosh(k\eta)}{k~\sinh(k)}.\end{align}
To evaluate $\widetilde{\mathcal{I}}_1$ (where $~x-\eta t -\xi \geq 0$), we notice that integral can be considered as a limit of a sequence of contour integrals in the upper half--plane. Each of the aforementioned integrals can be evaluated using the Residue theorem. We obtain that
\begin{align} \label{eq:6.11}
\widetilde{\mathcal{I}}_1&=\int_0^{r_0} d\xi \int_0^{\frac{x-\xi}{t}} d\eta~\chi(\xi,\eta) \sum_{m=0}^{\infty} ~\text{Res} \bigg\{ e^{ik(x-\eta t -\xi)}~\frac{\sinh(k(1-y))\cosh(k\eta)}{k~\sinh(k)}, ~\pi m i \bigg\} \nonumber \\
&=- \int_0^{r_0} d\xi \int_0^{\frac{x-\xi}{t}} d\eta~\chi(\xi,\eta) \sum_{m=1}^{\infty} ~e^{-\pi m(x-\eta t -\xi)}~\frac{\sin(\pi m y)\cos(\pi m \eta)}{\pi m} \nonumber \\
&~~+(1-y)\int_0^{r_0} d\xi \int_0^{\frac{x-\xi}{t}} d\eta~\chi(\xi,\eta).
\end{align}

\noindent Evaluating $\widetilde{\mathcal{I}}_2$ (where $x-\eta t - \xi<0$), we use a sequence of contours closed in the lower half--plane. Taking into account the direction on each contour we get 
\begin{align} \label{eq:6.12}
\widetilde{\mathcal{I}}_2 &=-\int_0^{r_0} d\xi \int_{\frac{x-\xi}{t}}^y d\eta~\chi(\xi,\eta) \sum_{m=1}^{\infty} ~\text{Res} \bigg\{ e^{ik(x-\eta t -\xi)}~\frac{\sinh(k(1-y))\cosh(k\eta)}{k~\sinh(k)}, ~-\pi m i \bigg\} \nonumber \\
&=\int_0^{r_0} d\xi \int_{\frac{x-\xi}{t}}^y d\eta~\chi(\xi,\eta) \sum_{m=1}^{\infty} ~e^{-\pi m|x-\eta t -\xi|}~\frac{\sin(\pi my)\cos(\pi m \eta)}{\pi m} .
\end{align}
Substituting (\ref{eq:6.11}) and (\ref{eq:6.12}) into (\ref{eq:6.10}) we obtain
\begin{align} \label{eq:6.13}
\frac{1}{2\pi i}~ \widehat{\mathbf{I}}_2 =&~(1-y)\int_0^{r_0} d\xi \int_0^{\frac{x-\xi}{t}} d\eta~\chi(\xi,\eta)\nonumber \\
&-~\int_0^{r_0} d\xi \int_{0}^y d\eta~\chi(\xi,\eta) ~\frac{x-\eta t -\xi}{|x - \eta t -\xi|}~\sum_{m=1}^{\infty} ~e^{-\pi m|x-\eta t -\xi|}~\frac{\sin(\pi my)\cos(\pi m \eta)}{\pi m} \nonumber \\
~&-~\frac{(1-y)}{2}\int_0^{r_0} d\xi \int_0^y d\eta~\chi(\xi,\eta).
\end{align}

\noindent To evaluate the integral $\widehat{\mathbf{I}}_4$, we notice that on the domain of integration $x-\eta t -\xi<0.$ Therefore, we can use a sequence of contours closed in the lower half--plane and have 
\begin{align} \label{eq:6.14}
 \frac{1}{2\pi i}~&\lim_{\delta_n\rightarrow 0} \int_0^{r_0}d\xi \int_y^1 d\eta~\chi(\xi,\eta)~\landdownint_{\ell_n} dk~ e^{ik(x-\eta t -\xi)}~\frac{\sinh(ky) \cosh(k(1-\eta))}{k~\sinh(k)}\nonumber \\
 &=-\int_0^{r_0} d\xi \int_y^1 d\eta~\chi(\xi,\eta) \sum_{m=1}^{\infty} ~\text{Res} \bigg\{e^{ik(x-\eta t -\xi)}~\frac{\sinh(ky)\cosh(k(1-\eta))}{k~\sinh(k)}, ~-\pi m i \bigg\} \nonumber \\
  &=-\int_0^{r_0} d\xi \int_y^1 d\eta~\chi(\xi,\eta) \sum_{m=1}^{\infty} e^{-\pi m|x-\eta t -\xi|}~\frac{\sin(\pi my)\cos(\pi m \eta)}{\pi m}.
 \end{align}

\noindent Combining formulas (\ref{eq:6.3}) and (\ref{eq:6.14}) and Lemma 6.1, we obtain
 \begin{align} \label{eq:6.15}
 \frac{1}{2 \pi i} ~\widehat{\mathbf{I}}_4&-~\frac{y}{2}\int_0^{r_0} d\xi \int_y^1 d\eta~\chi(\xi,\eta) \nonumber \\
 =&~  \int_0^{r_0} d\xi \int_y^1 d\eta~\chi(\xi,\eta) \sum_{m=1}^{\infty} e^{-\pi m|x-\eta t -\xi|}~\frac{\sin(\pi my)\cos(\pi m \eta)}{\pi m}
 \nonumber \\
 =&~ - \int_0^{r_0} d\xi \int_y^1 d\eta~\chi(\xi,\eta)~ \frac{x-\eta t -\xi}{|x-\eta t -\xi|}~ \sum_{m=1}^{\infty} e^{-\pi m|x-\eta t -\xi|}~\frac{\sin(\pi my)\cos(\pi m \eta)}{\pi m}.
 \end{align}
 Summing up formulae (\ref{eq:6.13}) and (\ref{eq:6.15}) yields the result of (\ref{eq:6.7}).

Now we show that formula (\ref{eq:6.7}) is valid for any $t>x/y$. Let $BL$ (the dashed line on Fig.\ref{fig:BKL}) be the straight line $x - \eta t - \xi=0$ corresponding to such $t$. The domain of integration of $\widehat{\mathbf{I}}_2$ consists of two trapezoidal subdomains, $AKL0$ and $KEDL$. In the sequel, it is convenient to use new notations
\begin{align} 
 \frak{V}(\xi,\eta)  =&~\frac{1}{2 \pi i}~\chi(\xi,\eta)~ \text{P.V.} \int_{-\infty}^{\infty} dk~e^{ik(x-\eta t -\xi)}~ \frac{\sinh(k(1-y))\cosh(k\eta)}{k~\sinh(k)},\label{eq:6.16}\\
\frak{W}_{\pm}(\xi,\eta) =&~\chi(\xi,\eta)~\sum_{m=1}^{\infty}~\text{Res}\bigg\{e^{ik(x-\eta t -\xi)}~\frac{\sinh(k(1-y))\cosh(k\eta)}{k~\sinh(k)}, ~\pm \pi m i \bigg\}, \label{eq:6.17}\\
 \frak{U} (\xi,\eta) =&~\chi(\xi,\eta)~\sum_{m=1}^{\infty}e^{-\pi m|x-\eta t -\xi|}~\frac{\sin(\pi m y) \cos(\pi m \eta)}{\pi m}. \label{eq:6.18}
\end{align}
To evaluate the integral over the trapezoid $AKL0$, we consider a sequence of expanding contours in the upper half--plane, and have
\begin{align} \label{eq:6.19}
 \int_0^{r_0} d\xi &\int_0^{\frac{x-\xi}{t}}d\eta~\frak{V}(\xi,\eta) = \int_0^{r_0} d\xi \int_0^{\frac{x-\xi}{t}}d\eta~\frak{W}_+(\xi,\eta) +(1-y)\int_0^{r_0} d\xi \int_0^{\frac{x-\xi}{t}} d\eta ~\chi(\xi,\eta) \nonumber \\
&=-\int_0^{r_0} d\xi \int_0^{\frac{x-\xi}{t}}d\eta~\frak{U}(\xi,\eta)+ (1-y)\int_0^{r_0} d\xi \int_0^{\frac{x-\xi}{t}}d\eta\chi(\xi,\eta).
\end{align}
We notice that this result coincides with (\ref{eq:6.11}).

To evaluate the integral over the trapezoid $KEDL$, we consider a sequence of expanding contours in the lower half--plane and arrive at the following result:
\be \label{eq:6.20}
\frac{1}{2\pi i} \int_0^{r_0} d\xi \int_{\frac{x-\xi}{t}}^y d\eta~\frak{V}(\xi,\eta)
=-\int_0^{r_0} d\xi \int_{\frac{x-\xi}{t}}^yd\eta~\frak{W}_-(\xi,\eta) 
=\int_0^{r_0} d\xi \int_{\frac{x-\xi}{t}}^yd\eta~\frak{U}(\xi,\eta)
\ee
which coincides with (\ref{eq:6.12}). Summing equations (\ref{eq:6.19}) and (\ref{eq:6.20}) and using Lemma 6.1, we obtain
\begin{align}\label{eq:6.21}
 \frac{1}{2\pi i}~\widehat{\mathbf{I}}_2 =& -\int_0^{r_0} d\xi \int_0^yd\eta ~\frac{x-\eta t - \xi}{|x - \eta t -\xi|}~~\frak{U}(\xi,\eta) +(1-y)\int_0^{r_0} d\xi \int_0^{\frac{x-\xi}{t}} d\eta~\chi(\xi,\eta)\nonumber \\
 &~-~\frac{(1-y)}{2}\int_0^{r_0} d\xi \int_0^y d\eta ~\chi(\xi,\eta).
 \end{align}

To evaluate $\widehat{\mathbf{I}}_4$ we notice that the domain of its integration is located above the line $x - \eta t - \xi = 0$. We consider a sequence of contours closed in the lower half--plane and get
\be \label{eq:6.22}
-~\frac{1}{2\pi i} \int_0^{r_0} d\xi \int_y^1d\eta~\frak{V}(\xi,\eta)
=\int_0^{r_0} d\xi \int_y^1 d\eta~\frak{W}_-(\xi,\eta)
=\int_0^{r_0} d\xi \int_y^1 d\eta~\frak{U}(\xi,\eta) .
\ee
Using (\ref{eq:6.22}) and Lemma 6.1, we obtain for the integral $\widehat{\mathbf{I}}_4:$
\be \label{eq:6.23}
 \widehat{\mathbf{I}}_4 = -2\pi i \int_0^{r_0} d\xi \int_y^1d\eta ~\frac{x-\eta t - \xi}{|x - \eta t -\xi|}
~~\frak{U}(\xi,\eta) +\pi i~ y \int_0^{r_0} d\xi \int_y^1 d\eta~\chi(\xi,\eta).
 \ee
 This formula is essentially the same as (\ref{eq:6.15}). Summing equations (\ref{eq:6.21}) and (\ref{eq:6.23}) together yields the result
 \begin{align} \label{eq:6.24}
 \widehat{\mathbf{I}}_2 + \widehat{\mathbf{I}}_4 =& -2\pi i \int_0^{r_0} d\xi \int_0^1 d\eta~\frac{x-\eta t -\xi}{|x - \eta t -\xi|}~\frak{U}(\xi,\eta) +\pi i~ y \int_0^{r_0} d\xi \int_y^1 d\eta~\chi(\xi,\eta )\nonumber\\
  &~ +2\pi i~(1-y)\int_0^{r_0} d\xi \int_0^{\frac{x-\xi}{t}} d\eta~\chi(\xi,\eta)~-~\pi i(1-y)\int_0^{r_0} d\xi \int_0^y d\eta ~\chi(\xi,\eta).
\end{align}
We notice that formulae (\ref{eq:6.16}) and (\ref{eq:6.24}) coincide. 

Case A is shown.

 \textbf{Case B.} We evaluate the integrals for $\widehat{\mathbf{I}}_2$ and $\widehat{\mathbf{I}}_4$ for $t =(x-r_0)/y$. A straight line $x - \eta t - \xi = 0$ passes through the points $B$ and $E$ with the coordinates $(\xi = r_0, \eta = y)$ (see Fig.\ref{fig:BMNP}). The domain of integration for $\widehat{\mathbf{I}}_2$ corresponds to the case $x-\eta t -\xi>0$. The integral along $\ell_n$ can be evaluated as a limit of a sequence of contour integrals in the upper half--plane. Using Lemma 6.1, we have
\begin{align}\label{eq:6.25}
\frac{1}{2\pi i}~ \widehat {\mathbf{I}}_2~ =&~ \frac{1}{2\pi i}~\lim_{\delta_n\rightarrow 0} \int_0^{r_0} d\xi \int_0^y d\eta~\chi(\xi,\eta)~\landdownint_{\ell_n} dk~ e^{ik(x-\eta t -\xi)}~\frac{\sinh(k(1-y)) \cosh(k\eta)}{k~\sinh(k)}\nonumber \\
&~-\frac{(1-y)}{2} \int_0^{r_0} d\xi\int_0^y d\eta ~\chi(\xi,\eta) \nonumber \\
 =&~\frac{(1-y)}{2} \int_0^{r_0} d\xi\int_0^y d\eta ~\chi(\xi,\eta) 
-\int_0^{r_0} d\xi \int_0^y d\eta~ \frac{x-\eta t -\xi}{|x-\eta t -\xi|}~ \frak{U}(\xi,\eta). \end{align}
To evaluate $\widehat{\mathbf{I}}_4$ we split the domain of integration into the trapezoidal subdomain $EDJF$ and the triangular subdomain $EGF$, which yields the decomposition:
\begin{align} \label{eq:6.26}
\frac{1}{2\pi i}~ \widehat{\mathbf{I}}_4~ =&~ ~ \widetilde {\mathcal{I}}_3 +  \widetilde {\mathcal{I}}_4+\frac{y}{2}\int_{0}^{r_0} d\xi \int_y^1 d\eta~\chi(\xi,\eta) \quad \text{ where,}\nonumber \\
 \widetilde {\mathcal{I}}_3 =&~ -\frac{1}{2\pi i}~\lim_{\delta_n \rightarrow 0}\int_0^{r_0} d\xi \int_y^{\frac{x-\xi}{t}} d\eta~\chi(\xi,\eta)~\landdownint_{\ell_n} dk~ e^{ik(x-\eta t -\xi)}~\frac{\sinh(ky) \cosh(k(1-\eta))}{k~\sinh(k)},\nonumber \\
 \widetilde {\mathcal{I}}_4 =& ~- \frac{1}{2\pi i}~\lim_{\delta_n \rightarrow 0}~\int_0^{r_0} d\xi \int_{\frac{x-\xi}{t}}^1 d\eta~\chi(\xi,\eta)~\landdownint_{\ell_n} dk~ e^{ik(x-\eta t -\xi)}~\frac{\sinh(ky) \cosh(k(1-\eta))}{k~\sinh(k)}.
\end{align}
To evaluate the $ \widetilde {\mathcal{I}}_3$ (where $x - \eta t - \xi >0$) we consider a sequence of contours in the upper half--plane and get
\begin{align}\label{eq:6.27}
 \widetilde {\mathcal{I}}_3 =&- \int_0^{r_0} d\xi \int_y^{\frac{x-\xi}{t}} d\eta~\chi(\xi,\eta)~\frak{W}_+(\xi,\eta)\nonumber \\
 =&- \int_0^{r_0} d\xi \int_y^{\frac{x-\xi}{t}} d\eta~\chi(\xi,\eta)~ \frak{U}(\xi,\eta)~-~ y\int_0^{r_0} d\xi \int_y^{\frac{x-\xi}{t}} d\eta~\chi(\xi,\eta).
\end{align}
To evaluate $\widetilde {\mathcal{I}}_4$ (where $x-\eta t -\xi<0$) we consider a sequence of contours in the lower half--plane and get
\be\label{eq:6.28}
 \widetilde {\mathcal{I}}_4=~-\int_0^{r_0} d\xi \int_{\frac{x-\xi}{t}}^1 d\eta~\chi(\xi,\eta)~\frak{W}_-(\xi,\eta) =-\int_0^{r_0} d\xi \int_{\frac{x-\xi}{t}}^1 d\eta~ \frac{x-\eta t -\xi}{|x-\eta t -\xi|}~\frak{U}(\xi,\eta).
\ee
Summing (\ref{eq:6.27}) and (\ref{eq:6.28}) and taking into account (\ref{eq:6.3}) yields
\begin{align}\label{eq:6.29}
\frac{1}{2\pi i} ~ \widehat{\mathbf{I}}_4 =& ~- \int_0^{r_0} d\xi \int_y^1 d\eta~ \frac{x-\eta t -\xi}{|x-\eta t -\xi|}~ \frak{U}(\xi,\eta)\nonumber \\ 
&-~ y\int_0^{r_0} d\xi \int_y^{\frac{x-\xi}{t}} d\eta~\chi(\xi,\eta) +~ 
\frac{y}{2} \int_0^{r_0} d\xi \int_y^1 d\eta~\chi(\xi,\eta).
\end{align}
Combining together formulae (\ref{eq:6.25}) for $\widehat{\mathbf{I}}_2$ and (\ref{eq:6.29}) for $\widehat{\mathbf{I}}_4$, we obtain the result of (\ref{eq:6.8}).

Now we show that the formula for $(~\widehat{\mathbf{I}}_2 + \widehat{\mathbf{I}}_4)$ is valid for any $t$: $x/y>t\geq (x-r_0)/y$. Let $BP$ (the dashed line on Fig.\ref{fig:BMNP}) be the straight  line $x - \eta t -\xi = 0$ corresponding to such $t$, i.e. located between $BD$ and $BF$. 

The domain of integration for $\widehat{\mathbf{I}}_2$ is a union of $(i)$ a rectangular subdomain $RDN0$, $(ii)$ a trapezoidal subdomain $RAMN$, and $(iii)$ a triangular subdomain $MEN$. For both subdomains, $RND0$ and $RAMN$, we consider sequences of contours of integration for $\widehat{\mathbf{I}}_2$ closed in the upper half--plane. Denoting the integral over the rectangle $RND0$ by $\widetilde{\mathcal{I}}_5$ we get
\be\label{eq:6.30}
\widetilde{\mathcal{I}}_5~=-\int_0^R d\xi \int_0^yd\eta ~\frak{U}(\xi,\eta) + (1-y)\int_0^R d\xi \int_0^yd\eta~\chi(\xi,\eta).
\ee
Denoting the integral over the trapezoid $RAMN$ by $\widetilde{\mathcal{I}}_6$, we get
\begin{align}\label{eq:6.31}
\widetilde{\mathcal{I}}_6~= &\int_R^{r_0} d\xi \int_0^{\frac{x-\xi}{t}}d\eta ~\frak{V}(\xi,\eta)\nonumber \\ 
=&-\int_R^{r_0} d\xi \int_0^{\frac{x-\xi}{t}}d\eta~\frak{U}(\xi,\eta)
~ +~ (1-y)\int_R^{r_0} d\xi \int_0^{\frac{x-\xi}{t}}d\eta~\chi(\xi,\eta).
\end{align}
For the triangular subdomain $MEN$ we consider a sequence of contours in the lower half--plane. Denoting this integral by $\widetilde{\mathcal{I}}_7$, we get
\be\label{eq:6.32}
\widetilde{\mathcal{I}}_7~= \int_R^{r_0} d\xi \int_{\frac{x-\xi}{t}}^yd\eta ~\frak{V}(\xi,\eta) =-\int_R^{r_0} d\xi \int_{\frac{x-\xi}{t}}^y d\eta~\chi(\xi,\eta) ~\frak{U}(\xi,\eta).
\ee
Inserting $(~\widetilde{\mathcal{I}}_5 + \widetilde{\mathcal{I}}_6 + \widetilde{\mathcal{I}}_7)$ into the formula for $\widehat{\mathbf{I}}_2$ and using Lemma 6.1, we obtain
\begin{align} \label{eq:6.33}
\widehat{\mathbf{I}}_2 =& -2\pi i \int_0^{r_0} d\xi \int_0^y d\eta~~\frac{x - \eta t -\xi}{|x - \eta t -\eta|}~\frak{U}(\xi,\eta) -\pi i~(1-y)\int_0^{r_0} d\xi \int_0^y d\eta~\chi(\xi,\eta)\nonumber \\
&~~+2\pi i ~(1-y)\int_0^R d\xi \int_0^yd\eta~\chi(\xi,\eta) + 2\pi i ~(1-y)\int_R^{r_0} d\xi \int_0^{\frac{x-\xi}{t}}d\eta~\chi(\xi,\eta).
\end{align}

\noindent Before we move to evaluation of $\widehat{\mathbf{I}}_4$, let us check that formula (\ref{eq:6.33}) is consistent with the previous results for $\widehat{\mathbf{I}}_2$. Namely, let $t$ change in such a way that the straight line $BP$ moves towards the line $BD$. In this case, $R\rightarrow 0$ and formula (\ref{eq:6.33}) becomes
\[ \frak{G}(x,y,t) + 2\pi i~(1-y) \int_0^{r_0} d\xi \int_0^{\frac{x-\xi}{t}}d\eta ~ \chi(\xi,\eta)-\pi i~(1-y) \int_0^{r_0} d\xi \int_0^y \chi(\xi,\eta),\]
which coincides with formula (\ref{eq:6.25}).

The domain of integration for $\widehat{\mathbf{I}}_4$ (Fig.\ref{fig:BMNP}) is a union of $(i)$ a rectangular subdomain $NEGT$, $(ii)$ a trapezoidal subdomain $NTJP$, and $(iii)$ a triangular subdomain $NPD$. For the triangular subdomain, the approximating contours of integration can be closed in the upper half--plane and for the remaining two subdomains the contours of integration can be closed in the lower half--plane. For the integral over the triangle $NPD$ denoted by $\widetilde{\mathcal{I}}_8$ we obtain
 \begin{align} \label{eq:6.34}
\widetilde{\mathcal{I}}_8=&- \int_0^N d\xi \int_y^{\frac{x-\xi}{t}}d\eta ~\frak{V}(\xi,\eta) \nonumber \\
=&-\int_0^{N} d\xi \int_y^{\frac{x-\xi}{t}}d\eta~\frak{U}(\xi,\eta) -y \int_0^N d\xi \int_y^{\frac{x-\xi}{t}} d\eta ~\chi(\xi,\eta).
\end{align}
For the integral over the trapezoidal $NTHP$ denoted by $\widetilde{\mathcal{I}}_9$ we obtain
 \be \label{eq:6.35}
\widetilde{\mathcal{I}}_9=- \int_0^N d\xi \int_{\frac{x-\xi}{t}}^1d\eta~\frak{V}(\xi,\eta)
~=\int_0^N d\xi \int_{\frac{x-\xi}{t}}^1d\eta~\frak{U}(\xi,\eta).
\ee
For the integral over the rectangular $NEGT$ denoted by $\widetilde{\mathcal{I}}_{10}$ we obtain
 \be \label{eq:6.36}
\widetilde{\mathcal{I}}_{10} =- \int_N^{r_0} d\xi \int_y^1d\eta ~\frak{V}(\xi,\eta)
~=\int_N^{r_0} d\xi \int_y^1d\eta~\frak{U}(\xi,\eta).
\ee
Inserting $(~\widetilde{\mathcal{I}}_8 + \widetilde{\mathcal{I}}_9 + \widetilde{\mathcal{I}}_{10})$ into the formula for $\widehat{\mathbf{I}}_4$ and using Lemma 6.1, we obtain
\begin{align} \label{eq:6.37}
 \widehat{\mathbf{I}}_4 =& -2\pi i \int_0^{r_0} d\xi \int_y^1d\eta ~\frac{x-\eta t - \xi}{|x - \eta t -\xi|}\frak{U}(\xi,\eta) + \pi i y\int_0^{r_0} d\xi \int_y^1 d\eta~ \chi(\xi,\eta)  \nonumber \\
&~~ -2\pi i~ y \int_0^R d\xi \int_y^{\frac{x-\xi}{t}} d\eta~ \chi(\xi,\eta).
 \end{align}

\noindent Let us verify that formula (\ref{eq:6.37}) is consistent with the previous results obtained for $\widehat{\mathbf{I}}_4$. First, let $t$ change in such a way that the straight line $BP$ moves towards the line $BD$. In this case, $R\rightarrow 0$ and formula (\ref{eq:6.37}) coincides with formula (\ref{eq:6.16}). Second, let $t$ change in such a way that the straight line $BP$ moves towards the line $BF$. In this case, $R\rightarrow r_0$, and formula (\ref{eq:6.37}) coincides with formula (\ref{eq:6.29}). 

Finally, summing equations (\ref{eq:6.33}) and (\ref{eq:6.37}) together yields the result:
 \begin{align} \label{eq:6.38}
 \widehat{\mathbf{I}}_2 + \widehat{\mathbf{I}}_4 =& -2\pi i \int_0^{r_0} d\xi \int_0^1 d\eta~\frac{x-\eta t -\xi}{|x - \eta t -\xi|}~\frak{U}(\xi,\eta) +\pi i~ y \int_0^{r_0} d\xi \int_y^1 d\eta~\chi(\xi,\eta)\nonumber \\
&~~+ 2\pi i ~(1-y) \int_0^R d\xi \int_0^y d\eta~\chi(\xi,\eta) + 2\pi i~(1-y)\int_R^{r_0} d\xi \int_0^{\frac{x-\xi}{t}} d\eta~\chi(\xi,\eta)\nonumber \\
&~~-\pi i~(1-y) \int_0^{r_0} d\xi \int_0^y d\eta~\chi(\xi,\eta) - 2\pi i~ y \int_0^R d\xi \int_y^{\frac{x-\xi}{t}} d\eta~\chi(\xi,\eta).
\end{align}

\noindent Taking into account that the value of $R$ corresponds to the $\xi$-coordinate at the point of intersection of the straight lines $x - \eta t -\xi = 0$ and $\eta = y$, i.e. $R = x-yt$, we arrive at formula (\ref{eq:6.7}).

 Case B is shown.

\indent \textbf{Case C.} We evaluate the integrals $\widehat{\mathbf{I}}_2$ and $\widehat{\mathbf{I}}_4$ for $t = x-r_0$. The domains of integration for both $\widehat{\mathbf{I}}_2$ and $\widehat{\mathbf{I}}_4$ are located below the straight line $x - \eta t -\xi =0$. Therefore, for both integrals the approximating contours of integration can be closed in the upper half--plane. Evaluating $\widehat{\mathbf{I}}_2$ and $\widehat{\mathbf{I}}_4$ yields
\begin{align}
\frac{1}{2\pi i}~ \widehat {\mathbf{I}}_2 
=&~-~\int_0^{r_0} d\xi \int_0^y d\eta~\frak{U}(\xi,\eta)+\frac{(1-y)}{2} \int_0^{r_0} d\xi\int_0^y d\eta ~\chi(\xi,\eta), \label{eq:6.39}\\
 \frac{1}{2 \pi i} ~\widehat{\mathbf{I}}_4 =&~ - \int_0^{r_0} d\xi \int_y^1 d\eta~\frak{U}(\xi,\eta) ~-\frac{y}{2}\int_0^{r_0} d\xi \int_y^1 d\eta~\chi(\xi,\eta). \label{eq:6.40}
 \end{align}
Collecting (\ref{eq:6.39}) and (\ref{eq:6.40}), we obtain the following result:
 \begin{align}\label{eq:6.41}
 \widehat{\mathbf{I}}_2 + \widehat{\mathbf{I}}_4 =&~  \pi i ~(1-y) \int_0^{r_0} d\xi \int_0^y d\eta~ \chi(\xi, \eta)~ - \pi i ~y \int_0^{r_0} d\xi \int_y^1 d\eta~ \chi(\xi,\eta) \nonumber \\
 &~-2\pi i \int_0^{r_0} d\xi \int_0^1 d\eta~ \frac{x-\eta t -\xi}{|x-\eta t -\xi|}~ \frak{U}(\xi,\eta).
\end{align}

On the next step we show that the formula for $(\widehat{\mathbf{I}}_2 + \widehat{\mathbf{I}}_4)$ is valid for any $t$: $(x-r_0)/y>t\geq x-r_0$. Let $BS$ (the dashed line on Fig.\ref{fig:BQS}) be the straight line $x - \eta t - \xi = 0$ located between $BF$ and $BH$.
The domain of integration for $\widehat{\mathbf{I}}_2$ is positioned below the line $x - \eta t - \xi = 0$.  Therefore the result for $\widehat{\mathbf{I}}_2$ coincides with (\ref{eq:6.25}).

The domain of integration for $\widehat{\mathbf{I}}_4$ is a union of $(i)$ a rectangular subdomain $VSJD$, $(ii)$ a trapezoidal subdomain $EQSV$, and $(iii)$ a triangular subdomain $QGS$. For the subdomains $VSJD$ and $EQSV$, the approximating contours of integration can be closed in the upper half--plane, for the subdomian $QGS$ the contours of integration can be closed in the lower half--plane. Evaluating $\widehat{\mathbf{I}}_4$ over the rectangle $VSJD$, we obtain
\be \label{eq:6.42}
\widetilde{\mathcal{I}}_{11} =- \int_0^V d\xi \int_y^1d\eta~\frak{V}(\xi,\eta) 
=-\int_0^V d\xi \int_y^1d\eta~\frak{U}(\xi,\eta)-y\int_0^V d\xi \int_y^1 d\eta~\chi(\xi,\eta).
\ee

\noindent Evaluating $\widehat{\mathbf{I}}_4$ over the trapezoid $EQSV$, we obtain
\be \label{eq:6.43}
\widetilde{\mathcal{I}}_{12} \equiv-\int_V^{r_0} d\xi \int_y^{\frac{x-\xi}{t}}d\eta~\frak{V}(\xi,\eta) 
=-\int_V^{r_0} d\xi \int_y^{\frac{x-\xi}{t}}d\eta~\frak{U}(\xi,\eta)-y\int_V^{r_0} d\xi \int_y^{\frac{x-\xi}{t}} d\eta~\chi(\xi,\eta).
\ee

\noindent Evaluating $\widehat{\mathbf{I}}_4$ over the triangle $QGS$, we obtain
 \be \label{eq:6.44}
\widetilde{\mathcal{I}}_{13} =- \int_V^{r_0} d\xi \int_{\frac{x-\xi}{t}}^1d\eta~\frak{V}(\xi,\eta)
=\int_V^{r_0} d\xi \int_{\frac{x-\xi}{t}}^1d\eta~\frak{U}(\xi,\eta).
\ee
Inserting $(~\widetilde{\mathcal{I}}_{11} + \widetilde{\mathcal{I}}_{12} +\widetilde{\mathcal{I}}_{13})$ into the formula for $\widehat{\mathbf{I}}_4$ and using Lemma 6.1, we obtain
\begin{align}\label{eq:6.45}
 \widehat{\mathbf{I}}_4 = &-2\pi i \int_0^{r_0} d\xi \int_y^1d\eta~ ~\frac{x-\eta t - \xi}{|x - \eta t -\xi|} ~\frak{U}(\xi,\eta) \nonumber \\ 
 &~~-2\pi i~y\int_0^V d\xi \int_y^1 d\eta~\chi(\xi,\eta) -2\pi i~ y \int_V^{r_0} d\xi \int_y^{\frac{x-\xi}{t}} d\eta~\chi(\xi,\eta)\nonumber \\
 &~~+\pi i~y \int_0^{r_0}d\xi \int_y^1 d\eta~ \chi(\xi,\eta).
 \end{align}
 Finally, summing equations (\ref{eq:6.25}) and (\ref{eq:6.45}) together we obtain
 \begin{align} \label{eq:6.46}
 \widehat{\mathbf{I}}_2 + \widehat{\mathbf{I}}_4 =& ~\frak{G}(x,y,t)+i\pi~(1-y) \int_0^{r_0} d\xi \int_0^y d\eta~\chi(\xi,\eta) -2\pi i~y \int_0^{V} d\xi \int_y^1 d\eta~\chi(\xi,\eta)\nonumber \\
 &-2\pi i ~y \int_V^{r_0} d\xi \int _y^{\frac{x-\xi}{t}} d\eta~ \chi(\xi,\eta) + \pi i ~y\int_0^{r_0} d\xi \int_y^1 d\eta~ \chi(\xi,\eta). \end{align}
 The value of $V$ is defined as the $\xi$-coordinate of the point of intersection of the straight lines $x - \eta t -\xi = 0$ and $\eta =1$, i.e. $V = x-t$. 
 
 Case C is shown.

 The theorem is proven. \end{proof}

\section{Conclusion}
The results of the research are devoted to the problem of stability of the fluid flow moving in a channel with flexible walls and interacting with the walls. The walls of the vessel are subject to traveling waves. Experimental data shows that the energy of the flowing fluid can be transferred and consumed by the structure (the walls), inducing ``traveling wave flutter.'' The problem of stability of fluid structure interaction splits into two parts: $(i)$ stability of the fluid flow in the channel with harmonically moving walls and $(ii)$ stability of solid structure participating in the energy exchange with the flow. Stability of fluid flow, the main focus of the research, is obtained by solving the initial boundary value problem for the \textit{stream function}. The boundary conditions reflect the axisymmetric geometry of the flow, and the absence of relative movement between the near boundary flow and the structure (``no--slip'' condition). The closed form solution is derived and represented in the flow of an infinite functional series.
\newpage
\noindent \textbf{Acknowledgment.} Partial support by the National Science Foundation grant \#DMS--1810826 is highly appreciated by the first author.
\\~\\
\noindent \textbf{Statement of funding.} No funding was requested for this research.
\\~\\
\noindent \textbf{Statement on data availability.} The present paper is concerned with an important question on stability of a fluid flow interaction with flexible channel walls. This research has a purely analytical nature, i.e. it is devoted to finding a representation for a solution of the initial boundary--value problem generated by Navier-Stokes equations in the case of a specifically chosen geometry of the channel and properties of the boundary. Due to the very nature if this research, we have used many mathematical tools (like techniques from the area of ordinary and partial differential equations, boundary value problems, integral transformations, improper integration, evaluation of contour integrals on the complex plane by using the Residue theorem, etc.). The obtained results are theoretical. In our forthcoming work, we are \textit{planning} to compare analytical results with the experimental data. However, for the present work no data was necessary.
\\~\\
\noindent \textbf{Competing interests.} The authors have no competing interests.
\\~\\
\noindent \textbf{Authors' Contributions.} Both authors contributed about equal amount of time and efforts in all stages of the manuscript writing and preparing it for submission.

\small
\bibliographystyle{papercitations}
\bibliography{Blood_Flow_Paper_CITATIONS}

\begin{thebibliography}{10}

\bibitem{Aittokallio1999}
Aittokallio, T., Gyllenberg, M., and Polo, O., 1999, {A model of a snorer's
  upper airway,} \textit{Turku Cent. Comput. Sci. Tech. Rep.}

\bibitem{Amabili2002}
Amabili, M., Pellicano, F., and Pa{\"{i}}doussis, M., 1999, {Non-linear
  dynamics and stability of circular cylindrical shells conveying flowing
  fluid,} \textit{J. Sound Vib.,} \textbf{225}, p. 655--699.

\bibitem{Ashley1950}
Ashley, H. and Haviland, G., 1950, {Bending vibrations of a pipeline containing
  flowing fluid,} \textit{J. Appl. Mech.,} \textbf{17}, p. 229--232.

\bibitem{Auregan1995}
Aur{\'{e}}gan, Y. and Depollier, C., 1995, {Snoring: linear stability analysis
  and in-vitro experiments,} \textit{J. Sound Vib.,} \textbf{188}, p. 39--53.

\bibitem{Beck1995}
Beck, R., Odeh, M., Oliven, A., and Gavriely, N., 1995, {The acoustic
  properties of snores,} \textit{Eur. Respir. J.,} \textbf{8}, p. 2120--2128.

\bibitem{Garrad1986}
Carpenter, P.~W. and Garrad, A.~D., 1986, {The hydrodynamic stability of flow
  over {K}ramer-type compliant surfaces. Part 2: flow-induced surface
  instabilities,} \textit{J. Fluid Mech.,} \textbf{170}, p. 199--232.

\bibitem{Case1960}
Case, K.~M., 1960, {Stability of inviscid plane Couette flow,} \textit{Phys.
  Fluids,} \textbf{3}, p. 143--148.

\bibitem{Case1962}
Case, K.~M., 1962, {Hydrodynamic stability and the initial-value problems,}
  \textit{Proc. Symp. Appl. Math.,} \textbf{13}, p. 25--33.

\bibitem{Drazin1981}
Drazin, P. and Reid, W., 1981, \textit{{Hydrodynamic Stability,}} Cambridge
  Univ. Press, Cambridge.

\bibitem{Ellis1993}
Ellis, P., Williams, J., and Shneerson, J., 1993, {Surgical relief of snoring
  due to palatal flutter: a preliminary report,} \textit{Ann. R. Coll. Surg.
  Engl.,} \textbf{75}, p. 286--90.

\bibitem{Grotberg1989}
Grotberg, J.~B. and Gavriely, N., 1989, {Flutter in collapsible tubes: a
  theoretical model of wheezes,} \textit{J. Appl. Physiol.,} \textbf{66}, p.
  2262--2273.

\bibitem{Heil2003}
Heil, M. and Jensen, O.~E., 2003, {Flows in deformable tubes and channels,}
  \textit{Ann. R. Coll. Surg. Engl.,} \textbf{75}, p. 15--49.

\bibitem{Huang1998}
Huang, L., 1998, {Reversal of the {B}ernoulli effect and channel flutter,}
  \textit{J. Fluids Struct.,} \textbf{12}, p. 131--151.

\bibitem{Huang2001}
Huang, L., 2001, {Viscous flutter of a finite elastic membrane in {P}oiseuille
  Flow,} \textit{J. Fluids Struct.,} \textbf{15}, p. 1061--1088.

\bibitem{Huang1999}
Huang, L. and Williams, J. E.~F., 1999, {Neuromechanical interaction in human
  snoring and upper airway obstruction,} \textit{J. Appl. Physiol.,}
  \textbf{86}, p. 1759--1763.

\bibitem{Jeffrey2006}
Jeffrey, A., 2006, \textit{{Complex Analysis and Applications,}} 2nd ed.,
  Chapman and Hall/CRC, Boca Raton, London.

\bibitem{Jensen2003}
Jensen, O.~E. and Heil, M., 2003, {High-frequency self-excited oscillations in
  a collapsible-channel flow,} \textit{J. Fluid Mech.,} \textbf{481}, p.
  235--268.

\bibitem{Karagiozis2008}
Karagiozis, K., Pa{\"{i}}doussis, M., Amabili, M., and Misra, A., 2008,
  {Nonlinear stability of cylindrical shells subjected to axial flow: Theory
  and experiments,} \textit{J. Sound Vib.,} \textbf{309}, p. 637--676.

\bibitem{Kumaran1996}
Kumaran, V., 1996, {Stability of inviscid flow in a flexible tube,} \textit{J.
  Fluid Mech.,} \textbf{320}, p. 1--17.

\bibitem{Kumaran1998}
Kumaran, V., 1998, {Stability of the flow of a fluid through a flexible tube at
  intermediate {R}eynolds number,} \textit{J. Fluid Mech.,} \textbf{357}, p.
  123--140.

\bibitem{Larose1997}
Larose, P.~G. and Grotberg, J.~B., 1997, {Flutter and long-wave instabilities
  in compliant channels conveying developing flows,} \textit{J. Fluid Mech.,}
  \textbf{331}, p. 37--58.

\bibitem{Liu2007}
Liu, Z., Luo, X., Lee, H., and Lu, C., 2007, {Snoring source identification and
  snoring noise prediction,} \textit{J. Biomech.,} \textbf{40}, p. 861--870.

\bibitem{Paidoussis1999}
Pa{\"{i}}doussis, M., 1999, {Aspirating pipes do not flutter at infinitesimally
  small flow,} \textit{J. Fluids Struct.,} \textbf{13}, p. 419--425.

\bibitem{Paidoussis2005a}
Pa{\"{i}}doussis, M., 2005, {Some unresolved issues in fluid-structure
  interactions,} \textit{J. Fluids Struct.,} \textbf{20}, p. 871--890.

\bibitem{Paidoussis1972}
Pa{\"{i}}doussis, M. and Denise, J., 1972, {Flutter of thin cylindrical shells
  conveying fluid,} \textit{J. Sound Vib.,} \textbf{20}, p. 9--26.

\bibitem{Paidoussis1993}
Pa{\"{i}}doussis, M. and Li, G., 1993, {Pipes conveying fluid: a model
  dynamical problem,} \textit{J. Fluids Struct.,} \textbf{7}, p. 137--204.

\bibitem{Paidoussis1985}
Pa{\"{i}}doussis, M. and Luu, T.~P., 1985, {Dynamics of a pipe aspirating fluid
  such as might be used in ocean mining,} \textit{J. Energy Resour. Technol.,}
  \textbf{107}, p. 250--255.

\bibitem{Paidoussis2005}
Pa{\"{i}}doussis, M., Semler, C., and Wadham-Gagnon, M., 2005, {A reappraisal
  of why aspirating pipes do not flutter at infinitesimal flow,} \textit{J.
  Fluids Struct.,} \textbf{20}, p. 147--156.

\bibitem{Pevernagie2010}
Pevernagie, D., Aarts, R.~M., and {De Meyer}, M., 2010, {The acoustics of
  snoring,} \textit{Sleep Med. Rev.,} \textbf{14}, p. 131--144.

\bibitem{Roach1995}
Roach, G.~F., 1995, \textit{{Green's Functions,}} 2nd ed., Cambridge Univ.
  Press, Cambridge.

\bibitem{Saff2003}
Saff, E.~B. and Snyder, A.~D., 2003, \textit{{Fundamentals of Complex Analysis
  with Applications to Engineering and Science,}} 3rd ed., Pearson Hall, Upper
  Saddle River, NJ, USA.

\bibitem{Shivamoggi1982}
Shivamoggi, B., 1982, {Stability of inviscid plane couette flow,} \textit{Acta
  Mech.,} \textbf{44}, p. 327--329.

\bibitem{Stakgold1998}
Stakgold, I., 2011, \textit{{Green's Functions and Boundary Value Problems,}}
  John Wiley {\&} Sons, Inc., Hoboken, NJ, USA.

\bibitem{Wang2007}
Wang, J., Tetlow, G., and Lucey, A., 2007, {Flow-structure interaction in the
  upper airway: Motions of a cantilevered flexible plate in channel flow with
  flexible walls,} \textit{Proc. 16th Australasian. Fluid Mech. Conf.,} p.
  342--345.

\bibitem{Whittaker2010}
Whittaker, R.~J., Heil, M., Jensen, O.~E., and Waters, S.~L., 2010, {Predicting
  the onset of high-frequency self-excited oscillations in elastic-walled
  tubes,} \textit{Proc. R. Soc. A Math. Phys. Eng. Sci.,} \textbf{466}, p.
  3635--3657.

\end{thebibliography}

}
\end{document}